\documentclass[12pt, a4paper]{article}
\pdfoutput=1
\usepackage[table]{xcolor}
\usepackage{amsmath}
\usepackage{amsfonts}
\usepackage{amssymb}
\usepackage{graphicx, rotating}
\usepackage{epstopdf}
\usepackage{epsfig}
\usepackage{latexsym}
\usepackage{graphicx}
\usepackage{color}
\usepackage{amsmath,amssymb}
\usepackage{cite}
\usepackage{bbold}
\usepackage{slashed}
\usepackage{hyperref}
\hypersetup{colorlinks, citecolor=bluscuro, linkcolor=black, urlcolor=bluscuro}
\definecolor{rossos}{cmyk}{0,1,1,0.55}
\definecolor{bluscuro}{rgb}{0.15, 0.2, .85}
\definecolor{bluchiaro}{cmyk}{1,.3,0.,0.1}
\usepackage{cancel}
\def\hhref#1{\href{http://arxiv.org/abs/#1}{#1}} 
\definecolor{oucrimsonred}{rgb}{0.6, 0.0, 0.0}
\definecolor{persianblue}{rgb}{0.11, 0.22, 0.73}
\definecolor{forestgreen}{rgb}{0.13,0.35,0.13}
 \hypersetup{colorlinks, citecolor=oucrimsonred, linkcolor=persianblue, urlcolor=oucrimsonred}
 \hypersetup{colorlinks, citecolor=oucrimsonred, linkcolor=persianblue, urlcolor=oucrimsonred}

\newcommand{\eq}[1]{Eq.~(\ref{#1})}
\newcommand{\lag}{\mathcal{L}}

\newcommand{\opf}{\mathcal{F}}
\newcommand{\ops}{\mathcal{S}}

\newcommand{\nn}{\nonumber}

\newcommand{\be}{\begin{equation}}
\newcommand{\ee}{\end{equation}}
\newcommand{\bea}{\begin{eqnarray}}
\newcommand{\eea}{\end{eqnarray}}
\newcommand{\bc}{\begin{center}}
\newcommand{\ec}{\end{center}}

\usepackage{adjustbox}
\usepackage{array}
\usepackage{booktabs}
\usepackage{multirow}

\newcolumntype{R}[2]{%
    >{\adjustbox{angle=#1,lap=\width-(#2)}\bgroup}%
    l%
    <{\egroup}%
}

\newcommand{\TeV}{\,\mathrm{TeV}}
\newcommand{\GeV}{\,\mathrm{GeV}}

\def\lra#1{\overset{\text{\scriptsize$\leftrightarrow$}}{#1}}

\begin{document}


\vspace*{-2cm}
\begin{flushright}
DESY 16-123 \\
CERN-TH-2016-154 \\
\today
\end{flushright}

\begin{center}
\vspace*{15mm}

\vspace{1cm}
{\Large \bf 
 The Last Gasp of Dark Matter Effective Theory
} \\
\vspace{1.4cm}

\bigskip

{\bf Sebastian Bruggisser$^a$, Francesco~Riva$^{b}$, Alfredo Urbano$^{b}$}
 \\[5mm]

{\it $^a$ DESY, Notkestrasse 85, D-22607 Hamburg, Germany}\\[1mm]
{\it $^b$ CERN, Theoretical Physics Department, Geneva, Switzerland}\\[1mm]

\end{center}
\vspace*{10mm} 
\begin{abstract}\noindent\normalsize
We discuss an interesting class of models, based on strongly coupled Dark Matter~(DM), where sizable effects can be expected in LHC missing energy (MET) searches, compatibly with a large separation of scales. In this case, an effective field theory (EFT) is appropriate (and sometimes necessary) to describe the most relevant interactions at the LHC. 
The selection rules implied by the structure of the new strong dynamics shape the EFT in an unusual way, revealing the  importance of higher-derivative interactions previously ignored.
We compare indications from relic density and direct detection experiments with consistent LHC constraints, and asses the relative importance of the latter.
Our analysis provides an interesting and well-motivated scenario to model MET at the LHC in terms of a handful  of parameters.
\end{abstract}

\vspace*{3mm}


\pagebreak
\section{Motivation}
Most of the information we have on Dark Matter (DM) is about exclusions and constraints. In fact, beside evidence for its existence through the gravitational force, DM has not been observed through any other interaction. Yet, these constraints have refined  our perspective on the dark sector, excluding baryonic DM, neutrinos and the prototype elctroweak WIMP.
Collider experiments add to the list. In principle, they constitute an important part of the DM search program, because uncertain astrophysical parameters play here a negligible r\^ole. Unfortunately, however, the information from collider constraints is at times analyzed in a way that hinders a transparent physical interpretation, so that a clear picture of what we have learned from colliders, is still missing.
On the one hand an effective field theory (EFT)  analysis provides an interesting tool to present these  constraints in a rather model-independent fashion, in terms of a handful of relevant parameters~\cite{Cao:2009uw,Bai:2010hh,Goodman:2010yf}. On the other hand, it is well known that the large kinematic range accessible at the LHC complicates a consistent EFT analysis, so that specific or simplified models seem to be necessary.
Here we take  the point of view that neither choice is ideal, but that they rather offer different languages to test different classes of theories. Indeed, the EFT allows, within its realm of validity,  to test different broad UV hypotheses. Whenever these hypotheses cannot be tested consistently, simplified models~\cite{ArkaniHamed:2007fw,deSimone:2014pda,Boveia:2016mrp,DeSimone:2016fbz}  can be more useful.\footnote{Notice that the simplified models proposed so far are themselves EFTs~\cite{Kahlhoefer:2015bea,Englert:2016joy}, so that the distinction is not so sharp: here we use the mass of the lightest mediator as discriminant.}
The goal of this article is to sharpen our perspective on where this line has to be drawn. So, instead of exposing situations in which the EFT description is inappropriate (see already Ref.~\cite{Bai:2010hh} or the more recent  \cite{Abdallah:2014hon} and references therein for discussions on this issue), we will identify and study the relevant cases where the EFT is useful and necessary.

Simple arguments of power-counting, symmetries and selection rules allow to relate broad properties of the Beyond the Standard Model (BSM) DM sector to specific characteristics of the EFT, without committing entirely to specific models, yet capturing the most relevant features~\cite{Cohen:1997rt,Giudice:2007fh,Remedios}. This is enough to reveal that strongly coupled theories give sizable  signatures, observable in non-resonant processes, well below the threshold of resonant production of new (mediator) states (see e.g. Ref.~ \cite{Contino:2016jqw}). This is the perfect target for analyses of  LHC DM searches based on an EFT parametrization.

Strongly coupled models have certainly received comparatively little attention in this context, a few reasons that come to mind are the difficulty of performing perturbative calculations at strong coupling (see however~\cite{Kribs:2016cew} for a review of lattice techniques in this context), as well as the observations that both the SM and the DM sectors appear to be weakly coupled at low energies (for DM this is made explicit by the WIMP miracle). Finally, strong coupling might seem at odds with the fact that both the SM fields and the DM that is searched at colliders are inherently light, much lighter than the characteristic scale mediating the SM-DM interaction.
Interestingly, all these obstacles are naturally overcome in the presence of \emph{approximate symmetries}. A well-known example are the pions of QCD that, although inherently strongly coupled, they are light and allow at low energy for a perturbative weakly-coupled description through the chiral Lagrangian. This is a consequence of (non-linearly realized) approximate chiral symmetry.

In a companion paper~\cite{companion}, we discuss similar situations, characterized by approximate symmetries in the context of DM, and show that a generic and complete description of strongly interacting DM might involve effective interactions in the Lagrangian captured either by operators of dimension $D = 6$ or by operators of $D = 8$. The approximate symmetries being considered correspond in practice to the case where DM is a pseudo-Nambu--Goldstone Boson (PNGB) -- if DM is a scalar -- or to the case where DM is a composite fermion or a Goldstino -- if DM is a Dirac or Majorana fermion respectively.

Two important novelties characterize the discussion in Ref.~\cite{companion}. First of all is the use of power-counting arguments to estimate the size of coefficients in the EFT expansion, in relation with generic microscopic properties (such as couplings and symmetries). This creates a hierarchy between different coefficients that, in some cases, can go as far as overcoming the suppression associated with the EFT energy expansion: $D=8$ effects can dominate over $D=6$ ones. In particular, in well motivated scenarios, some high-derivative operators, that have not been considered previously in the literature, dominate the collider phenomenology, still within the validity of the EFT description (i.e. $D>8$ operators are irrelevant).

In this work, we study in detail different phenomenological aspects of these scenarios, and discuss the consequences of these power-counting rules for DM searches.
We compare direct detection (DD) experiments, expectations from the relic density (RD), with LHC constraints from searches of mono-jets and missing energy, and discuss  the reliability of our estimates. 
Indeed, we show explicitly how a strong coupling implies that the EFT description can be used consistently in the context of the LHC, thus avoiding the criticism related to LHC and the DM EFT~(see also~\cite{Racco:2015dxa}). 
On the other hand, we will discuss how, despite the EFT being suitable for LHC analyses, the presence of effects that have different energy-growing behavior 
 compromises the comparison with RD and DD experiments: fast growing effects will dominate at large LHC-energy, but they will be subdominant at small RD and DD energy. 
 
 In section~\ref{sec:BSMperspective} we review the effective Lagrangian for DM and its symmetry structure; in section~\ref{sec:analysis} we discuss, in turn, constraints from the LHC, the RD, and DD experiments. Appendix~\ref{app:Notation} contains an extensive comparison of our notation (based on Weyl spinors) with the traditional one based on Dirac spinors, as well as additional details of our computations; Appendix~\ref{sec:event-gen} describes the collider analysis.

\section{An EFT for Strongly Interacting DM}\label{sec:BSMperspective}

The EFT is typically associated with an expansion in inverse powers of $M$, the physical mass scale characteristic of a new sector, so that 
\begin{equation}\label{EFT}
{\cal L}=\sum_{i,D}c_i\frac{_D{\cal O}_i}{M^{D-4}}
\end{equation}
where $_D{\cal O}_i$ is a field operator, of mass-dimension $D$, involving light fields only, which in our case correspond to the SM fields and the DM. Here, $c_i$'s are the Wilson coefficients that scale as
\begin{equation}\label{scalingwithcoupling}
c_i\sim (\textrm{coupling})^{n_i-2},
\end{equation}
where $n_i$ is the number of fields in the operator $_D{\cal O}_i$. This behavior \eq{scalingwithcoupling} can be established unambiguously from a bottom-up perspective by restoring the appropriate dimensions in powers of~$\hbar\neq 1$ in the Lagrangian\footnote{It is easy to see that ${\cal L}\sim\hbar^1$ and fields scale as $\Phi\sim\hbar^{1/2}$, while couplings scale generically as $g_\Phi\sim\hbar^{-1/2}$, so that the genuinely dimensionless expansion parameter is $g_\Phi\Phi$.} \cite{Cohen:1997rt,Giudice:2007fh,Contino:2016jqw}. In practice, we are mostly interested in operators contributing directly to $SM+SM\to DM+DM$, most of which contain 4 fields only: in this case $c_i\sim(\textrm{coupling})^2$.

The expansion in \eq{EFT} is valid only if the condition
\begin{equation}\label{EFTval}
E/M\ll1
\end{equation}
is fulfilled, where $E$ is the relevant energy of the experiment. Now, the problem is that, from a low-energy perspective, we only have access to the combination $\delta\equiv c_iE^{D-4}/M^{D-4}$. So, for an experiment with a given sensitivity to $\delta$,  the EFT validity condition \eq{EFTval} is fulfilled  only for large enough $c_i\gg \delta$ or, in other words (cf. \eq{scalingwithcoupling}) only in theories with a \emph{large enough coupling}~\cite{Contino:2016jqw}. This fact, in combination with the bias that the DM sector be weakly coupled, has led great part of the DM community to distrust analyses based on DM EFT. 

In this article we discuss theories with large Wilson coefficients $c_i$, because the underlying dynamics is strong, so that sizable effects are compatible with the EFT assumption.
Discussing strongly coupled theories does not necessarily imply explicit calculations in explicit or simplified models: broad BSM assumptions can be captured by an adequate power counting, such as in \eq{scalingwithcoupling}, which allows to estimate the size of EFT Wilson coefficients. 
This estimate is discussed in a companion paper~\cite{companion}; here we recall the underlying assumptions:\begin{itemize}

\item[\emph{i)}]  The SM couples to a new sector, characterized by {\bf one mass scale $M\gg m_{\rm DM}$}, corresponding to the physical masses of resonances in this sector, and {\bf one coupling $g_*$} that characterizes the self-coupling of this sector and its coupling  to the SM and to DM.\footnote{ This simplistic picture with one BSM scale, one BSM coupling, and symmetries, is certainly reductive w.r.t. realistic scenarios, but it represents a good approximation even in systems with multiple scale/couplings, in the limit where one coupling/scale is much larger than the others.} 

\item[\emph{ii)}] The new sector respects the following (approximate or exact) SM symmetries, namely {\bf gauge}   $SU(3)_C\times SU(2)_L \times U(1)_Y$, {\bf CP, Flavour $U(1)^5$, custodial $SU(2)_L\times SU(2)_R$, Baryon $U(1)_B$} and {\bf individual Lepton number $U(1)_{L_{1,2,3}}$}. We will also assume that the new sector respects a DM stabilizing symmetry and that, if DM is a Dirac fermion, it respects DM {\bf chiral} symmetry.

\item[\emph{iii)}] In addition to these linearly realized symmetries (which are manifest even in the SM), spontaneous symmetry breaking can account for the existence of other symmetries in the broken phase, preserved by the new sector. At high energy such symmetries will be manifest.

The spontaneous breaking of a global symmetry $G$ to a subgroup $H\subset G$ delivers naturally light (P)NGBs, whose leading interactions are characterized by higher derivatives~\cite{Coleman:1969sm}. This is a consequence of {\bf nonlinearly realized $G/H$.}

In a similar way, the {\bf spontaneous breaking of supersymmetry} delivers a light (Majorana) fermion: the Goldstino. In the limit where all other supersymmetric particles are heavy, Goldstino interactions arise first at the level of $D=8$ operators and involve higher derivatives \cite{goldstino}.

\end{itemize}

These assumptions are respected by the new sector alone, but can be broken by the couplings of the new sector to either the SM or DM. For instance, the SM introduces more than one coupling (violating \emph{i)}) and the hypercharge coupling $g^\prime$ breaks custodial symmetry (violating \emph{ii)}).
Similarly, a DM mass breaks chiral symmetry if DM is a Dirac fermion (violating \emph{ii)}) and it breaks the non-linearly realized symmetries if it's a scalar or Majorana fermion (violating \emph{iii)}).

The way this  can be put into use is the following. We write the most general effective Lagrangian capturing interactions between the SM and DM (both in the fermion and in the scalar case), including operators up to $D=8$ (the reason for this will become clear a posteriori -- see also discussion in \cite{companion}). Then, interactions that preserve all the relevant symmetries can be thought as genuinely originating from the strong sector, and the power counting of their coefficients $c_i$ will genuinely follow \eq{scalingwithcoupling}, in terms of the strong coupling $c_i\simeq g_*^{n_i-2}$. On the other hand, terms that break one of the symmetries in \emph{ii), iii)} will pay a price proportional to the associated symmetry breaking parameter. For instance, for scalars of mass-squared $m^2$, terms that break the associated non linearly-realized $G/H$ symmetry will be proportional to $m^2/M^2$, while for  Dirac (Majorana) fermions, terms that break chiral symmetry (non-linearly realized SUSY), will be suppressed by $m/M$.
This is the crucial ingredient for our {\bf high-energy/strong-coupling, low-energy/weak-coupling} dichotomy: terms that preserve the symmetries have more derivatives and grow at high-energy, while they become small at low-energy, where symmetry breaking effects (which have less energy dependence) can be comparable.
\begin{table}[htp]
\begin{center}
\begin{tabular}{|c|c|c||c|c|c|}
\hline\hline
  \multicolumn{6}{|c|}{\textbf{Fermionic dark matter}}   \\
  \hline
    \multicolumn{3}{|c||}{\textbf{Unsuppressed}} & \multicolumn{3}{c|}{\textbf{Suppressed}}    \\
  \hline
  {\color{blue}{Name}}  &  {\color{blue}{Operator}} & {\color{blue}{Wilson coeff.}} & {\color{blue}{Name}} & {\color{blue}{Operator}} & {\color{blue}{Wilson coeff.}}  \\
\hline\hline
 $_6\mathcal{F}_{\psi}^V$ & $\chi^{\dag}\bar{\sigma}^{\mu}\chi \psi^{\dag}\bar{\sigma}_{\mu}\psi$ & $c_{\psi}^V g_*^2/M^2$ &  \multicolumn{3}{c}{} \\ \hline
 \multicolumn{3}{c||}{} &  $_6\mathcal{F}_{H}^S$ & $\chi\chi H^{\dag}H$ & $c_H^S y_t^2m_{\chi}/M^2$  \\
  \cline{4-6}
 \multicolumn{3}{c||}{} & $_6\mathcal{F}_B^{dip}$  & $\chi\sigma^{\mu\nu}\chi B_{\mu\nu}$ & $c_B^{dip}g_*m_{\chi}/M^2$ \\
\cmidrule{1-6}\morecmidrules\cmidrule{1-6}
 $_8\mathcal{F}_{\psi}^V$ & $\chi^\dagger\bar\sigma^\mu\partial^{\nu}\chi \psi^\dagger\bar\sigma_\mu D_\nu\psi$  & $C_{\psi}g_*^2/M^4$ & $_8\mathcal{F}_{\psi}^{\slashed{s}}$ & $\chi\chi\psi\psi H$ & $C_{\psi}^{\slashed{s}}g_*^2y_{\psi}m_{\chi}/M^4$ \\
 \hline
  $_8\mathcal{F}_{\psi}^{V\,\prime}$ & $\chi^\dagger\bar\sigma^\mu \chi D_{\nu}\psi^\dagger \bar\sigma_\mu D^\nu\psi$   & $C^{\prime}_{\psi}g_*^2/M^4$ & $_8\mathcal{F}_{\psi}^{\slashed{s}\,\prime}$ & $\chi\psi\psi\chi H$ & $C_{\psi}^{\slashed{s}\,\prime}g_*^2y_{\psi}m_{\chi}/M^4$  \\
  \hline
$_8\mathcal{F}_V$ & $\chi^\dagger\bar\sigma^{\mu}\partial^{\nu}\chi V_{\mu\rho}^aV^{a\,\rho}_\nu$ & $C_Vg_*^2/M^4$ & $_8\mathcal{F}_V^{\slashed{s}}$ & $\chi\chi V_{\mu\nu}^aV^{a\,\mu\nu}$ & $C_V^{\slashed{s}}g_*^2m_{\chi}/M^4$  \\
\hline
 $_8\mathcal{F}_H$ & $\chi^{\dag}\bar{\sigma}^{\mu}\partial^{\nu}\chi D_{\mu}H^{\dag}D_{\nu}H$ & $C_Hg_*^2/M^4$ & \multicolumn{3}{c}{} \\
  \cline{1-3}
\end{tabular}
\end{center}
\caption{{\it Effective operators characterizing $DM-SM$ interactions at $D = 6$ and specific to $2\to2$ processes for $D = 8$, and the largest possible coefficients allowed by our power-counting rules, as discussed in Ref.~\cite{companion}. Operator nomenclature as follows: the subscript denotes what particles DM couples to and the supscript refers to the particular properties of the operator, $\not \!s$ denotes  symmetry breaking effects while $S,V,T$ betray the structure of a scalar, vector or tensor mediator respectively,  $^{dip}$ for dipole-type operators. In the text $c_i$($C_i$) is the Wilson coefficient of the $D = 6$ ($D = 8$) operator $_6\opf_i$ ($_8\opf_i$).}}
\label{tab:FermionicDM}
\end{table}
\begin{table}[htp]
\begin{center}
\begin{tabular}{|c|c|c||c|c|c|}
\hline\hline
  \multicolumn{6}{|c|}{\textbf{Scalar dark matter}}   \\
    \hline
    \multicolumn{3}{|c||}{\textbf{Unsuppressed}} & \multicolumn{3}{c|}{\textbf{Suppressed}}    \\
  \hline
  {\color{blue}{Name}}  &  {\color{blue}{Operator}} & {\color{blue}{Wilson coeff.}} & {\color{blue}{Name}} & {\color{blue}{Operator}} & {\color{blue}{Wilson coeff.}}  \\
\hline\hline
 $_6\mathcal{S}_{\psi}^V$ & $\phi^{\dag}\lra{\partial}_{\mu}\phi \psi^{\dag}\bar{\sigma}^{\mu}\psi$ & $c_{\psi}^Vg_*^2/M^2$ & $_6\mathcal{S}_{\psi}^{\slashed{s}}$ & $|\phi|^2\psi\psi H$ & $c_{\psi}^{\slashed{s}}g_*^2 y_{\psi}/M^2$ \\ 
  \hline
 $_6\mathcal{S}_{H}^S$ & $\partial_{\mu}\phi^{\dag}\partial^{\mu}\phi |H|^2$ & $c_H^Sg_*^2/M^2$ &  $_6\mathcal{S}_H^{\slashed{s}}$ & $|\phi|^2|H|^2$ & 
 $c_H^{\slashed{s}}g_*^2m_{\phi, H}^2/M^2$ \\  \hline
 $_6\mathcal{S}_B^{dip}$ & $\partial_{\mu}\phi^{\dag}\partial_{\nu}\phi B^{\mu\nu}$ & $c_B^{dip}g_*/M^2$ &  \multicolumn{3}{c}{} \\ 
\cmidrule{1-3}\morecmidrules\cmidrule{1-3}
 $_8\mathcal{S}_{\psi}^T$ & $\partial^{\mu}\phi^{\dag}\partial^{\nu}\phi \psi^{\dag}\bar{\sigma}_{\mu}D_{\nu}\psi$ & $C_{\psi}^T g_*^2/M^4$  & \multicolumn{3}{c}{} \\ \hline
 \multicolumn{3}{c||}{} & $_8\mathcal{S}_{\psi}^{S}$ & $\partial^{\mu}\phi^{\dag}\partial_{\mu}\phi \psi\psi H$ & $C_{\psi}^Sg_*^2 y_{\psi}/M^4$ \\
 \hline
 $_8\mathcal{S}_V^S$  & $\partial^{\mu}\phi^{\dag}\partial_{\mu}\phi V_{\rho\nu}^a V^{a\,\rho\nu}$ & $C_V^Sg_*^2/M^4$ & $_8\mathcal{S}_V^{\slashed{s}}$ & $|\phi|^2V_{\mu\nu}^a V^{a\,\mu\nu}$ & $C_V^{\slashed{s}}g_*^2m_{\phi}^2/M^4$ \\ 
 \hline
$_8\mathcal{S}_V^T$ & $\partial^{\mu}\phi^{\dag}\partial^{\nu}\phi V^a_{\mu\rho}V_{\nu}^{a\,\rho}$ & $C_V^Tg_*^2/M^4$ &  \multicolumn{3}{c}{} \\  \cline{1-3}
$_8\mathcal{S}_H^S$ & $\partial^{\mu}\phi^{\dag}\partial_{\mu}\phi D^{\nu}H^{\dag}D_{\nu}H$ & $C_H^S g_*^2/M^4$ &  \multicolumn{3}{c}{} \\  \cline{1-3}
$_8\mathcal{S}_H^T$ & $\partial^{\mu}\phi^{\dag}\partial^{\nu}\phi D_{\{\mu}H^{\dag}D_{\nu\}}H$ & $C_H^Tg_*^2/M^4$ &  \multicolumn{3}{c}{} \\  \cline{1-3} 
\end{tabular}
\end{center}
\caption{{\it Same as table~\ref{tab:FermionicDM} but for scalar DM, with operators denoted $\ops$.
 }}
\label{tab:ScalarDM}
\end{table}

We summarize the results (explained in more detail in Ref.~\cite{companion}), in table~\ref{tab:FermionicDM} when DM is a fermion and in table~\ref{tab:ScalarDM} when it is a scalar, separating suppressed and unsuppressed effects, and using a notation based on Weyl spinors, where the coefficients $c_i$ are matrices mixing different chiralities (Appendix~\ref{app:Notation} is dedicated to a comparison with Dirac notation). Tables~\ref{tab:FermionicDM} and~\ref{tab:ScalarDM}  report the maximal possible value of the Wilson coefficient, where $c,C\simeq O(1)$ but, depending on the particular choice of $G/H$ if DM is a scalar, and depending on whether it is Majorana or Dirac DM if it is a fermion, some terms might be forbidden $c,C=0$. The most interesting cases are~\cite{companion}:
\begin{itemize}

\item If DM is a Goldstino~(see e.g. \cite{Cheung:2010mc} and the discussion in \cite{companion}), $c_\psi^V=0$, and the only genuinely strong interactions are at $D = 8$~\cite{goldstino,Clark:1997aa,Sasha}.

\item If DM is a (real) scalar from an abelian SSB pattern $U(1)/{\cal Z}_2$~\cite{Weinberg:2013kea,Kilic:2009mi,Antipin:2014qva}, then the $D = 6$ Lagrangian is penalized: $c_\psi^V=c_H^S=c_B^{dip}=0$ vanish, and $c_{\psi}^{\slashed{s}}$, $c_{H}^{\slashed{s}}$ are further suppressed by $m_\phi^2/M^2$.

\item \label{itemist}For scalar DM, $c_\psi^V$, $c_B^{dip}$ can be non-vanishing only if (complex) DM originates from a non-abelian SSB, such as $SU(2)/U(1)$~\cite{Belyaev:2010kp,Hietanen:2013fya,Bhattacharya:2013kma,Carmona:2015haa,Kribs:2016cew}. On the other hand  $c_H^S$ $c_{\psi}^{\slashed{s}}$, $c_{H}^{\slashed{s}}$ are further suppressed, unless the generators associated with DM and those associated with the Higgs do not commute, such as in $SO(6)/SO(5)$ \cite{Gripaios:2009pe,Frigerio:2012uc,Marzocca:2014msa} or larger \cite{Chala:2012af}. 

\item For simplicity, tables~\ref{tab:FermionicDM} and~\ref{tab:ScalarDM} make the optimistic assumptions that SM fermions are fully composite; in case they are only partially composite~\cite{Kaplan:1991dc,Agashe:2004rs}, operators involving SM fermions $\psi$, are suppressed by the degree of compositeness $\xi_\psi^2$.\footnote{\label{fn1}In fact in composite Higgs models based on partial compositeness, a favorable situation is when only the right-handed top quark is fully composite \cite{Pomarol:2012qf,Pomarol:2008bh}, while the mixing of lighter fermions to the strong sector is suppressed; see Ref.~\cite{Haisch:2015ioa}.}

\item Finally, operators involving gauge-boson field strengths $V_{\mu\nu}$ can be sizable only if the SM gauge boson realize the paradigm of deformed symmetry -- \emph{Remedios} -- of Ref.~\cite{Remedios}. If instead the transverse polarizations of gauge bosons are elementary, operators involving two field strengths will be suppressed by $\sim g_V^2/g_*^2$ (and additionally by $\sim g_*^2/16\pi^2$ if the underlying theory is minimally coupled \cite{Giudice:2007fh}).
\end{itemize}

A few further remarks are necessary. First of all, for generic Wilson coefficients $c_i$, tables~\ref{tab:FermionicDM} and~\ref{tab:ScalarDM} include all $D = 6$ (or smaller, depending on notation) operators connecting the SM with a pair of DM fields; partial integration, field redefinitions (that eliminate operators proportional to the leading equation of motion), Bianchi and Fiertz identities, have been used to focus on a smaller set of operators, in agreement with previous literature~\cite{Duch:2014xda,Foadi:2008qv}. 
The same is true for operators classified as $D = 8$, although in this case we have focussed on direct contributions to $2\to2$ scatterings, which typically constitute the most favorable processes to test SM-DM interactions~ (see Ref.~\cite{companion} for more details on these criteria). If the gauge vectors are composite, however, the  additional unsuppressed structures
\begin{align}
_8\lag_{\rm eff}^{DM} &= 
C^{mono}_{\psi}\frac{g_*^3}{M^4} \chi^\dagger\bar\sigma_\mu\chi \,\psi^\dagger\bar\sigma_\nu T^a\psi V^{a\,\mu\nu}
+C^{mono}_{H}\frac{g_*^3}{M^4}\chi^\dagger\bar\sigma_\mu\chi \,H^\dagger\lra {D_\nu}\tau^a H\,\,W^{a\,\mu\nu}
\label{ferm8jet}
\end{align}
for fermion DM, and
\begin{align}
_8\lag_{\rm eff}^{DM} &= 
C^{mono}_{\psi}\frac{g_*^3}{M^4} \phi^\dagger\lra \partial_\mu\phi \,\psi^\dagger\bar\sigma_\nu T^a\psi V^{a\,\mu\nu}
+C^{mono}_{H}\frac{g_*^3}{M^4}\phi^\dagger\lra \partial_\mu\phi \,H^\dagger\lra {D_\mu}\tau^a HW^{a\,\mu\nu}\,.
\label{scal8jet}
\end{align}
for scalar DM,
contribute directly to $SM+SM\to DM+DM+SM$ and could play a r\^ole in DM searches at the LHC, as we will discuss in the next section.

Secondly, our construction assumes that the SM itself emerges from the new strongly interacting sector, along the lines of Refs.~\cite{Giudice:2007fh,Remedios}, a possibility that has only received a partially rigorous phenomenological treatment. Leaving a detailed analysis of this for future work, we envisage here that a small suppression of the SM couplings to the new sector, and an enhancement of the DM ones, might suppress effects in $SM\to SM$ processes in favor of $SM\to DM$ ones (see \cite{Chala:2015ama} for a thorough discussion of this in explicit $Z^\prime$ models)

Finally, the reader might wonder why we are not discussing (linearly realized) supersymmetry, as a \emph{raison d'\^etre } for naturally light scalars. The main reason is that our working hypothesis includes in the light spectrum only the SM fields and DM. A spectrum with only a light scalar DM ($m_\phi\sim$ few GeV) and multi-TeV fermionic partners implies however a precise cancellation between the SUSY-preserving and SUSY-breaking mass terms: a tuning analogous in spirit to the $\mu$-$B_\mu$ problem in the MSSM with a multi-TeV Higgsino. 
We believe therefore that this scenario might be a better target for SUSY simplified models, with a complete chiral multiplet in the light spectrum, rather than our EFT description.\footnote{Moreover, since SUSY does not commute with custodial symmetry, it is unlikely that the Higgs  takes part in strong supersymmetric dynamics in the few TeV range.}

\section{Comparison with Experiments}\label{sec:analysis}
In this section we discuss the implications of our arguments  for DM searches, focussing first on collider experiments (which motivates our analysis in the first place) and turning to DD and the RD below. A global perspective on the results is given in section~\ref{sec:results}, to which the reader can skip if not interested in a schematic discussion of the analysis (details are postponed to Appendices~\ref{app:Notation} and~\ref{sec:event-gen}).
We focus on LHC processes involving light quarks and gluons, which provide at present the best sensitivity in cases where the DM couples to colored particles.
Electroweak-DM processes were instead studied  in Refs.~\cite{Cotta:2012nj,Davidson:2014eia,Crivellin:2015wva,Brooke:2016vlw} in the EFT framework. It will be interesting to extend this to our strongly coupled perspective, that might betray a well-motivated relation between the DM  sector and the mechanism of electroweak symmetry breaking. 
It is plausible that a large DM-Higgs coupling might have more important effects than a weak DM coupling to fermions, and overcome the prejudice that QCD processes have better sensitivity.
An additional example is provided by the operators $_6\ops_H^S$ and $_6\mathcal{S}_{\psi,H}^{\slashed{s}}$ which, as studied in \cite{Frigerio:2012uc},  reveal the importance of symmetry breaking effects for RD and DD computations. We leave a thorough analysis of this to future work.

\subsection{Monojet at LHC}\label{sec:LHC}

At the beginning of section \ref{sec:BSMperspective} we have highlighted the difficulties of ensuring that DM analyses are consistent with the EFT assumption. These difficulties were of two kind: the conceptual need of relying on BSM assumptions to discuss EFT validity (we have addressed this point above), and the technical need of performing DM searches in a way that keeps track of the information about the relevant energy of the process. 
We do this using a procedure in which  signal samples are repeatedly generated and analyzed  with different upper cuts $E=\sqrt{\hat s}<M_{cut}^i$ in the center-of-mass energy of the process (see Refs.~\cite{Biekoetter:2014jwa,Contino:2016jqw} but in particular Ref.~\cite{Racco:2015dxa} that discusses this in the context of DM).\footnote{Ref.~\cite{Busoni:2013lha} proposed a similar method to estimate a posteriori whether a given measurement based on the EFT parametrization is reliable.
Moreover, as noted already in the literature, there are different energy scales that can be associated with the hard process. Here we chose the largest, which is $\hat s$ and leads to the most conservative analysis. Other choices (e.g. $t$), might give stronger constraints, but always imply specific assumptions about the UV physics (for instance \cite{Papucci:2014iwa}), which are not necessarily met in our generic analysis. }
For each cut $M_{cut}^i$, a signal $\sigma^{ref}$ is generated for a given operator for reference values of the parameters $M=1$ TeV, $g_*=1$, etc , and then rescaled accordingly: for instance the signal of operator $_6\opf^V_{\psi}$ scales as $\sigma= g_*^4(1\textrm{TeV}/M)^4\sigma^{ref}$.
 
 In this way, for each $M_{cut}^i$, we can put  constraints on  EFTs which themselves satisfy $M\gtrsim M_{cut}^i$, in a consistent way. Here we chose to saturate $M= M_{cut}^i$ for illustration, although larger values can be used to increase the reliability of our constraints~\cite{Contino:2016jqw}.
This allows us to extract  for every operator $_D{\cal O}_i$ consistent constraints in the plane $(M,c_i)$ where $c_i$ is the Wilson coefficient, or similarly in the plane $(M,g_*)$, using the power-counting of $c(g_*)$ described in the previous section. 
The important aspect is that the constraints obtained with this additional cut are always conservative, independently of the specific UV completion, and also for $M$ well within the kinematic range accessible at LHC. This follows from the fact that the signal cross section in the complete theory $\sigma_{true}$ splits as $\sigma_{true}=\sigma_{true}(E>M_{cut})+\sigma_{true}(E<M_{cut})$, where $\sigma_{true}(E<M_{cut})\approx \sigma_{EFT}(E<M_{cut})$ is well approximated by the EFT, while $\sigma_{true}(E>M_{cut})$ always contributes positively to the cross section in a given kinematic region, so that 
\begin{equation}
\sigma_{true}>\sigma_{EFT}(E<M_{cut})\,,
\end{equation}
and the EFT is always conservative, see Ref.~\cite{Racco:2015dxa}.

We use this technique to recast the results of (multi- and) mono-jet searches in association with missing transverse energy at the LHC~\cite{Aad:2011xw,Aad:2015zva,Chatrchyan:2011nd}. In practice, we compare the data of Ref.~\cite{Aad:2015zva} with different signal simulations, analyzed  assuming different values of the cutoff~$M_{cut}$. 
Technical details of this recast study (which contains some novelties w.r.t. previous literature, due to the presence of multiple hard jets in the original analysis of Ref.~\cite{Aad:2015zva}), are discussed in Appendix \ref{sec:event-gen}. The results are shown in Figs.~\ref{fig:SNR}-\ref{fig:scalars}, and discussed in section~\ref{sec:results} below.

  \subsection{Relic density}\label{sec:RD}
Here we discuss the information that can be extracted on the new sector, from analyses of the RD, according to the standard freeze-out paradigm. In the early universe, DM particles are kept in thermal equilibrium through their interactions with the SM  thermal  bath, until the thermal kinetic energy of lighter particles drops below kinematic thresholds and the expansion of the Universe dilutes the number density
of heavier  particles  in such a way that their annihilation processes become less and less frequent. 
 Eventually, heavier particles freeze-out and their number density, no longer
altered  by  interaction  processes,  remains  constant.  
The physics of the freeze-out is well-known, and can be summarized by 
\begin{equation}\label{eq:freezeout}
\Omega_{\rm DM}h^2 \approx \frac{10^{-26}\,{\rm cm}^3/{\rm s}}{\langle \sigma v_{\rm rel}\rangle}\,,\quad \quad
\Omega_{\rm DM}h^2 |_{obs}= 0.1199\pm0.0027(68\%~\textrm{C.L.\cite{Ade:2013zuv}})
\end{equation}
where $\langle \sigma v_{\rm rel}\rangle$ is the thermally-averaged annihilation cross section times relative velocity $v_{\rm rel}$,
$\Omega_{\rm DM} \equiv \rho_{\rm DM}/\rho_{c}$ is  the  ratio  between  the  energy  density  of  DM  and  the  critical
energy density of the Universe, $h \equiv H_0/(100\,{\rm km}/{\rm s}/{\rm Mpc})$ is the reduced value of the present Hubble parameter~$H_0$.

The cross section can be written generically as $\langle \sigma v_{\rm rel}\rangle \sim \alpha_{\rm DM}^2/m_{\rm DM}^2$, where $\alpha_{\rm DM}$ has dimensions of a coupling and is evaluated at the relevant scale for freeze-out, given by the total 
energy in the center of mass frame,\footnote{Note that the validity of the EFT during the freeze-out epoch  requires $2m_{\rm DM}/\sqrt{1- v_{\rm rel}^2/4} < M$, a condition that is always fulfilled according to our construction.}
\begin{equation}\label{eq:EFTRelicDensity}
\sqrt{s} = \frac{2m_{\rm DM}}{\sqrt{1- v_{\rm rel}^2/4}}\approx 2 m_{\rm DM},
\end{equation}
where (at  freeze-out) $v_{\rm rel}\simeq 1/3$. Then \eq{eq:freezeout} reads
$
\Omega_{\rm DM}h^2 \approx 0.1 \times \left(
{0.01}/{\alpha_{\rm DM}}
\right)^2 \times \left(
{m_{\rm DM}}/{100\,{\rm GeV}}
\right)^2$,
an expression that lies at the heart of the so-called \emph{WIMP miracle} paradigm: a typical weak-scale coupling $\alpha_{\rm DM}\approx \alpha_{em}$ together with a DM mass of the order of the electroweak scale  correctly reproduces the observed abundance in \eq{eq:freezeout}. For our discussion it is important 
that the RD probes low-energy scales - as exemplified in \eq{eq:EFTRelicDensity} - where even the inherently-strong (but irrelevant) higher dimension interactions of the previous section appear to be weak~\cite{companion}. Indeed, a na\"{\i}ve estimate (to be refined later) for unsuppressed $D = 6$ operators (scaling as $\sim g_*^2/M^2$) implies that $\alpha_{\rm DM}^2\approx g_*^2 s/M^2$ and 
\begin{equation}\label{eq:freezeoutStrong}
\textrm{\bf D=6:}\quad\quad\Omega_{\rm DM}h^2 \approx 0.1 \times 
\left(
\frac{4\pi}{g_*}
\right)^4 \times \left(
\frac{5\,{\rm GeV}}{m_{\rm DM}}
\right)^2\times
\left(
\frac{M}{3\,{\rm TeV}}
\right)^4~.
\end{equation}
This  clearly shows that a strongly coupled light DM belonging to a new sector around the TeV scale 
fits  well the observed abundance. 

Given that at freeze-out $v_{\rm rel}\simeq 1/3$ is a relatively small number, the estimate in \eq{eq:freezeoutStrong} can be refined with a simple analysis of the non-relativistic limit of DM-DM scattering, searching for velocity-suppressed effects.
An expansion of the amplitude in partial waves implies
 $\sigma v_{\rm rel} = \sum_L a_Lv_{\rm rel}^{2L}=a_{L=0} + a_{L=1}v_{\rm rel}^2 + a_{L=2}v_{\rm rel}^4 +  \mathcal{O}(v_{\rm rel}^6)$, where the $L=0$, $L=1$ and $L=2$ terms define respectively  s-wave, p-wave and d-wave.\footnote{
 The correspondence between 
total orbital angular momentum $L$ and velocity $v_{\rm rel}$ can be understood considering the representation of a plane wave propagating with absolute momentum $|\vec{p}|$ in direction $\theta$ in terms of Legendre polynomials $P_L$ and spherical Bessel functions $j_L$,
\begin{equation}
\langle \vec{p},\theta |\vec{x} \rangle = e^{i|\vec{p}||\vec{x}|\cos\theta} = 
\sum_{L=0}^{\infty}i^L (2L+1)j_L\left(|\vec{p}||\vec{x}|\right)P_L(\cos\theta)~.
\end{equation}
The argument of $j_L$ is proportional to the velocity, and for small values $j_L(y) \approx {y^L}/{2^L\Gamma(1+L)} $, with $\Gamma(z)$ the Gamma function; hence the expansion in powers of $v_{\rm rel}^{2L}$ in the non-relativistic limit.
}
After thermal averaging at temperature $T$ one obtains,
\begin{equation}
\langle\sigma v_{\rm rel}  \rangle = a_{L=0} + 6\,a_{L=1}\frac{T}{m_{\rm DM}}+ 60\,a_{L=2}\frac{T^2}{m_{\rm DM}^2}+\cdots
\end{equation}
Since $T\approx m_{\rm DM}/25$, this implies that processes in p-wave are  suppressed to $\sim 25\,\%$ w.r.t. s-wave, while d-wave to $10\,\%$.

Whether or not an operator can mediate s-wave scattering, can be inferred from the $C$ and $P$ properties of the state it sources: $C=(-1)^{L+S}$, $S$ being the total spin, and $P=(-1)^{L+1}$ for fermions, $P=(-1)^{L}$ for bosons (see Ref.~\cite{Kumar:2013iva} for a review in the context of DM). 
In our basis only the following (non-relativistic, in Dirac notation\footnote{When discussing RD and DD, we use Dirac, rather than Weyl notation, in order to comply with the relevant literature on the subject; a detailed translation can be found in Appendix~\ref{app:Notation}.}) DM bilinear operators can source a state with the appropriate $C$ and $P$ quantum numbers corresponding to $L=0$:
\begin{align}\label{s-wave}
&\textrm{Fermionic DM:}\quad\bar{\pmb{\chi}}\gamma^i\pmb{\chi}\,,\quad \bar{\pmb{\chi}}\gamma^0\gamma^5\pmb{\chi}\,,\quad i\bar{\pmb{\chi}}\gamma^5\pmb{\chi}\,,\quad\quad\textrm{Scalar DM:}\quad \partial_\mu\phi^\dagger\partial^\mu\phi\,,\quad\phi^\dagger\phi\,.
\end{align}
Moreover, it is important to remark that the contributions from operators involving the SM bilinears $\bar\Psi\gamma^0\Psi$ vanish for charge conservation, while $\bar\Psi\gamma^5\gamma^0\Psi$ vanish in the massless quark limit (since in that case chiral symmetry is exact and the associated current is conserved). 
Finally, the traceless part of operators involving the tensor structures  $\bar{\pmb{\chi}}\gamma^\mu\partial^\nu\pmb{\chi}$ and $\partial^\mu\phi^\dagger\partial^\nu\phi$ has total angular momentum $J=S+L=2$ and implies that for scalars ($S=0$) the corresponding amplitudes are d-wave suppressed, while for fermions ($S=0,1$) they are at least p-wave suppressed.
These remarks are sufficient to keep track of p- and d-wave suppressions in DM annihilation and we summarize their impact in tables~\ref{tab:Fit} and \ref{tab:FitScalar}. 

We can now repeat the estimate of \eq{eq:freezeoutStrong} for  $D=8$ operators, using $\alpha_{\rm DM}\simeq g_*^2s^2/M^4$ and taking into account the fact that most of them suffer a p-wave suppression,
\begin{equation}\label{eq:freezeoutStrong2}
\textrm{\bf D = 8:}\quad\quad\Omega_{\rm DM}h^2 \approx 0.1 \times 
\left(
\frac{4\pi}{g_*}
\right)^4 \times \left(
\frac{60\,{\rm GeV}}{m_{\rm DM}}
\right)^6\times
\left(
\frac{M}{1\,{\rm TeV}}
\right)^8~.
\end{equation}
Interactions mediated uniquely by $D = 8$ operators are, as expected, rather week  during the freeze-out era and tend to overproduce DM unless $M$ is small enough or the DM is heavy enough. This  should however be taken with a grain of salt. First of all, as we will discuss later, as long as these symmetries are not exact,  symmetry breaking effects play an important r\^ole in explicit models. Secondly, information from the RD depends on the detailed cosmological history (e.g. late-time entropy releases can dilute DM 
overdensity) and should be taken as an indication to orient collider searches, rather than a constraint on the model.

In what follows, we provide detailed expressions of all the relevant cross sections.

\begin{table}[htp]
\begin{center}
\begin{tabular}{|c|c|c|c|c|}
\hline\hline
  \multicolumn{5}{|c|}{\textbf{Fermionic dark matter}}   \\
  \hline\hline 
\multicolumn{2}{|c|}{\textbf{DM-SM interaction}}  & \textbf{RD} & \textbf{DD} & \textbf{LHC}  \\  [1 pt] \cline{1-2}
 Effective operator & chiral decomposition  & $\sigma v_{\rm rel}$ & $\sigma_{\rm SI}$ & $\mathcal{A}$   \\ 
  \hline\hline 
    \multicolumn{5}{|c|}{{\color{blue}{Composite fermions}}}   \\
    \hline
 \multirow{4}{*}{$\frac{g_*^2}{M^2} \chi^\dagger\bar\sigma^\mu\chi \psi^\dagger\bar\sigma_\mu\psi$}  & $(\bar{\pmb{\chi}}\gamma^{\mu}\pmb{\chi})(\bar{\Psi}\gamma_{\mu}\Psi) $ & 
    \multirow{2}{*}{$g_*^4m_{\chi}^2/M^4$}  & $g_*^4\mu_N^2/M^4$ & \multirow{4}{*}{$\frac{g_*^2E^2}{M^2}$}  \\  [1 pt] \cline{2-2}\cline{4-4}
  & $(\bar{\pmb{\chi}}\gamma^{\mu}\pmb{\chi})(\bar{\Psi}\gamma_{\mu}\gamma^5\Psi) $ & & $\times$ & \\   [1 pt] \cline{2-3}\cline{4-4}
  & $(\bar{\pmb{\chi}}\gamma^{\mu}\gamma^5\pmb{\chi})(\bar{\Psi}\gamma_{\mu}\Psi)$  & \multirow{2}{*}{$g_*^4m_{\chi}^2v_{\rm rel}^2/M^4$} &  
  $\times$ &  \\  [1 pt] \cline{2-2}\cline{4-4}
    & $(\bar{\pmb{\chi}}\gamma^{\mu}\gamma^5\pmb{\chi})(\bar{\Psi}\gamma_{\mu}\gamma^5\Psi)$  & & $\times$ &  \\
      \hline\hline
          \multicolumn{5}{|c|}{{\color{blue}{Goldstino dark matter}}}   \\
          \hline
           \multirow{4}{*}{$\frac{g_*^2}{M^4}\chi^\dagger\bar\sigma^\mu\partial^{\nu}\chi \psi^\dagger\bar\sigma_\mu D_\nu\psi$}  &  
           $(\bar{\pmb{\chi}}\gamma^{\mu}\partial^{\nu}\pmb{\chi})
(\bar{\Psi}\gamma_{\mu}\lra{\partial_{\nu}}\Psi) $ & \multirow{4}{*}{$g_*^4m_{\chi}^6v_{\rm rel}^2/M^8$}   & $\frac{g_*^4 \mu_N^2m_{\chi}^2m_N^2}{M^8}$ & \\  [1 pt] \cline{2-2}\cline{4-4}
           &  $(\bar{\pmb{\chi}}\gamma^{\mu}\partial^{\nu}\pmb{\chi})
(\bar{\Psi}\gamma_{\mu}\gamma^5\lra{\partial_{\nu}}\Psi) $ &  & $\times$ & \multirow{5}{*}{$\frac{g_*^2E^4}{M^4}$}  \\  [1 pt] \cline{2-2}\cline{4-4}
           & $(\bar{\pmb{\chi}}\gamma^{\mu}\gamma^5\partial^{\nu}\pmb{\chi})
(\bar{\Psi}\gamma_{\mu}\partial_{\nu}\Psi + h.c.) $ &   & $\times$ &  \\ [1 pt] \cline{2-2}\cline{4-4}
           & $(\bar{\pmb{\chi}}\gamma^{\mu}\gamma^5\partial^{\nu}\pmb{\chi})
(\bar{\Psi}\gamma_{\mu}\gamma^5\partial_{\nu}\Psi + h.c.) $ & & $\times$ &  \\ [1 pt] \cline{1-4}
    \multirow{2}{*}{$\frac{g_*^2}{M^4}\chi^\dagger\bar\sigma^\mu \chi [(D_{\nu}\psi^\dagger)
           \bar\sigma_\mu (D^\nu\psi)]$}           & $(\bar{\pmb{\chi}}\gamma^{\mu}\gamma^5\pmb{\chi}) [(\partial^{\nu}\Psi)\gamma_{\mu}(\partial_{\nu}\Psi)]$ &  
           \multirow{2}{*}{$g_*^4m_{\chi}^6v_{\rm rel}^2/M^8$}   & $\times$  &  \\ [1 pt] \cline{2-2}\cline{4-4}
           & $(\bar{\pmb{\chi}}\gamma^{\mu}\gamma^5\pmb{\chi}) [(\partial^{\nu}\Psi)\gamma_{\mu}\gamma^5(\partial_{\nu}\Psi)]$ & & $\times$  &  \\ [1 pt] \cline{1-4}
       $\frac{g_*^2}{M^4}\chi^\dagger\bar\sigma^{\mu}\partial^{\nu}\chi V_{\mu\rho}^aV^{a\,\rho}_\nu$ & 
       $\bar{\pmb{\chi}}\gamma^{\mu}(\partial^{\nu}\pmb{\chi})V_{\mu\rho}^aV^{a\,\rho}_\nu$  & $g_*^4m_{\chi}^6v_{\rm rel}^2/M^8$ & $\frac{g_*^4 \mu_N^2m_{\chi}^2m_N^2}{M^8}$ & \\    
  \hline\hline           
          \multicolumn{5}{|c|}{{\color{blue}{Composite gluons}}}   \\
    \hline
      $\frac{g_*^2 m_\chi }{M^4}\chi\chi V_{\mu\nu}^aV^{a\,\mu\nu}$ & $\bar{\pmb{\chi}}\pmb{\chi} V_{\mu\nu}^aV^{a\,\mu\nu} $ & $g_*^4 m_{\chi}^6 v_{\rm rel}^2/M^8$ & 
      $\frac{g_*^4 \mu_N^2m_{\chi}^2m_N^2}{M^8}$ & $\frac{g_*^2m_{\chi}E^3}{M^4}$    \\   \hline
  \multirow{2}{*}{$\frac{g_*^2}{M^4}\chi^\dagger\bar\sigma^{\mu}\partial^{\nu}\chi V_{\mu\rho}^aV^{a\,\rho}_\nu$}  & $\bar{\pmb{\chi}}\gamma^{\mu}\lra{\partial^{\nu}}\pmb{\chi} V_{\mu\rho}^aV^{a\,\rho}_\nu$ &  
  $g_*^4 m_{\chi}^6 v_{\rm rel}^2/M^8$ & $\frac{g_*^4 \mu_N^2m_{\chi}^2m_N^2}{M^8}$ &  \multirow{2}{*}{$\frac{g_*^2E^4}{M^4}$}   \\  [1 pt] \cline{2-2}\cline{3-3}\cline{4-4}
    & $\bar{\pmb{\chi}}\gamma^{\mu}\gamma^5\lra{\partial^{\nu}}\pmb{\chi} V_{\mu\rho}^aV^{a\,\rho}_\nu$ & 0 & $\times$ &  \\  
  \hline
\end{tabular}
\end{center}
\caption{{\it 
Summary table in the case of fermionic DM.
We list the effective operators relevant for our phenomenological analysis, together with their chiral decomposition (see Appendix~\ref{app:Notation} for details).
For each operator, in the last three columns  we highlight 
the parametric dependence - as a function of DM mass $m_{\chi}$, coupling $g_*$, and effective scale $M$ - of the annihilation 
cross section times relative velocity $\sigma v_{\rm rel}$ (relevant for the computation of RD), 
 spin-independent (unsuppressed) DM-nucleon elastic scattering cross section $\sigma_{\rm SI}$ (relevant for DD),
and scattering amplitude for DM production at the LHC.
 }}
\label{tab:Fit}
\end{table}%
\begin{table}[htp]
\begin{center}
\begin{tabular}{|c|c|c|c|c|}
\hline\hline
  \multicolumn{5}{|c|}{\textbf{Scalar dark matter}}   \\
  \hline\hline 
\multicolumn{2}{|c|}{\textbf{DM-SM interaction}}  & \textbf{RD} & \textbf{DD} &  \textbf{LHC}  \\  [1 pt] \cline{1-2}
 Effective operator & chiral decomposition  & $\sigma v_{\rm rel}$ & $\sigma_{\rm SI}$ & $\mathcal{A}$  \\ 
  \hline\hline 
    \multicolumn{5}{|c|}{{\color{blue}{Complex scalar from $SU(2)\to U(1)$}}}   \\
    \hline
     \multirow{2}{*}{$c^V_{\psi}\frac{g_*^2}{M^2} \phi^\dagger\lra \partial_\mu\phi \,\psi^\dagger\bar\sigma^\mu\psi$}  & 
     $(\phi^\dagger \lra{\partial_{\mu}}\phi)\bar{\Psi}\gamma^{\mu}\gamma^5\Psi$ & \multirow{2}{*}{$g_*^4m_{\chi}^2v_{\rm rel}^2/M^4$}
   & $\times$ & \multirow{2}{*}{$\frac{g_*^2 E^2}{M^2}$} \\  [1 pt] \cline{2-2}\cline{4-4}
  & $(\phi^\dagger \lra{\partial_{\mu}}\phi)\bar{\Psi}\gamma^{\mu}\Psi$ &  & $g_*^4\mu_N^2/M^4$ & \\  
      \hline\hline
 \multicolumn{5}{|c|}{{\color{blue}{Real scalar from $U(1)\to \mathcal{Z}_2$}}}   \\
          \hline
      \multirow{2}{*}{$C^T_{\psi}\, \frac{g_*^2}{M^4}\partial^\mu\phi^\dagger\partial^\nu\phi \psi^\dagger\bar\sigma_\mu D_\nu\psi$}  & 
      $\partial^{\{\mu}\phi^\dagger\partial^{\nu\}}\phi (\bar{\Psi}\gamma_{\mu}\lra{D_{\nu}}\Psi)$
      & \multirow{2}{*}{$g_*^4m_{\chi}^6v_{\rm rel}^4/M^8$}
   & $\frac{g_*^4 \mu_N^2m_{\phi}^2m_N^2}{M^8}$  &  \multirow{4}{*}{$\frac{g_*^2 E^4}{M^4}$}  \\  [1 pt] \cline{2-2}\cline{4-4}
  & $\partial^{\{\mu}\phi^\dagger\partial^{\nu\}}\phi (\bar{\Psi}\gamma_{\mu}\gamma^5\lra{D_{\nu}}\Psi)$ &  & $\times$ & \\  [1 pt] \cline{1-4}   
        $C^S_{V}\frac{g_*^2}{M^4}\partial^\mu\phi^\dagger\partial_\mu\phi V_{\rho\nu}^aV^{a\,\rho\nu}$  &  -  &  $g_*^4m_{\chi}^6/M^8$  & 
        $\frac{g_*^4 \mu_N^2m_{\phi}^2m_N^2}{M^8}$  & \\  [1 pt] \cline{1-4}    
     $C^T_{V}\frac{g_*^2}{M^4}\partial^\mu\phi^\dagger\partial^\nu\phi V_{\mu\rho}^aV^{a\,\rho}_\nu$  &  -  &  $g_*^4m_{\chi}^6v_{\rm rel}^4/M^8$  & $\frac{g_*^4 \mu_N^2m_{\phi}^2m_N^2}{M^8}$  & \\ 
           \hline             
\end{tabular}
\end{center}
\caption{{\it 
Same as in Table~\ref{tab:Fit}, but in the case of scalar DM.
 }}
\label{tab:FitScalar}
\end{table}%

\subsubsection*{Fermionic DM}

\noindent
{\bf D=6.} 
At the leading order in the $1/M$ expansion we find only the operator $_{6}\mathcal{F}_{\psi}^{V}$, capturing DM interactions with light quarks. In the limit $m_{\psi} = 0$, in agreement with  the scaling in \eq{eq:freezeoutStrong}
\begin{align}\label{eq:D6FermionRelic}
&\left.\sigma v_{\rm rel}\right|_{[_{6}\mathcal{F}_{\psi}^{V}]_{AA}^{\mathcal{D}}} =  \left.\sigma v_{\rm rel}\right|_{[_{6}\mathcal{F}_{\psi}^{V}]_{AV}^{\mathcal{D}}} = 
 \frac{c^{V\,2}_{\psi} g_*^4 m_{\chi}^2 v_{\rm rel}^2}{2\pi M^4}\,,\nn\\
&\left.\sigma v_{\rm rel}\right|_{[_{6}\mathcal{F}_{\psi}^{V}]_{VV}^{\mathcal{D}}} =  \left.\sigma v_{\rm rel}\right|_{[_{6}\mathcal{F}_{\psi}^{V}]_{VA}^{\mathcal{D}}} = 
 \frac{c^{V\,2}_{\psi} g_*^4 m_{\chi}^2 (6 + v_{\rm rel}^2)}{2\pi M^4}\,,
\end{align}
where the different results correspond to different chiral structures ($AA$, $VV$, $AV$, $VA$) for the coefficients $c^{V}_{\psi}$, see Appendix~\ref{app:Notation}. 
In Eq.~(\ref{eq:D6FermionRelic}) we 
added to the operator name $[_{6}\mathcal{F}_{\psi}^{V}]$ two additional indices outside the square bracket.
The lower one refers to the chiral structure, the upper one to the nature - Dirac ($\mathcal{D}$) or Majorana ($\mathcal{M}$) - of the DM particle. 
The cross sections in Eq.~(\ref{eq:D6FermionRelic}) refer to  Dirac DM. 
 If DM is a Majorana particle, the vector bilinear $\bar{\pmb{\chi}}\gamma^{\mu}\pmb{\chi}$ identically vanishes since $\pmb{\chi}^C = \pmb{\chi}$: in this case
  the annihilation cross sections corresponding to the operators $[_{6}\mathcal{F}_{\psi}^{V}]_{VV}$ and $[_{6}\mathcal{F}_{\psi}^{V}]_{VA}$ vanish while that for $[_{6}\mathcal{F}_{\psi}^{V}]_{AV}$ and $[_{6}\mathcal{F}_{\psi}^{V}]_{AA}$ must be multiplied by a factor of $4$ since the number of diagrams in the scattering amplitude -- fermions being equivalent to anti-fermions -- doubles.\\

\noindent
{\bf D=8.} 
We consider the   $D = 8$  operators in table~\ref{tab:FermionicDM} mediating DM interactions with quarks and gluons. For DM interactions with gluons, $V=G$, the relevant operators are $_{8}\mathcal{F}_{G}^{\not \,s}$, which breaks chiral symmetry and is suppressed by the DM mass, and  $_{8}\mathcal{F}_{G}$, 
which comes in two different chiral structures ($V$ and $A$ -- we refer, again, to Appendix~\ref{app:Notation} for expressions that include the explicit chiral structure).
Note that in our analysis we do not include operator involving dual field strengths; as a consequence, 
CP invariance restricts the spinor contraction $\chi\chi$ in $_{8}\mathcal{F}_{G}^{\not \,s}$ to the chiral combination $[_{8}\mathcal{F}_{G}^{\not \,s}]_V$ without $\gamma^5$.
Furthermore, in $_{8}\mathcal{F}_{G}$ we consider the traceless combination $G^a_{\mu\rho}G_{\nu}^{a\,\rho} \to G^a_{\mu\rho}G_{\nu}^{a\,\rho} - 
\frac{g_{\mu\nu}}{4}G^a_{\rho\sigma}G^{a\,\rho\sigma}$ in order to single out a genuine tensor structure in the gluon field.
 We find
  \begin{align}\label{eq:RelicGluon1}
   \left.\sigma v_{\rm rel}\right|_{[_{8}\mathcal{F}_{G}^{\not \,s}]_V^{\mathcal{D}}} = \frac{16C_{V}^{\not \,s\,2}g_*^4 m_{\chi}^6 v_{\rm rel}^2}{\pi M^8}~,\quad\quad
    \left.\sigma v_{\rm rel}\right|_{[_{8}\mathcal{F}_{G}]_V^{\mathcal{D}}} = \frac{14 C_V^2 g_*^4 m_{\chi}^6 v_{\rm rel}^2}{3\pi M^8}~,
   \end{align}
while the axial combination
\begin{equation}\label{eq:RelicGluon2}
\left.\sigma v_{\rm rel}\right|_{[_{8}\mathcal{F}_{G}]_A^{\mathcal{D}}} =  \frac{2C_V^2 g_*^4m_{\chi}^6 v_{\rm rel}^4}{5\pi M^8}~,
\end{equation}
is d-wave suppressed as can be understood from our arguments below \eq{s-wave} (see for more details the discussion before \eq{referto} in the appendix).
As mentioned above, for Majorana fermions the cross sections are four times larger.
Notice that the  structure $\chi\chi V_{\mu\nu}^aV^{a\,\mu\nu}$ 
 exhibits the same functional dependence on DM mass and coupling if compared with
 the genuine $D=8$ operator $[_{8}\mathcal{F}_{G}]_V$. This is due to the fact that
the former has an explicit mass suppression due to the chiral breaking in the DM sector, while the derivative couplings in the latter pick up, when contracted with the corresponding external momenta, the DM mass. 

For DM interactions with fermions, we focus on the operators $_{8}\mathcal{F}^{V}_{\psi}$ and $_{8}\mathcal{F}^{V\,\prime}_{\psi}$, and neglect operators suppressed by the small SM Yukawas (we have checked that for the case of operators proportional to the top or bottom quark Yukawa, the LHC is not able to provide constraints consistent with the EFT expansion, even in the limit $g_*\sim 4\pi$, for this reason we will not discuss these operators in what follows). Moreover since the most interesting scenario to consider these operators is that in which the DM is a Goldstino, which is a Majorana fermion, it is worth in this case focussing on Majorana DM.
Then 
 \begin{equation}\label{eq:Goldstino3}
   \left.\sigma v_{\rm rel}\right|_{[_{8}\mathcal{F}^{V}_{\psi}]_{VV}^{\mathcal{M}}} =   \left.\sigma v_{\rm rel}\right|_{[_{8}\mathcal{F}^{V}_{\psi}]_{VA}^{\mathcal{M}}}  =
   2 \left.\sigma v_{\rm rel}\right|_{[_{8}\mathcal{F}^{V}_{\psi}]_{VV}^{\mathcal{M}}} = 
    2 \left.\sigma v_{\rm rel}\right|_{[_{8}\mathcal{F}^{V}_{\psi}]_{VA}^{\mathcal{M}}} =  \frac{8C_{\psi}^{2}g_*^4 m_{\chi}^6 v_{\rm rel}^2}{\pi M^8}~.
   \end{equation}
and
\begin{equation}
\left.\sigma v_{\rm rel}\right|_{[_{8}\mathcal{F}^{V\prime}_{\psi}]_{AV}^{\mathcal{M}}} =\left.\sigma v_{\rm rel}\right|_{[_{8}\mathcal{F}^{V\prime}_{\psi}]_{AA}^{\mathcal{M}}} = 
 \frac{32C_{\psi}^{\prime\,2}g_*^4 m_{\chi}^6 v_{\rm rel}^2}{\pi M^8}~.
\end{equation}

\subsubsection*{Scalar DM}

\noindent
{\bf D=6.} At the lowest order, interactions with SM quarks arise from the effective operator $_6{\cal S}^V_{\psi}$ in table~\ref{tab:ScalarDM}, 
which appears in two versions, depending on the SM fermion chiral structure (Eq.~(\ref{eq:ComplexScalarEff}) in Appendix~\ref{app:Notation}). We neglect the Yukawa-suppressed effect  $_6\mathcal{S}_\psi^{\slashed{s}}$ which (as we commented for the fermion case) cannot be accessed at the LHC; the important impact of this operator for the RD and DD experiments has been discussed in \cite{Frigerio:2012uc} in the context of the $SO(6)/SO(5)$ model.

From our arguments around \eq{s-wave}, the effects of $_6{\cal S}^V_{\psi}$ are always p-wave suppressed for DM-annihilation in the early Universe. Explicitly, 
 in the limit of vanishing fermion masses,
\begin{equation}\label{eq:D6ScalarRelic}
 \left.\sigma v_{\rm rel}\right|_{[_6{\cal S}^V_{\psi}]_A^{\mathcal{C}}} = 
 \left.\sigma 
 v_{\rm rel}\right|_{[_6{\cal S}^V_{\psi}]_V^{\mathcal{C}}} = \frac{c^{V\,2}_{\psi}g_*^4 m_{\phi}^2 v_{\rm rel}^2}{2\pi M^4 }\,.
\end{equation}
Note that the operator $_6{\cal S}^V_{\psi}$ is non-vanishing only for complex scalar DM.
In Eq.~(\ref{eq:D6ScalarRelic}), we added an extra upper index to distinguish between real ($\mathcal{R}$) and complex ($\mathcal{C}$) scalar DM. 

Finally, it is worth emphasizing that  $_6{\cal S}^V_{\psi}$ gives the same cross sections obtained for the operators $[_6{\cal F}^V_{\psi}]_{AA,AV}$: in the computation of the RD the only difference between scalar and fermionic case is the number of internal degrees of freedom, which equals to 4  for a Dirac fermion and 2 for a complex scalar. \\

\noindent
{\bf D=8.} We consider the   $D = 8$  operators in table~\ref{tab:ScalarDM} mediating DM interactions with quarks and gluons.
As far as interactions with gluons are concerned, we find again two structures, $_8\mathcal{S}_G^S$ and $_8\mathcal{S}_G^T$.
As before, in $_8\mathcal{S}_G^T$ the traceless combination $G^a_{\mu\rho}G_{\nu}^{a\,\rho} \to G^a_{\mu\rho}G_{\nu}^{a\,\rho} - 
\frac{g_{\mu\nu}}{4}G^a_{\rho\sigma}G^{a\,\rho\sigma}$ is understood.
We find 
\begin{equation}
\left.\sigma v_{\rm rel} \right|_{_8\mathcal{S}_{G}^S}  = \frac{C_G^{S\,2}g_*^4 m_{\phi}^6(4 + 5v_{\rm rel}^2)}{4\pi M^8}~,~~~~
\left.\sigma v_{\rm rel} \right|_{_8\mathcal{S}_{G}^T}  = \frac{2 C_G^{T\,2} g_*^4 m_{\phi}^6 v_{\rm rel}^4}{15\pi M^8}~.
\end{equation}
If compared with the fermionic case in Eq.~(\ref{eq:RelicGluon1}), note that the operator  $_8\mathcal{S}_G^S$, responsible for DM interactions with the scalar gluon current 
$G_{\mu\nu}^aG^{a\,\mu\nu}$, does not suffer from a p-wave suppression. This is a consequence of CP properties of the initial scalar state. The tensor structure $_8\mathcal{S}_{G}^T$, on the contrary, exhibits the expected d-wave suppression.

Interactions with SM fermions are described by the operator $_8\mathcal{S}_{\psi}^T$.
Assuming CP invariance, two  combinations are possible - V and A - depending on the chiral structure in the SM current, 
\begin{equation}\label{superspacial}
\left.\sigma v_{\rm rel} \right|_{[_8\mathcal{S}_{\psi}^T]_V}  =
\left.\sigma v_{\rm rel} \right|_{[_8\mathcal{S}_{\psi}^T]_A} =
  \frac{C_{\psi}^{T\,2} g_*^4 m_{\phi}^6 v_{\rm rel}^4}{10 \pi M^8}~.
  \end{equation}
 The annihilation cross sections are d-wave suppressed, as expected given the tensor nature of the SM current in the final state (see our arguments around \eq{s-wave} and Appendix~\ref{app:Notation}).
 
\subsection{Direct  Detection}

DD of DM can occur through elastic scattering between an incident DM
particle  and  a  nucleon $N$ in a nucleus at  rest  inside  the  detector. 
The nuclear recoil energy $E_{\rm NR} = \frac{\mu_N^2 v^2}{m_N}(1-\cos\theta)$ for which the experiment has maximum sensitivity is of order $E_{\rm NR} \approx 3\div25$ keV, where $v\equiv |\vec{v}|=|\vec{v}_{DM,in}- \vec{v}_{N,in}|$ is the relative incoming velocity 
between DM and nucleon, $\mu_N \equiv {m_{\rm DM} m_N}/({m_N + m_{\rm DM}})$ the reduced mass and $\theta$ the scattering angle. 
Then, assuming a Maxwellian DM velocity distribution with $v_0 = 220$ km/s, escape velocity $v_{esc} = 544$ km/s, and Earth average velocity $245$ km/s, gives a lower bound on the DM mass $m_{\chi} \gtrsim 8$ GeV that can be detected in these experiments. Given how stringent DD constraints can be, this motivates LHC searches for light DM particles that is the primary focus of this article.

The typical velocities $v\sim 10^{-3}$ are well within the realm of non-relativistic mechanics, where the kinematics of DM-nucleon scattering can be described by the following quantities: the DM and nucleon mass, $m_{\rm DM}$, $m_n \approx m_p\equiv m_N= 1$ GeV (and their combination, $\mu_N$); the vectors for velocity $\vec{v}$ and momentum transfer $\vec{q}$, and the (parity even) pseudovectors  for  nucleon and DM spin $\vec{s}_N$, $\vec{s}_{\chi}$, when DM is a fermion. As a matter of fact, amplitudes that are proportional to  $\vec{v}$ or $\vec{q}$ 
are suppressed and subleading  in  DD experiments, and we will neglect them.\footnote{See however Appendix~\ref{app:GoldstiniEffOp} for a discussion about their impact if compared with LHC searches.} On the other hand, amplitudes proportional to  $\vec{s}_N$ contribute only to spin-dependent detection experiments, whose constraints are reputedly weaker than spin-independent ones, and will also be ignored. 
Finally, the $SM+DM\to SM+DM$ amplitudes we are interested in, are at most linear in the DM spin $\vec{s}_{\chi}$, but rotational invariant combinations that include $\vec{s}_{\chi}$ necessary include a product with one of the above-cited vectors, and will hence also be neglected. 

We are therefore interested in amplitudes that depend at leading order only on the DM and nucleon masses and on none of the above-mentioned (pseudo) vectors.
Now,  all relativistic operators include, in the non-relativistic limit, a matrix element where the contribution from the DM bilinear transforms as a scalar (e.g. the time-component of $\bar\chi\sigma^\mu\chi$) and can potentially have unsuppressed interactions. It is easy to see, however, that all structures that in Dirac notation contain a $\gamma_5$ will be odd under parity. In terms of the above-mentioned quantities this can be associated only with the product of a vector ($\vec{v}$ or $\vec{q}$) and a pseudovector ($\vec{s}_N$ or $\vec{s}_{\chi}$) and is therefore always suppressed (see Refs.~\cite{Fan:2010gt,Fitzpatrick:2012ix} and~\cite{DelNobile:2013sia} for a review). For this reason DD experiments have their largest impact on structures without $\gamma_5$ and we shall limit our discussion to those.\footnote{
As discussed in Ref.~\cite{D'Eramo:2016atc}, renormalization group (RG) effects from the scale $M$ down  to  the  scale  at  which  the  experimental  measurements  are  performed - that is roughly given by $\mu_N$ in DD experiments - mix effective operators among themselves (in fact no symmetry, but only selection rules reflecting the UV particle content,  forbid the presence of a vector over an axial structure); for instance, the operator 
 $\bar{\pmb{\chi}}\gamma^{\mu}\pmb{\chi} \bar{\Psi}\gamma_{\mu}\gamma^5\Psi$ flows to $\bar{\pmb{\chi}}\gamma^{\mu}\pmb{\chi} \bar{\Psi}\gamma_{\mu}\Psi$ thus generating a sizable coupling with the vector quark current in the nucleon. In the strongly coupled models we consider here, we expect multiple resonances at the cut-off $M$ with different properties (similarly to QCD), so that in general all structure can be generated in the UV; for this reason we focus on the leading order contributions and neglect RG effects.
 }

 Then it is very easy to estimate the size of the unsuppressed contributions  to the  spin-independent cross section. For $D = 6$ operators we find generically
  \footnote{In the text we refer to the
computation of the DM-nucleon (rather than DM-nucleus) elastic cross section, since the experimental results in~\cite{LUX,Akerib:2016lao} are normalized w.r.t. this quantity.}
\begin{equation}\label{eq:D6SpinIndependent}
\textrm{\bf D=6:}\quad\quad\sigma_{\rm SI} \sim \frac{g_*^4\mu_N^2}{\pi M^4}\approx 
10^{-38}~{\rm cm}^2\times \left(\frac{g_*}{4\pi}\right)^4\times \left(\frac{3\,{\rm TeV}}{M}\right)^4~,
\end{equation}
up to numerical $O(1)$ factors that depend on the particular nucleon constituents the DM couples to.
Note that in Eq.~(\ref{eq:D6SpinIndependent}) the dependence from the DM mass cancels out in $\mu_N$ for $m_{\rm DM} \gg m_N \simeq 1$ GeV, a good approximation 
already for $m_{\rm DM} \gtrsim 10$ GeV.
  For relatively small DM masses above the threshold $m_{\rm DM}\gtrsim 8$ GeV, and $M$ in the multi-TeV regime,
   this estimate shows that the presence of a strong coupling is not compatible with the typical ballpark of cross sections tested by
    DD experiments~\cite{LUX}, which imply
\begin{equation}\label{eq:LUXestimate}
\sigma_{\rm SI}|_{obs}\lesssim 10^{-45}~{\rm cm}^2\,.
\end{equation}
This estimate strengthens the connection between light DM - possibly below the kinematic threshold for DD experiments - and strong coupling 
naturally provided by the framework of approximate symmetries~\cite{companion}.
A similar estimate for $D = 8$ operators, whose contribution to the amplitude is suppressed by an additional $m_{\rm DM}m_N/M^2$, gives
\begin{equation}\label{eq:D8SpinIndependent}
\textrm{\bf D=8:}\quad\quad\sigma_{\rm SI} \sim \frac{g_*^4\mu_N^2m_{\rm DM}^2m_N^2}{\pi M^8}\approx 
10^{-44}~{\rm cm}^2 \times \left(\frac{g_*}{4\pi}\right)^4 \times \left(\frac{m_{\rm DM}}{60~{\rm GeV}}\right)^2
\times \left(
\frac{1~{\rm TeV}}{M}
\right)^8~.
\end{equation}
Similarly to calculations for the RD, DD experiments have access to $D = 8$ operators only for rather small $M$ 
or large enough DM mass. 
We summarize these estimates in tables~\ref{tab:Fit} and \ref{tab:FitScalar} and in what follows we provide more refined expressions.
We parallel the sterile presentation from the previous section on RD, and include more details in Appendix~\ref{app:Notation}.
In particular, for completeness, in Appendix~\ref{app:GoldstiniEffOp} we compare mono-jet LHC searches and RD with 
 spin-dependent and suppressed spin-independent DD cross sections. 

\subsubsection*{Fermionic DM}

\noindent
{\bf D=6.}  For operator $_6\mathcal{F}^{V}_{\psi}$, only the  $VV$ structure - non vanishing in the case of Dirac DM - gives relevant constraints, with
\begin{equation}\label{eq:DominantSpinIndependent}
\left.\sigma_{\rm SI}\right|_{[_6\mathcal{F}^{V}_{\psi}]_{VV}^{\mathcal{D}}} = \frac{9g_*^4\mu_N^2}{\pi M^4}~,
\end{equation}
where we assumed equal coupling with up- and down-type quarks, $c_{u}^V = c_{d}^{V} = 1$.
Note that this cross section falls in the ballpark estimated in Eq.~(\ref{eq:D6SpinIndependent}).\\

\noindent
{\bf D=8.}  Similarly, for $_8\mathcal{F}_{\psi}^V$ we consider only the chiral structure that has unsuppressed effects in DD (without $\gamma_5$ in 
Dirac notation). In order to make contact with previous literature on the subject~\cite{Drees:1993bu}, we rewrite it as\footnote{Here we use the equivalence resulting from the equations of motion and add a mass-suppressed operator that is always suppressed at high-energy; at low-energy, however, it has an $O(1)$ effect on the result, captured by the last 3 terms in \eq{eq:GoldstinoDDXSection}.}
\begin{equation}\label{eq:Dim8FermionDD}
C_{\psi}\,\frac{g_*^2}{M^4}\left[\bar{\pmb{\chi}}\gamma^{\mu}(\partial^{\nu}\pmb{\chi}) \right]
\left(\bar{\Psi}\gamma_{\mu}\lra{\partial_{\nu}}\Psi\right)  = 
C_{\psi}\,\frac{g_*^2}{M^4}
\left\{
\mathcal{O}_{\mu\nu}^{\psi}\left[\bar{\pmb{\chi}}\gamma^{\mu}(\partial^{\nu}\pmb{\chi}) \right]
- \frac{1}{2}m_{\Psi}m_{\chi}\bar{\pmb{\chi}}\pmb{\chi}\bar{\Psi}\Psi
\right\}+O\left(\!\frac{1}{M^6}\!\right)
\end{equation}
where 
\begin{equation}
\mathcal{O}_{\mu\nu}^{\psi} \equiv \frac{1}{2}
\left[
\bar{\Psi}\gamma_{\mu}(\partial_{\nu}\Psi) 
-(\partial_{\nu}\Psi)\gamma_{\mu}\Psi
+\bar{\Psi}\gamma_{\nu}(\partial_{\mu}\Psi) - (\partial_{\mu}\bar{\Psi})\gamma_{\nu}\Psi + im_{\Psi}g_{\mu\nu}\bar{\Psi}\Psi
\right]~,
\end{equation}
defines the so-called quark twist-2 operator whose nuclear matrix element $\langle N| \mathcal{O}_{\mu\nu}^{\psi} |N \rangle$ is known and can be parametrized in terms of the second
moments $\psi_{(2)}$ of the parton distribution functions and the coefficients $f_{T_{q}}^{(N)} \equiv \frac{\langle N| m_{\Psi}\bar{\Psi}\Psi |N \rangle}{m_N}$ describing 
the light quark contribution to the nucleon mass.  We obtain
\begin{equation}\label{eq:GoldstinoDDXSection}
\left.\sigma_{\rm SI}\right|_{[_{8}\mathcal{F}^{V}_{\psi}]_{VV}^{\mathcal{M}}} = \frac{g_*^4\mu_N^2m_{\chi}^2m_N^2}{4\pi M^8}
\left\{
\sum_{\psi = u,d,s}
\left[
3(\psi_{(2)}(\mu_0) +\bar{\psi}_{(2)}(\mu_0)) + \left(f_{T_{u}}^{(N)} + f_{T_{d}}^{(N)} + f_{T_{s}}^{(N)}\right)
\right]
\right\}^2
\end{equation}
where we assumed $C_{u} = C_{d} = C_{s} = 1$. In our analysis we use the numerical values quoted in~\cite{Hisano:2015rsa}, evaluated at the scale $\mu_0 = m_Z$.\footnote{Formally, we have 
\begin{equation}
\psi_{(2)}(\mu_0) +\bar{\psi}_{(2)}(\mu_0) \equiv \int_0^1dx x[\psi(x,\mu_0) +\bar{\psi}(x,\mu_0)]~,
\end{equation}
where $\psi(x,\mu_0)$ and $\bar{\psi}(x,\mu_0)$ are the  parton distribution functions of quark and antiquark in the nucleon at the scale
$\mu_0$.}  
Eq.~(\ref{eq:GoldstinoDDXSection}) is valid in the case of Majorana DM (in~\cite{Drees:1993bu} the operator in Eq.~(\ref{eq:Dim8FermionDD}) was studied 
in the context of effective neutralino-quark interactions). 
This is the relevant case in our analysis, since we consider the operator $_8\mathcal{F}_{\psi}^V$ in the context of Goldstino DM.

The operator $_8{\cal F}_{\psi}^{\prime}$, on the other hand, is expected to play a r\^ole mostly in the Goldstino case, where lower derivative terms are forbidden. In this situation, however, it is always associated with the $\gamma_5$ structure (since for a Majorana fermion, like the Goldstino, $\bar{\pmb{\chi}}\gamma_\mu\pmb{\chi}=0$) and is therefore subdominant in DD experiments.

For interactions with gluons, the effect of the mass-suppressed operator $_8\opf_G^{\not \,s}$ is well known; with our power-counting we find
\begin{equation}\label{usualfgg}
\left.\sigma_{\rm SI}\right|_{[_8\opf_{G}^{\not \,s}]_V^{\mathcal{D}}} =
\frac{1}{4}\left.\sigma_{\rm SI}\right|_{[_8\opf_{G}^{\not \,s}]_V^{\mathcal{M}}} =
 \frac{C_{G}^{\slashed{s}\,2} g_*^4 \mu_N^2 m_{\chi}^2 m_N^2}{\pi M^8}
\left[
\frac{8\pi}{9\alpha_s}f_{T_G}^{(N)}
\right]^2~.
\end{equation}
Let us now consider the operator $_8\opf_{G}$, absent in previous literature. 
There are two different chiral structures in the DM current, but only the one without the $\gamma^5$ - marked with a sub-index $V$ in the following -  gives a stringent spin independent constraint.
As already discussed for the computation of the RD, 
we take the traceless part of  $G_{\mu\rho}^a G^{a\,\rho}_\nu$  in order to couple DM with a genuine tensor current.
In the jargon of DD, the resulting combination is known as gluon twist-2 operator 
$\mathcal{O}_{\mu\nu}^{G} \equiv G_{\mu\rho}^a G^{a\,\rho}_\nu -\frac{1}{4}g_{\mu\nu}G^a_{\rho\sigma}G^{a,\rho\sigma}$.
 The nuclear matrix element  $\langle N| \mathcal{O}_{\mu\nu}^{G} |N \rangle$  at zero momentum transfer is known in terms of the second moment of the 
 gluon parton distribution function $g_{(2)}(\mu_0)$ (evaluated at the scale $\mu_0 = m_Z$) \cite{Hisano:2015rsa}. 
We find (see as usual Appendix~\ref{app:Notation} for a more detailed computation)
\begin{equation}\label{eq:GoldstinoDDXSection2}
\left.\sigma_{\rm SI}\right|_{[_{8}\mathcal{F}_{G}]_V^{\mathcal{D}}} = 
\frac{1}{4}\left.\sigma_{\rm SI}\right|_{[_{8}\mathcal{F}_{G}]_V^{\mathcal{M}}} = 
\frac{C_G^2 g_*^4\mu_N^2m_{\chi}^2m_N^2}{\pi M^8}\left[ \frac{3}{2}g_{(2)}(\mu_0)
\right]^2~.
\end{equation}

\subsubsection*{Scalar DM}

\noindent
{\bf D=6.} At the lowest order, interactions with SM quarks arise from the effective operator $_6{\cal S}^V_{\psi}$ in table~\ref{tab:ScalarDM}, 
which appears in two versions, depending on the SM fermion chiral structure (Eq.~(\ref{eq:ComplexScalarEff}) in Appendix~\ref{app:Notation}). For the vector interaction, we find 
\begin{equation}
\left.\sigma_{\rm SI}\right|_{[_{6}\mathcal{S}_{\psi}^V]_V^{\mathcal{C}}}  = \frac{9 g_*^4\mu_N^2}{\pi M^4}~,
\end{equation}
where we assumed equal unit coupling with up- and down-type quarks.\\

\noindent
{\bf D=8.} At $D = 8$, we have interactions with quarks and gluons, the former being captured by the operator $_8\mathcal{S}_{\psi}^T$.
The computation of $\sigma_{\rm SI}$ closely follow what already discussed for the operator $_8\mathcal{F}_{\psi}^V$
since it exploits the properties of the quark twist-2 operator.
We find 
 \begin{eqnarray}
\left.\sigma_{\rm SI}\right|_{[_8\mathcal{S}_{\psi}^T]_V^{\mathcal{C}}} &=& \frac{1}{4} \left.\sigma_{\rm SI}\right|_{[_8\mathcal{S}_{\psi}^T]_V^{\mathcal{R}}} \\
&=&
 \frac{g_*^4\mu_N^2m_{\phi}^2m_N^2}{\pi M^8}
\left\{
\sum_{\psi = u,d,s}
\left[
\frac{3}{4}(\psi_{(2)}(\mu_0) +\bar{\psi}_{(2)}(\mu_0)) + \frac{1}{2}\left(f_{T_{u}}^{(N)} + f_{T_{d}}^{(N)} + f_{T_{s}}^{(N)}\right)
\right]
\right\}^2~.\nn
 \end{eqnarray}
 Finally, scalar DM interactions with gluons are described by the operators $_8\mathcal{S}_{G}^S$ and $_8\mathcal{S}_{G}^T$.
 We find
 \begin{eqnarray}
\left.
\sigma_{\rm SI}\right|_{[_8\mathcal{S}_{G}^S]^{\mathcal{C}}} &=&
 \frac{1}{4}\left.\sigma_{\rm SI}\right|_{[_8\mathcal{S}_{G}^S]^{\mathcal{R}}} =  \frac{g_*^4 \mu_N^2 m_{N}^2 m_{\phi}^2}{\pi M^8}\left[
\frac{8\pi}{9\alpha_s}f_{T_G}^{(N)}
\right]^2~,\\
\left.
\sigma_{\rm SI}\right|_{[_8\mathcal{S}_{G}^T]^{\mathcal{C}}}  &=&
\frac{1}{4}
\left.\sigma_{\rm SI}\right|_{[_8\mathcal{S}_{G}^T]^{\mathcal{R}}} =
\frac{g_*^4\mu_N^2m_N^2m_{\phi}^2}{\pi M^8}\left[
\frac{3}{4}g_{(2)}(\mu_0)
\right]^2~.
\end{eqnarray}

\subsection{Summary of Results}\label{sec:results}
From the discussions in section \ref{sec:BSMperspective} and the construction of the effective Lagrangian in Ref.~\cite{companion}, we can  compile a list of fundamental questions that deserve a phenomenological discussion. For fermions:
\begin{itemize}
\item[{\bf (F1)}] What are the  $(g_*,M)$ constraints if  DM and  the SM fermions are composite: $_6\opf^V_{\psi}$ ?
\item[{\bf(F2)}] What reach in $(g_*,M)$ have DM experiments, if DM is a Goldstino: $_8\opf_{\psi}^{(\prime)}$ ?
\item[{\bf(F3)}] What if the gluons are composite (according to Ref.~\cite{Remedios}):  $_8\opf_{G}$ ?
\end{itemize}
Moreover it would be interesting to verify the consistency of the assumptions that led to our construction:
\begin{itemize}
\item[{\bf(F4)}] $D=8$ are small unless a symmetry differentiates them from $D=6$: $_6\opf^V_{\psi}$ vs. $_8\opf^V_{\psi}$
\item[{\bf(F5)}] Chiral symm. breaking operators can be neglected at LHC for light DM: $_8\opf_{G}^{\not \,s}$ vs. $_8\opf_{G}$
\item[{\bf(F6)}] If gluons are composite, can the operators of \eq{ferm8jet} dominate: $_6\opf^V_{\psi}$ vs $_8\opf^{mono}_{\psi}$?
\end{itemize}
Similar questions can be asked for scalar DM. In particular we are interested in 
\begin{itemize}
\item[{\bf(S1)}] What are the  $(g_*,M)$ constraints if  DM is a multi-component PNGB and  the SM fermions are composite: $_6\ops^V_{\psi}$ ?
\item[{\bf(S2)}] What reach in $(g_*,M)$ have DM experiments, if DM is a PNGB of an abelian SSB pattern: $_8\ops_{G}^{S,T}$ and $_8\ops^T_{\psi}$ ?
\end{itemize}

\begin{figure}[!htb!]
 \begin{center}
\fbox{F1 -- Composite fermions}\vspace{-0.4cm}
\end{center}
\minipage{0.5\textwidth}
  \includegraphics[width=.975\linewidth]{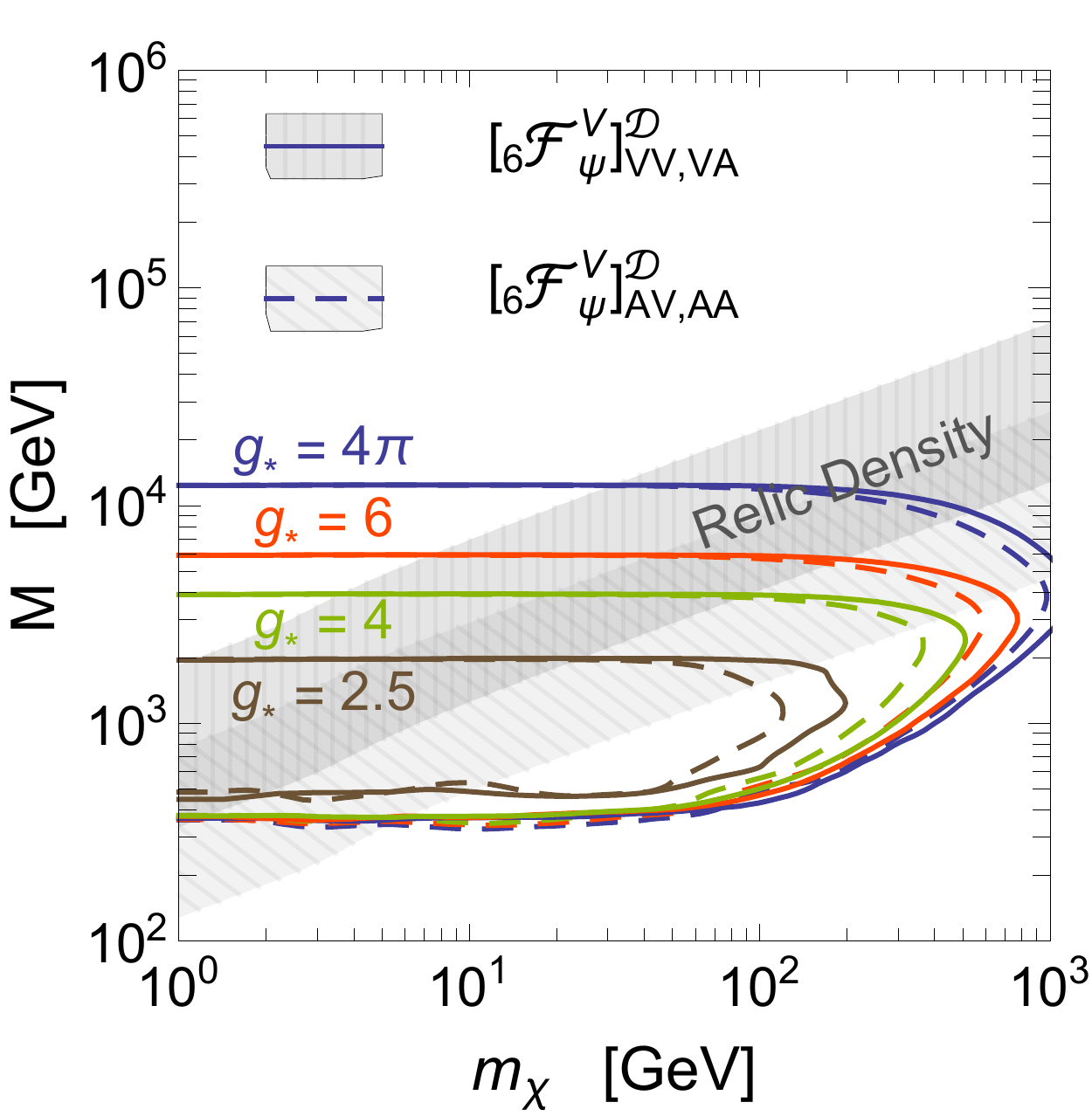}
\endminipage\hfill
\minipage{0.5\textwidth}
  \includegraphics[width=.975\linewidth]{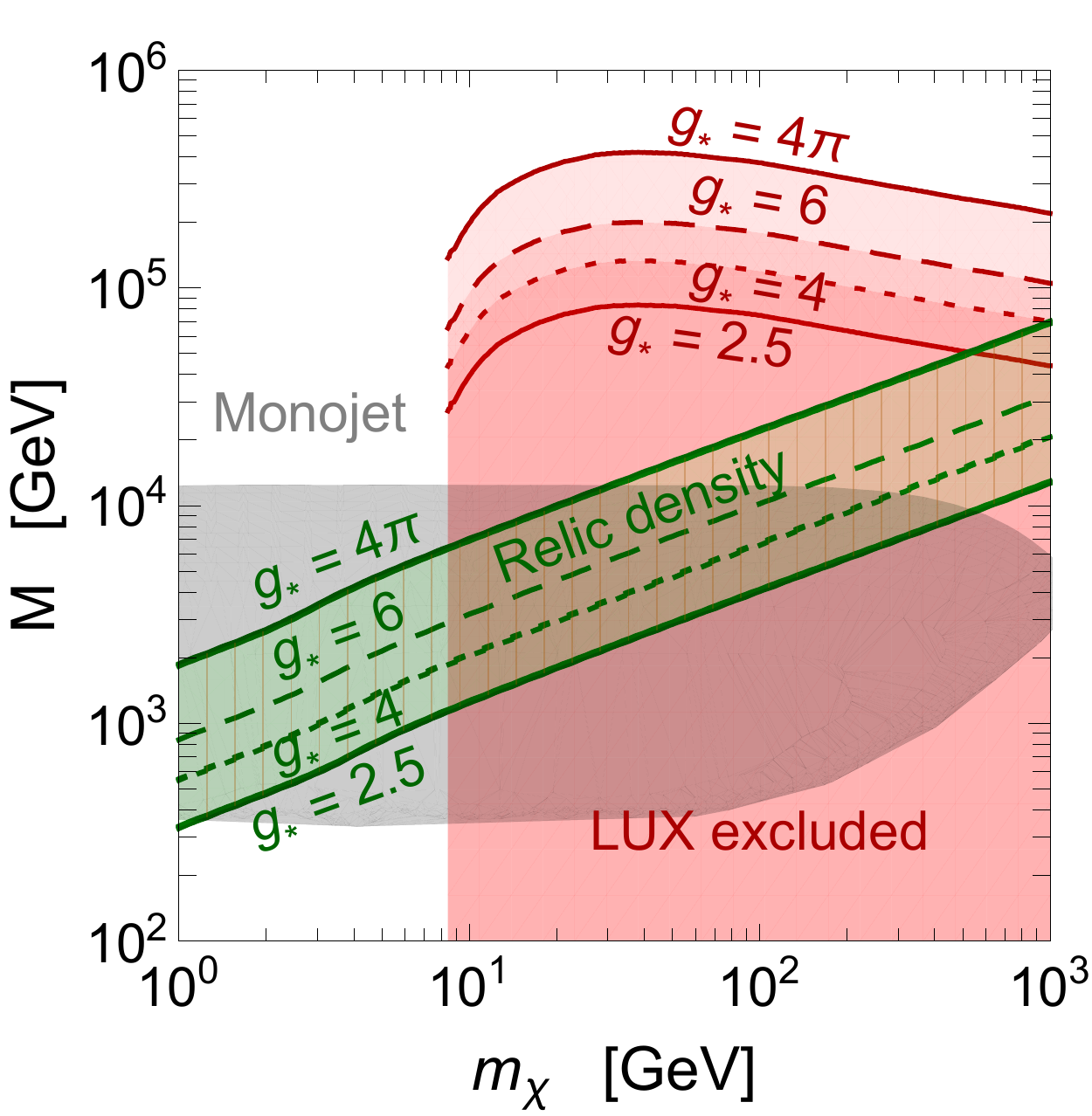}
\endminipage
\caption{\em 
Constraints on DM mass versus new physics scale $M$ for different values of $g_*$  in a model with composite fermions (including fermionic DM), operator $_6\opf^V_{\psi}$. LEFT: LHC constraints as colored solid (dashed) lines for the  $VV$ and $VA$ ($AA$ and $AV$) chiral structure; shadows correspond to RD constraints for different couplings and chiral structure, light for $AV$, $AA$ and dark for $VV$, $VA$. RIGHT: constraints from DD and RD for the $VV$ structure (other structures give weaker constraints); shadowed the LHC bounds. 
 }
\label{fig:SNR}
\end{figure}

\begin{figure}[!htb!]
 \begin{center}
\fbox{F2 -- Goldstino DM}\vspace{-0.4cm}
\end{center}
\minipage{0.5\textwidth}
\endminipage\hfill
\minipage{0.5\textwidth}
  \includegraphics[width=.975\linewidth]{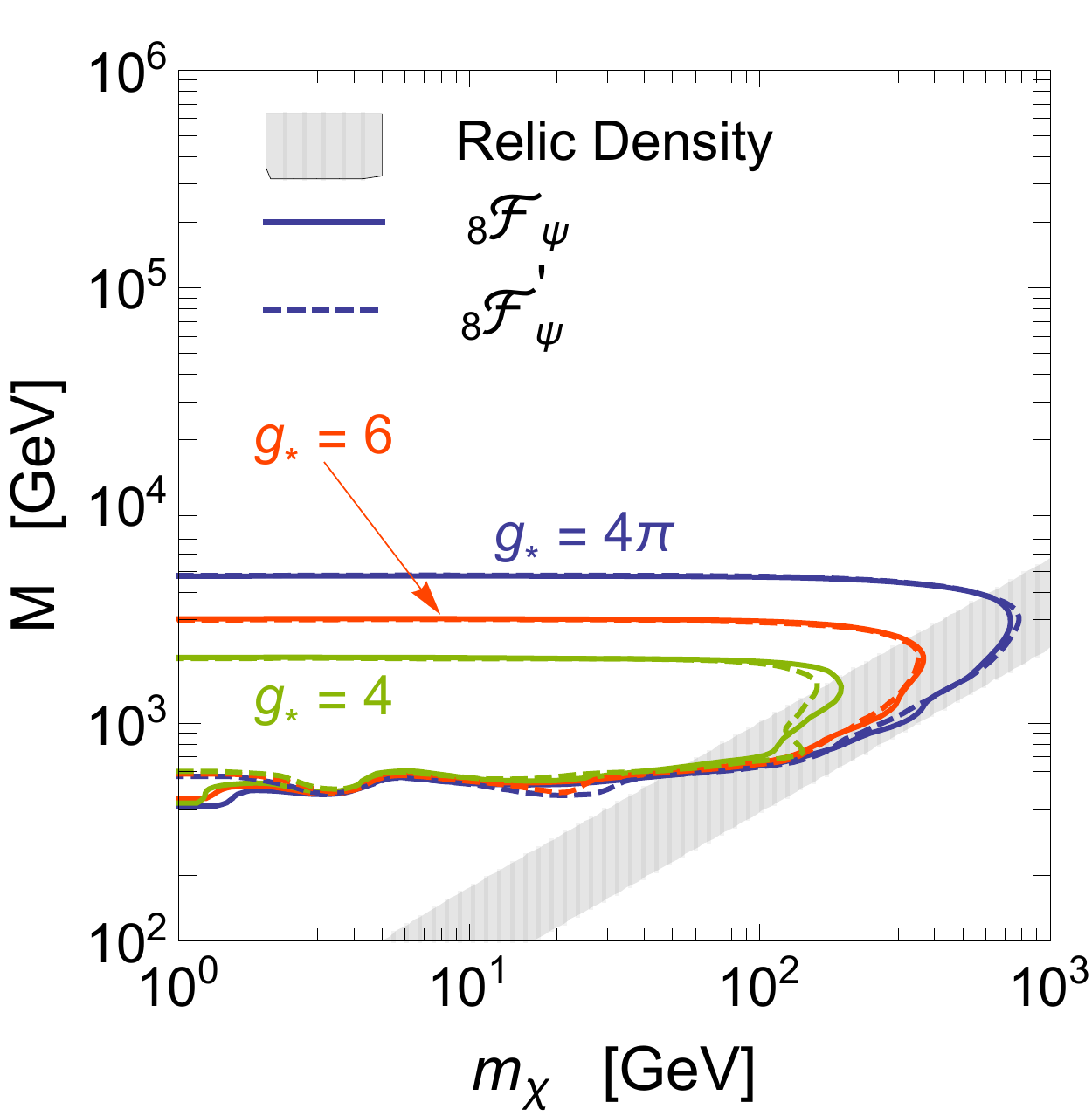}
\endminipage 
  \minipage{0.5\textwidth}
  \includegraphics[width=.975\linewidth]{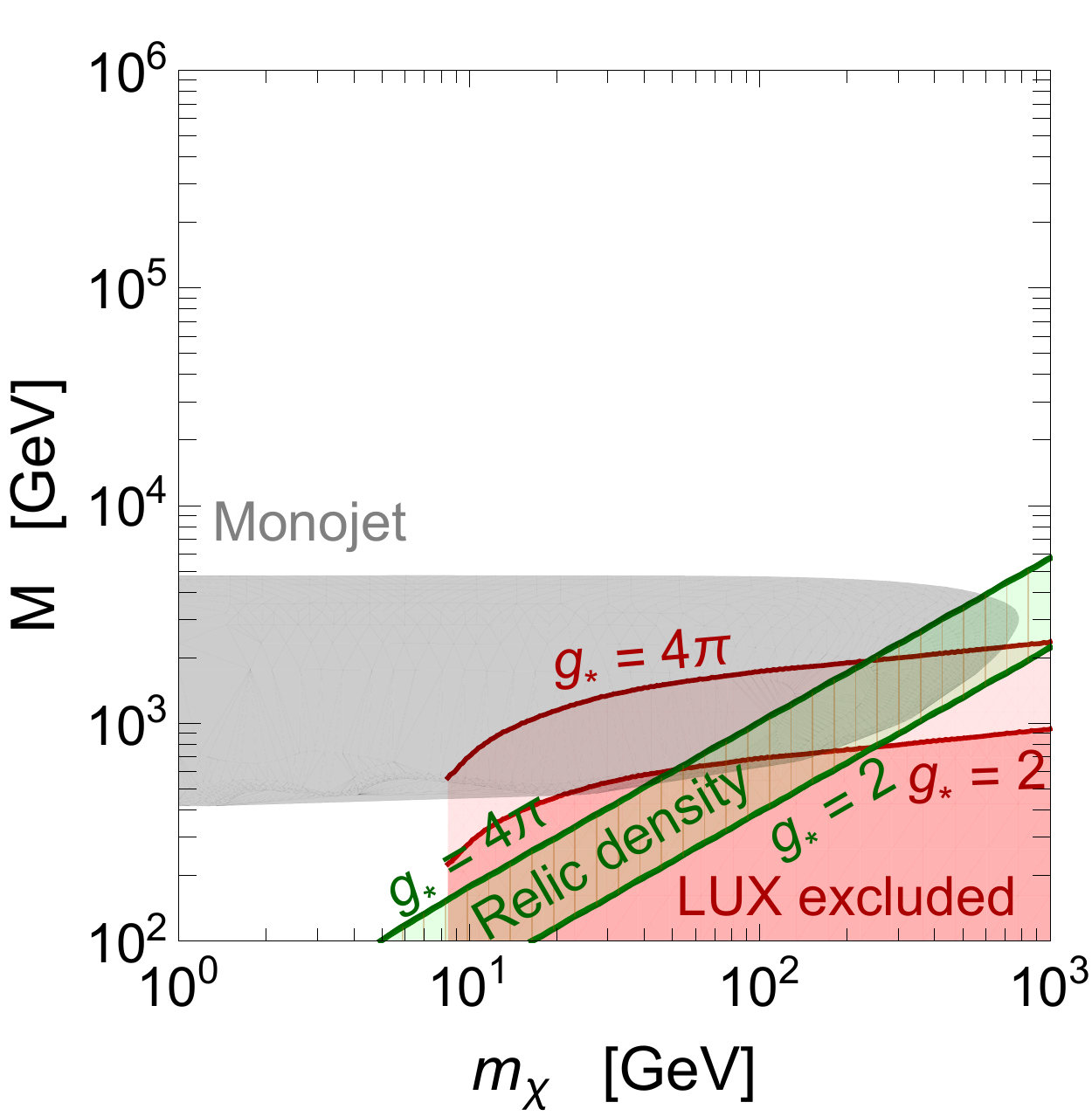}
\endminipage\hfill
\caption{\em 
Same as Fig.~\ref{fig:SNR}, but for the $D = 8$ operator $_8\opf_\psi$ ($_8\opf_\psi^\prime$) in solid (dashed). All constraints apply to Majorana DM, relevant for the Goldstino case.}
\label{fig:goldstino}
\end{figure}

\noindent
{\bf F1 -- Composite fermions}\\
The case of composite fermions $_6\opf_\psi^V$ has already been discussed in previous literature (see in particular Ref.~\cite{Racco:2015dxa} that uses our same power-counting); we show our results in Fig.~\ref{fig:SNR}. It is interesting that EFT-consistent LHC searches exactly exclude the light-mass region where the RD is correctly reproduced, and where DD experiments have no access (recall also that DD constraints are mostly relevant  for the VV chiral structure). This figure also shows in practice the necessity for a strong coupling: constraints consistent with the EFT assumption only exist for a coupling $g_*\gtrsim 2$~\cite{companion}. Notice that the particular chiral structure of these higher dimension operators (solid versus dashed lines) has little impact on the constraints away from threshold: this motivated our choice of Weyl basis in the first place.\footnote{In fact different chiral structures can  introduce  factors of 2 in the relevant cross sections even at high-$E$. Such factors of 2 are anyway not taken into account in our generic power-counting and, moreover, we expect that in strongly coupled scenarios different chiral structures are generated at the same time. So, in situations were LHC constraints differ substantially for different chiral structures, we include here only the strongest ones.} Finally, in this scenario (differently from what we will discuss below) a single interaction dominates at all energies so that the complementarity between LHC, RD and DD constraints, is solid. This is confirmed by the fact that, in this scenario, the effects of $D = 8$ operators is small (question {\bf F4}), as can be seen by comparing the  constraints in Fig.~\ref{fig:SNR} (on a $D = 6$ operator) with those in Fig.~\ref{fig:goldstino} (on a $D = 8$ operator with the same field content): the latter are always poorer, meaning that their effect is smaller.
\\

\noindent
{\bf F2 -- Goldstini}\\  
The case of Goldstini is novel to our analysis, we will therefore highlight differences w.r.t. the previous case.
Non-linearly realized supersymmetry suppresses $D = 6$ ($_6\opf_\psi^V$) w.r.t. $D = 8$ ($_8\opf_\psi^{(\prime)}$) operators; the former are suppressed by  $\sim m_\chi^2/M^2$ once explicit SUSY breaking\footnote{In the scenario of Ref.~\cite{Cheung:2010mc}, ${\cal N}=1$ SUSY is not strictly speaking broken; SUSY breaking effects refer here to explicit departure from the sequestered limit with ${\cal N}>1$ supersymmetries.} effects that give DM a mass are taken into account. 

\begin{figure}[!htb!]
 \begin{center}
\fbox{F3 -- Fermionic DM with Composite Gluons}\vspace{-0.4cm}
\end{center}
  \minipage{0.5\textwidth}
  \includegraphics[width=.975\linewidth]{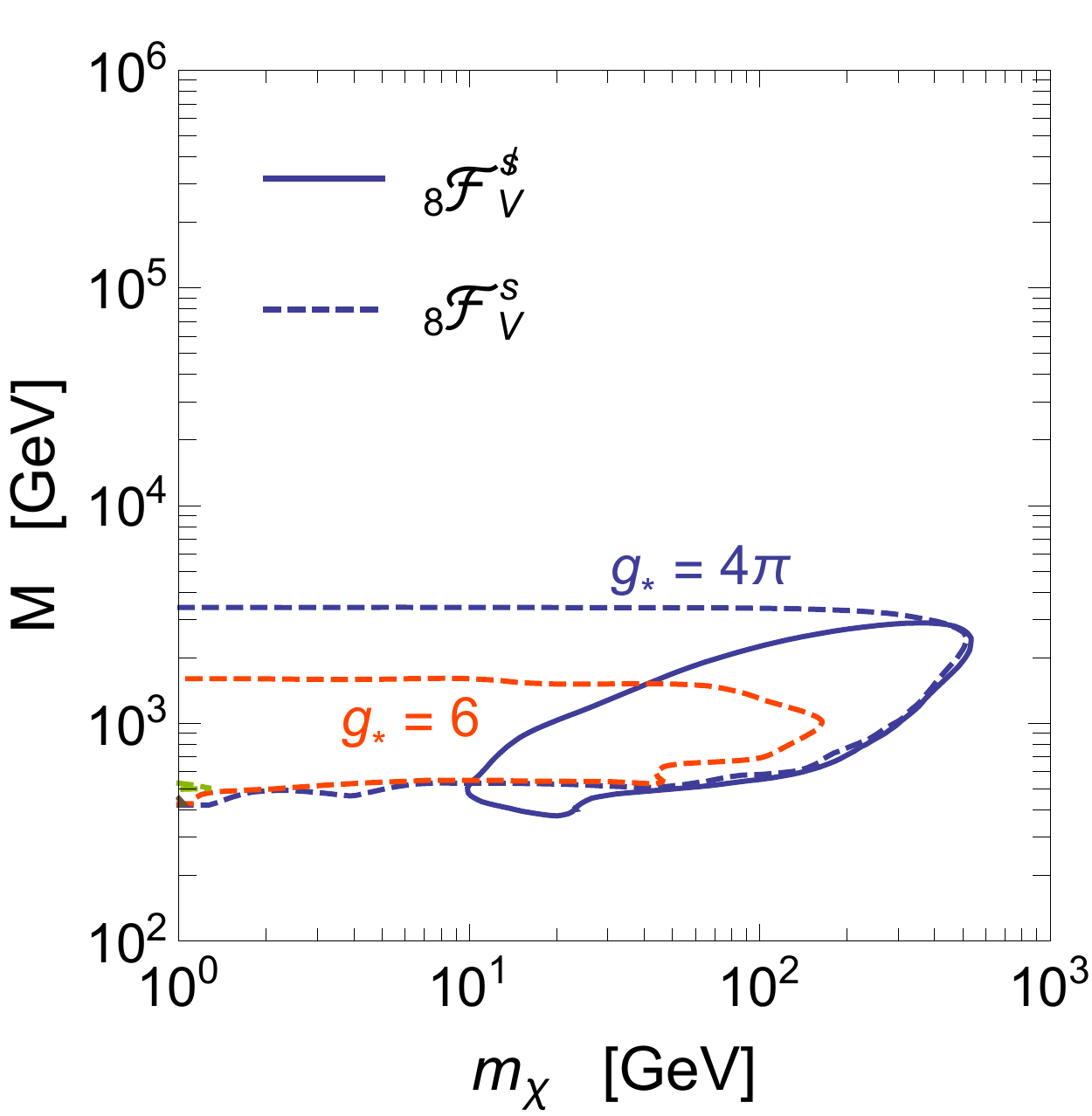}
\endminipage\hfill
  \minipage{0.5\textwidth}
  \includegraphics[width=.975\linewidth]{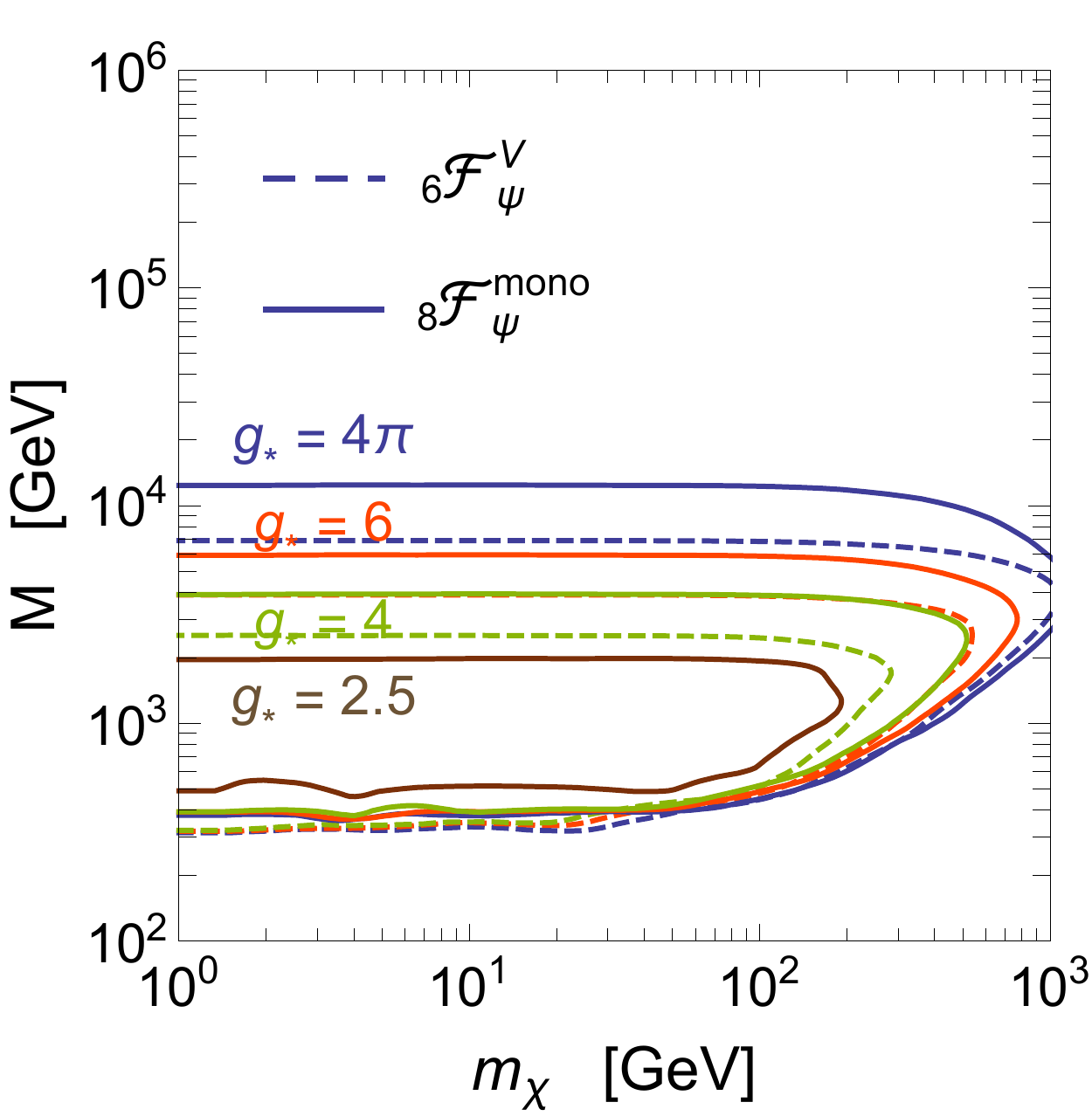}
\endminipage\hfill
\caption{\em Same labelling as Fig.~\ref{fig:SNR}. LEFT: comparison of constraints  from the chirality-breaking, mass-suppressed operator $_8\opf_{G}^{\not \,s}$ (solid) with the unsuppressed $_8\opf_{G}$ (dashed).
 RIGHT: comparison between constraints on $_6\opf^V_{\psi}$ (solid), where an additional mono-jets is emitted as ISR, versus constraints on $_8\opf^{mono}_{\psi}$ where a hard jet is emitted from the new strong interaction directly.
 }
\label{fig:gluon}
\end{figure}

Beside the poorer collider reach on this scenario (Fig.~\ref{fig:goldstino}), the most important aspect of the Goldstino case, concerns DM complementarity.  At the LHC, for $E\gg m_\chi$ the $D = 8$ effects, scaling in the amplitude as $\sim g_*^2 E^4/M^4$, dominate over the SUSY breaking $D = 6$ ones, scaling as $\sim g_*^2m_{\chi}^2E^2/M^4$ -- see also the left panel of  Fig.~\ref{fig:gluon} for a quantitative analysis of this statement. This is not the case however at the lower energies $E^2\sim m_\chi^2$ (c.f. \eq{eq:EFTRelicDensity}),  relevant for the computation of the RD: there the two contributions are comparable. This is exacerbated by the fact that in DM annihilation the $D = 8$ operators are always p-wave suppressed, while the SUSY-breaking $D = 6$ ones are s-wave, and hence enhanced. This implies that the error that we are committing when extracting RD constraints from the $D = 8$ operator, are at least of order 100\%. This is key to interpret the results from Fig.~\ref{fig:goldstino} which shows constraints on the single operator $_8\opf_\psi$ in the Goldstino scenario: the RD band has to be taken as a rough indication only.

Similar arguments apply to DD experiments, whose characteristic energy scale can be as low as the momentum transfer $|\vec{q}| = \sqrt{2 E_{\rm NR} M_{\rm Xe}} \simeq 0.05$ GeV, 
with typical values $E_{\rm NR} = 10$ keV, $M_{\rm Xe} = 120$ GeV for a target atom of Xenon.
This is particularly visible for the operator $_8\opf_\psi^\prime$, that receives LHC constraints comparable to $_8\opf_\psi$, but in DD experiments its energy-dependence  extends as low as $E\sim|\vec{q}|$ and makes the  effects of this  operator unobservable (see discussion below \eq{eq:GoldstinoDDXSection}). It is clear that in this situation DD are more likely to see the effects of the SUSY breaking operator $g_*^2m_{\chi}^2 (_6\opf_\psi^V)/M^4$. 
Of course it is not always the case that DD experiments are irrelevant for $D = 8$ operators. Indeed, in the non-relativistic limit, momentum and mass provide independent energy scales and, while $_8\opf_\psi^\prime$ is sensitive to the former, $_8\opf_\psi$ is to the latter. So, with $E\sim m_\chi$  this situation is similar to that of RD discussed above: DD constraints on the effects generated by the SUSY preserving $_8\opf_\psi$, are comparable to those from the SUSY breaking $g_*^2 m_{\chi}^2{_6\opf_\psi^V}/M^4$.

In summary, in well-defined scenarios with unsuppressed $D = 8$ and suppressed $D = 6$ effects, DD experiments and the RD are typically  sensitive to different effects than collider experiments (either so, or RD and DD experiments receive equivalent contributions from other operators, that are instead suppressed at colliders): the low/high energy complementarity that makes the comparison possible is in this case weak; we nevertheless show the largest constraint you can obtain from DD experiments in Fig.~\ref{fig:goldstino}. 
Given these uncertainties, in addition to those associated with the possibility of a non-standard thermal history, we can conclude that the LHC provides important information on this model with $D = 8$ strong interactions.
\\

\noindent
{\bf F3 -- Fermion DM with Composite Gluons}\\
In most BSM models, the transverse polarizations of vector bosons are assumed to be elementary and associated with the SM couplings $g,g^\prime,g_s$. In this case, operators $_8\opf_{G}^{\not \,s}$ and $_8\opf_{G}$ are always suppressed by $\alpha_{em}$ or $\alpha_{s}$ and their impact for LHC DM searches always negligible. Ref.~\cite{Remedios} proposes however a scenario, based on deformed symmetries, where these operators are sizable. The constraints from DM searches on the possibility that DM be strongly coupled to gluons with deformed symmetry, is shown in Fig.~\ref{fig:gluon}.

In this context we take the opportunity to discuss a few consistency aspects of our analysis (questions {\bf F5}, {\bf F6}). First of all, we compare in the right panel of Fig.~\ref{fig:gluon} the collider constraints from the operators with and without the mass suppression associated with chiral symmetry breaking, $_8\opf_{G}^{\not \,s}$ versus $_8\opf_{G}$. This clearly shows that for the interesting region with small $m_{\rm DM}$, chiral symmetry breaking operators play a negligible r\^ole - confirming {\bf F5}. 

Secondly, strongly interacting gluons \cite{Remedios} have large multipole interactions (associated with $G_{\mu\nu}^a$) but small monopole interactions (associated with the covariant derivative). In this extreme situation, it is not clear  whether emitting an additional jet directly from the new physics vertex (an effect that is captured for instance by the $D = 8$ operator $_8\opf^{mono}_{\psi}$), might not have a larger probability than emitting an initial state radiation (ISR) gluon from the quark in $_6\opf^V_{\psi}$ -- question {\bf F6}. The relative cross sections for $\sigma(pp\to \chi\chi+j)$ scale  as
\begin{equation}
\frac{\left.\sigma \right|_{_{6}\mathcal{F}^{V}_{\psi}} }{\left.\sigma \right|_{_{8}\mathcal{F}^{mono}_{\psi}} }\sim \frac{g_s^2}{g_*^2}\frac{E^4}{M^4}\,,
\end{equation}
and whether or not it is necessarily  larger than unity, depends on the details of the analysis.
We perform this comparison in the right panel of Fig.~\ref{fig:gluon} (dashed versus solid curves), showing that the two are in fact comparable. This exposes a limitation of describing inclusive observables in extreme strongly coupled situations with a parametrization designed for the $2\to2$ process $SM+SM\to DM+DM$. In what follows we shall assume that the SM couplings to the strong mediator sector are slightly suppressed w.r.t. to the DM couplings, a requirement that might also justify the absence of any departure in pure $SM+SM\to SM+SM$ amplitudes in  high-momentum  distributions. Then the parametrization discussed so far in terms of 4-point amplitudes is consistent.
\\

\begin{figure}[!htb!]
 \begin{center}
\fbox{S1 -- Complex Scalar PNGB DM}\vspace{-0.4cm}
\end{center}
\minipage{0.5\textwidth}
  \includegraphics[width=.975\linewidth]{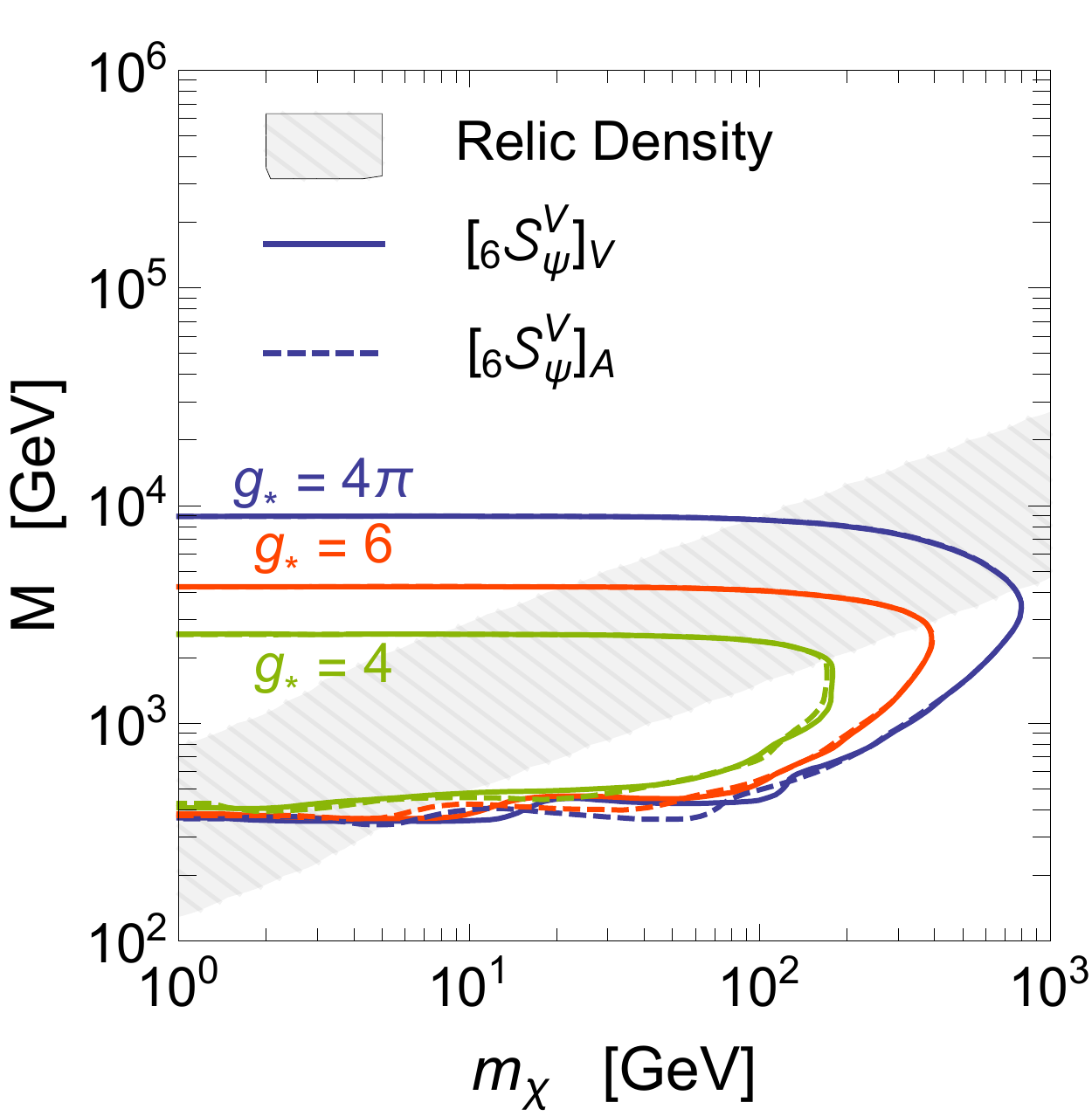}
\endminipage\hfill
  \minipage{0.5\textwidth}
  \includegraphics[width=.975\linewidth]{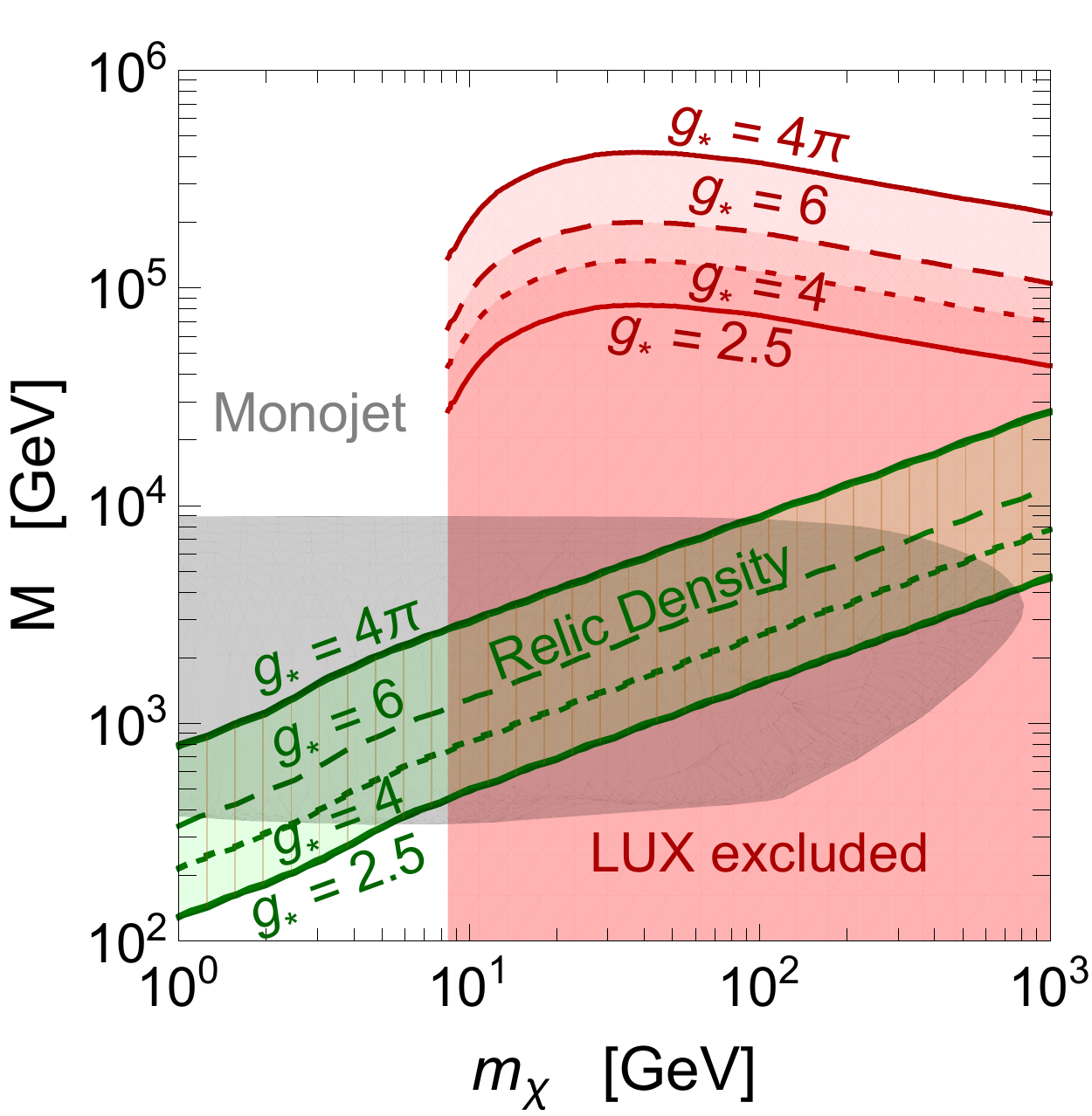}
\endminipage\hfill
\caption{\em Operator $[_6\ops_\psi^V]_V$($[_6\ops_\psi^V]_A$) in solid (dashed) with $\phi$ a complex scalar -- labelling as in Fig.~\ref{fig:SNR}, DD constraints only for the vector structure $[_6\ops_\psi^V]_V$.
 }
\label{fig:scalars}
\end{figure}

\noindent
{\bf S1 -- Complex Scalar PNGB DM}\\
In this case the leading interactions between DM and the proton constituents are captured by the $D = 6$ operator $_6\ops_\psi^V$. The same operator describes well the DM interactions at the energies relevant for computation of the RD and DD. So that the comparison between collider constraints and indications from the RD and DD experiments in Fig.~\ref{fig:scalars} is solid (notice however that while different chiral structures have equivalent LHC limits, those from the RD and DD do depend on the chiral structure).
\\

\noindent
{\bf S2 -- Real Scalar PNGB DM}\\
If the DM is instead associated with a single degree of freedom, as in the case of the abelian SSB $U(1)\to Z_2$, or  $SO(6)/SO(5)$, the first strong interactions with  (composite) SM fermions arise at $D = 8$: we show the constraints from $_8\ops_\psi^T$ and $_8\ops_G^{S,T}$ (for this interaction to be large vectors are also assumed to be composite) in the right panel of Fig.~\ref{fig:scalars}.

This scenario shares some similarities with the Goldstino case discussed above, but it's interesting to appreciate the differences. In particular in the case that the coupling to gluons dominates, the high energy regime is dominated by $_8\ops_G^{S,T}$, while at lower energy (for the RD and DD), the symmetry breaking operator $_8\ops^{\not \,s}_{G}$ contributes effects of the same order. More precisely: $_8\ops_G^S$ has s-wave annihilation like $_8\ops^{\not \,s}_{G}$, but $_8\ops^{T}_{G}$ has only a d-wave annihilation, so that in fact it is subleading to the symmetry breaking effects $_8\ops^{\not \,s}_{G}$ during freeze-out.
If instead the couplings to fermions are more important, then we have an interesting twist in the $U(1)/{\cal Z}_2$ case. Here the effects that break the non-linearly realized $U(1)$ symmetry are captured by $_6\ops^{\not \,s}_{\psi}$, which accidentally is further suppressed by the small SM Yukawas. For this reason, differently from the Goldstino case, symmetry breaking effects do not play an important r\^ole at freeze-out and the computation of the RD based on $_8\ops_\psi^T$ are to be taken more seriously; for DD, on the other hand, quark-mass suppressed effects in the nucleon matrix element are enhanced and eventually appear proportional to $m_N\approx$ 1 GeV, introducing an error $\sim m_N/m_\phi$ in the DD constraint based on $_8\ops_\psi^T$ only.
 In the $SO(6)/SO(5)$ case (see the third bullets of page~\pageref{itemist} or the table in \cite{companion}), the operators $_6\ops^{\not \,s}_{\psi}$ and $_6\ops^{S}_{H}$ are not suppressed by the mass and do play a r\^ole in the computation of  RD, as discussed in detail in \cite{Frigerio:2012uc}.
 \begin{figure}[!htb!]
 \begin{center}
\fbox{S2 -- Real Scalar PNGB DM}\vspace{-0.4cm}
\end{center}
\minipage{0.5\textwidth}
  \includegraphics[width=.975\linewidth]{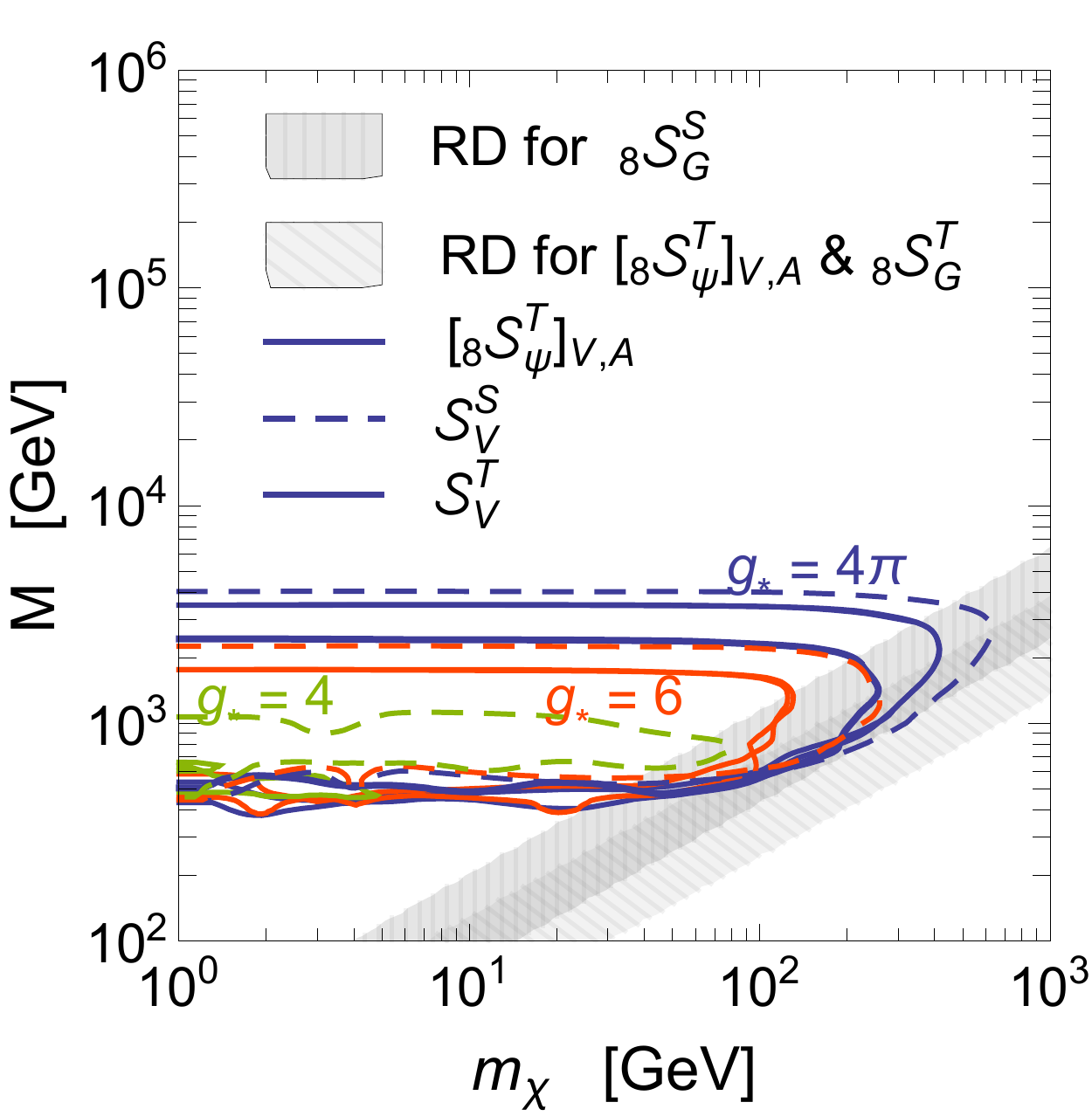}
\endminipage\hfill
  \minipage{0.5\textwidth}
  \includegraphics[width=.975\linewidth]{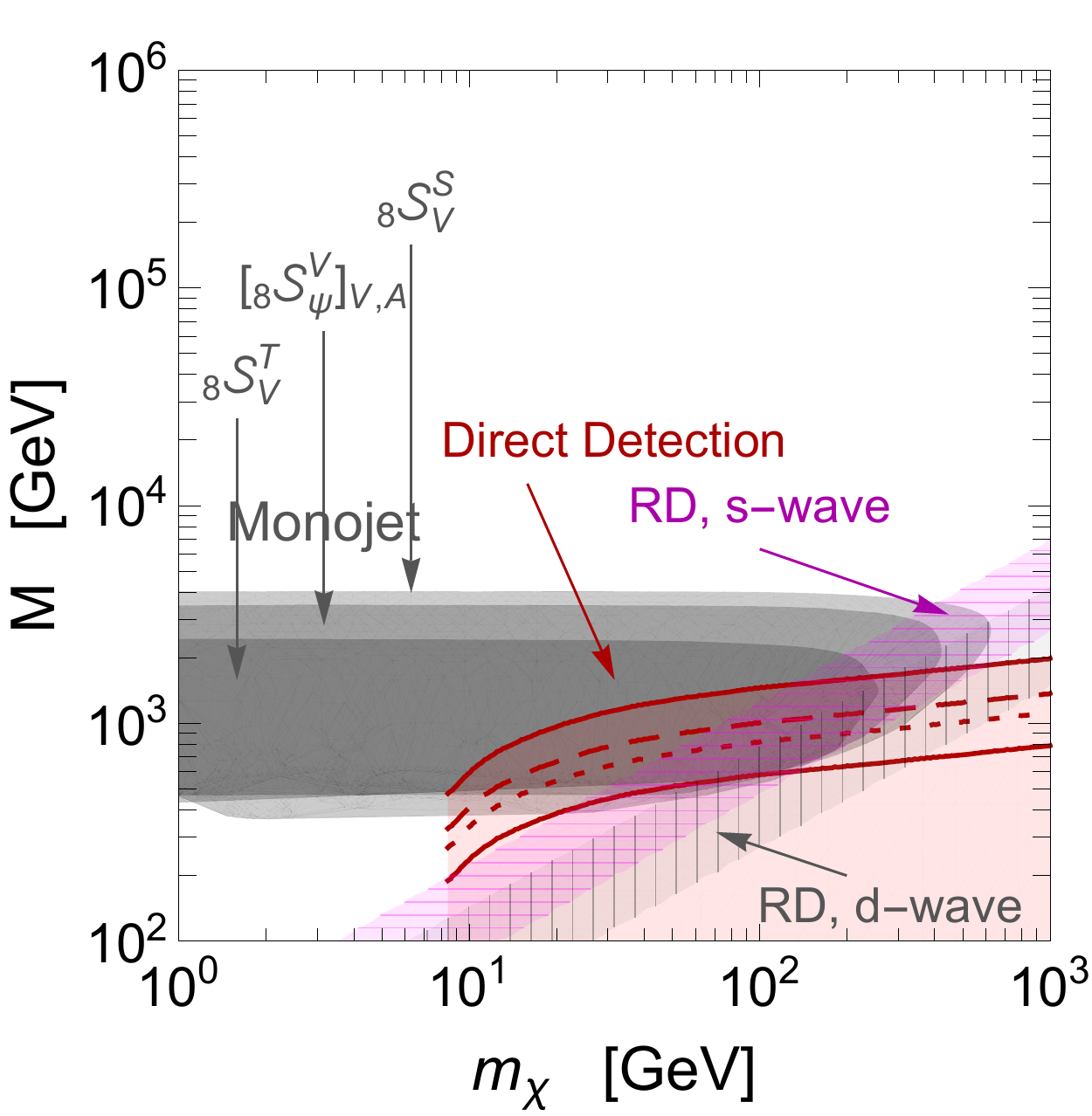}
\endminipage\hfill
\caption{\em Constraints on various $D = 8$ operators for a real scalar PNGB. LEFT: LHC constraints. RIGHT: for the RD operator $_8\ops^{S}_{G}$($_8\ops^{T}_{G}$) annihilate in s-(d-)wave; DD bounds are universal for all operators, except  $[_8\ops^{V}_{\psi}]_A$. Color code as previous figures. }
\label{fig:scalars2}
\end{figure}

\section{Conclusions and Outlook}
We have studied scenarios for LHC DM searches based on underlying strongly coupled dynamics. In these situations, the difficulty of performing perturbative calculations imply that an EFT describing the low-energy degrees of freedom is in fact necessary. Moreover, despite the large separation of scales $E^{LHC}\ll M$ required for the EFT to be predictive, these scenarios can produce sizable effects (enhanced by the strong coupling), visible in LHC searches based on missing energy. The crucial ingredient that guarantees the existence of a weakly coupled SM at low energy, and the realization of the WIMP miracle, compatibly with an underlying strong coupling, is \emph{approximate global symmetry}. We have identified 4 unique scenarios where light composite states (recognized  with DM) can emerge from the strong dynamics that match to the EFT we consider: scalar PNGB of a non-linearly realized abelian ({\bf S2}) or non-abelian  ({\bf S1})  symmetry, composite fermions with chiral symmetry  ({\bf F1})  and Goldstini of non-linearly realized supersymmetry, spontaneously broken by strong dynamics  ({\bf F2}). 

Cases {\bf S2} and {\bf F2} are particularly interesting and novel to our discussion, since the associated operators are characterized by higher derivatives ($D = 8$), while lower-dimension effects are suppressed by powers of the small DM mass. At the LHC, the large energy $E\gg m_{\rm DM}$ implies that mass suppressed effects are subdominant. For computations of the RD, on the other hand, $E\approx  m_{\rm DM}$, so that symmetry breaking and symmetry preserving effects turn out to be often comparable: DM complementarity, thought as a comparison of different constraints individually for each operator, is here lost, since different effects dominate at different energies. This is even more relevant when constraints from DD are included. Depending on the chiral structure of a given operator, its contribution to the amplitudes relevant for DD might be proportional to the momentum transfer, and hence  suppressed. In other instances, this contribution might scale like the DM rest mass, and arguments similar to RD apply for DD.
Our discussion shows the limits of comparing experimental constraints from widely separated regimes on individual operators, without a solid BSM perspective.
The LHC reach on these models is represented in Figs.~\ref{fig:scalars2}  and \ref{fig:goldstino}, where constraints from RD and DD have to be taken as a rough indication only: symmetry breaking effects introduce an $O(1)$ uncertainty. Given this uncertainty and our ignorance about the thermal history of the universe, we conclude that the LHC is providing important information on these models, constraining an interesting region of parameter space associated with very large couplings.

Cases {\bf S1} and {\bf F1} are instead closer to what has been studied in previous literature, in terms of their effective Lagrangian.
Yet, the power counting that we have introduced allows to relate specific UV assumptions with the size of coefficients of EFT operators and understand what effects can be expected large. In particular, sizable couplings to gluons are only possible if these have strong multipolar interactions, following the construction of Ref.~\cite{Remedios}. Moreover, it highlights the necessity of strong coupling, as can be seen in most of our figures, where constraints  consistent with the EFT description exist only for very large couplings (see also \cite{companion}).\footnote{Notice that in practice in the limit $g_*\simeq 4\pi$, the cutoff $\sqrt{s}<M$ is equivalent to the constraint obtained using partial wave unitarity arguments~\cite{Cornwall:1974km,Bell:2016obu}; our construction, however, provides a well-defined hypothesis testing and a consistent physical picture of the meaning of unitarity breakdown.}
In these cases the comparison of LHC  mono-jet searches with DD and RD experiments is rather solid, since symmetry breaking effects are typically further suppressed by small Yuakawa couplings. We show the results in Figs.~\ref{fig:scalars} and \ref{fig:SNR}.
First of all, our analysis reveals the interesting fact that even inherently strongly coupled DM can reproduce the correct RD, as long as an approximate symmetry forbids the lower-dimension interactions.\footnote{On the same lines Ref.~\cite{Hochberg:2014dra} suppressed $2\to 2$ interactions in favor of $2\to 3$ which also allows a strong coupling to be compatible with the observed RD.} Secondly, it shows that LHC experiments are pounding the very region of parameter space where the RD is correctly reproduced.

In general, our scenario (in particular the novel {\bf S2} and {\bf F2}) provide an interesting modeling of missing transverse energy processes for the LHC (similar in that sense to the spirit of simplified models), which is  inspired by an explicit UV realization but captured by a simple bottom-up EFT parametrization.

Natural extensions of our study, which we plan to entertain in the near future, include studies of mono-$W$ \cite{Bell:2015rdw}, mono-$Z$ \cite{Bell:2012rg}, mono-$h$~\cite{Carpenter:2013xra} and mono-$\gamma$ \cite{Fox:2011fx}  processes with the EFT classification proposed in \cite{companion} and discussed in section~\ref{sec:BSMperspective}. In particular it is already interesting to recall that the operators $\phi^\dagger\lra \partial_\mu\phi \,H^\dagger\lra {D^\mu} H $ and $\chi^\dagger\bar\sigma^\mu\chi H^\dagger\lra D_\mu H$  have been discarded from our analysis, as they violate custodial symmetry: this represent another instance where the hierarchy of operators that is traditionally assumed (associated uniquely to the $1/M$ expansion) is compromised in the presence of approximate symmetries.
Another avenue that we intend to pursue stems from arguments based on the analicity of scattering amplitudes, together with crossing symmetry and unitarity, that allow to extract some information, based on prime principles,  about the coefficients of operators  with a particularly soft IR behaviour. In our case this takes the form of positivity constraints on some of the coefficients of the EFT operators; from a practical point of view this corresponds to a theoretical prior  in which to perform the statistical analysis. 

\section*{Acknowledgments}
We are mostly indebted with Rakhi Mahbubani for her help in the collider analysis and her comments or suggestions (including the title!).
We also acknowledge important conversations with Mikael Chala, Roberto Contino, Felix Kahlhoefer, Matthew McCullough, Alex Pomarol, Davide Racco, Riccardo Rattazzi, Andrea Wulzer.

\appendix

\section{Notation and conventions}\label{app:Notation}
This appendix is dedicated to expanding on our notation in terms of Weyl spinors and comparing with previous literature. We also include details relative to the computation of RD and direct detection cross sections that have been neglected in the main text. 

\subsection{Fermions in two-component notation}\label{sec:AppNota}

The building blocks of the  Dirac Lagrangian $\mathcal{L}_{\Psi} = \bar{\Psi}i\gamma^{\mu}(\partial_{\mu}\Psi) - m_{\Psi}\bar{\Psi}\Psi$ in two-component spinor notation are
\begin{equation}
\Psi =
\left(
\begin{array}{c}
  \xi_{\alpha}   \\
  \eta^{\dag\dot{\alpha}}
\end{array}
\right)~,~~~~\bar{\Psi} =
\left(
\begin{array}{cc}
  \eta^{\alpha}  & \xi^{\dag}_{\dot{\alpha}}    
\end{array}
\right)~,~~~~
\gamma_{\mu} =
\left(
\begin{array}{cc}
  0    &  (\sigma_{\mu})_{\alpha\dot{\beta}}   \\
  (\bar{\sigma}_{\mu})^{\dot{\alpha}\beta}    &  0       
\end{array}
\right)~.
\end{equation}
where the two Weyl spinors $\xi_{\alpha}$ and $\eta^{\dag\dot{\alpha}}$
transform under the $(1/2, 0)$ and $(0,1/2)$ representations of the Lorentz group $SO(1,3) \simeq SL(2,\mathbb{C})$.
The Pauli matrices $\sigma_{i=1,2,3}$ define $\sigma^{\mu} = (\mathbb{1}_{2}, \vec{\sigma})$ 
and $\bar{\sigma}^{\mu} = (\mathbb{1}_{2}, -\vec{\sigma})$, so that 
\begin{equation}
\mathcal{L}_{\Psi} = \xi^{\dag}i\bar{\sigma}^{\mu}(\partial_{\mu}\xi) +
\eta^{\dag}i\bar{\sigma}^{\mu}(\partial_{\mu}\eta)  -m_{\Psi}(\xi\eta + \eta^{\dag}\xi^{\dag})~,
\end{equation}
with Lorentz-invariant spinor contractions $\xi\eta \equiv \xi^{\alpha}\eta_{\alpha} = \xi^{\alpha}\epsilon_{\alpha\beta}\eta^{\beta} = -\eta^{\beta}\epsilon_{\alpha\beta}\xi^{\alpha} =
\eta^{\beta}\epsilon_{\beta\alpha}\xi^{\alpha} = \eta^{\beta}\xi_{\beta} = \eta\xi$. For two four-component spinors $\Psi_{1,2}$, the relevant fermion bilinears  in four- and two-component notation are
\begin{equation}\label{eq:Dictionary}
\bar{\Psi}_1 P_L \Psi_2 = \eta_1\xi_2~,~~~    \bar{\Psi}_1 P_R \Psi_2 = \xi^{\dag}_1\eta^{\dag}_2~,~~~ 
 \bar{\Psi}_1\gamma^{\mu}P_L\Psi_2 = \xi^{\dag}_1\bar{\sigma}_{\mu}\xi_2~,~~~ 
 \bar{\Psi}_{1}\gamma_{\mu}P_R\Psi_2 = - \eta_2^{\dag}\bar{\sigma}_{\mu}\eta_2
\end{equation}

\subsection{Fermionic DM: Effective operators in matrix form}\label{app:GoldstiniEffOp}
The comparison between Weyl and Dirac notation is captured by the definition  
\begin{equation}
{\textbf{Weyl:}}~~\chi=(\xi_1, \eta_1)\,~~ \psi=(\xi_2, \eta_2)\quad {\textbf{Dirac:}}~~\pmb{\chi} = \left(
\begin{array}{c}
  \xi_{1\alpha}   \\
  \eta_1^{\dag\dot{\alpha}}
\end{array}
\right)~~~\Psi= \left(
\begin{array}{c}
  \xi_{2\alpha}   \\
  \eta_2^{\dag\dot{\alpha}}
\end{array}
\right)~,
\end{equation}
where in our notation, the two components of $\xi$ and $\psi$ are summed as elements of a vector, thinking of the Wilson coefficients as matrices, while in Dirac notation they are grouped into a 4-component vector. This will become clearer in the case by case analysis that follows.\\

\noindent
$\boxed{
_6\mathcal{F}^{V}_{\psi} }$
There are four possible $D=6$ operators coupling the two fermions
\begin{eqnarray}\label{not}
c_{\psi}[\xi^{\dag}\bar{\sigma}^{\mu}\xi][
\psi^{\dag}\bar{\sigma}^{\mu}\psi]  &\equiv&
\left(
\begin{array}{cc}
 \xi_1^{\dag}\bar{\sigma}^{\mu}\xi_1  &    \eta_1^{\dag}\bar{\sigma}^{\mu}\eta_1
\end{array}
\right)
\underbrace{\left(
\begin{array}{cc}
 c_{\psi}^{11}  &  c_{\psi}^{12}   \\
 c_{\psi}^{21} & c_{\psi}^{22}       
\end{array}
\right)}_{ c_{\psi} \in \mathbb{R}}
\left(
\begin{array}{c}
  \xi_2^{\dag}\bar{\sigma}_{\mu}\xi_2   \\
 \eta_2^{\dag}\bar{\sigma}_{\mu}\eta_2 
\end{array}
\right)\\
&=&c_{VV}^{\,\psi} \left[
\bar{\pmb{\chi}}\gamma^{\mu}\pmb{\chi}
\right]\left[
\bar{\Psi}\gamma_{\mu}\Psi
\right]
+ c_{VA}^{\,\psi} \left[
\bar{\pmb{\chi}}\gamma^{\mu}\pmb{\chi}\right]\left[
\bar{\Psi}\gamma_{\mu}\gamma^5\Psi
\right] \nn\\
&&+ c_{AV}^{\,\psi} \left[
\bar{\pmb{\chi}}\gamma^{\mu}\gamma^5\pmb{\chi}
\right]\left[
\bar{\Psi}\gamma_{\mu}\Psi
\right]+ c_{AA}^{\,\psi} \left[
\bar{\pmb{\chi}}\gamma^{\mu}\gamma^5\pmb{\chi}
\right]\left[
\bar{\Psi}\gamma_{\mu}\gamma^5\Psi
\right]~,\nonumber
\end{eqnarray}
where we defined the four linear independent combinations
\begin{eqnarray}
c_{VV}^{\,\psi} \equiv \frac{1}{4}\left(
c_{\psi}^{11} - c_{\psi}^{12} - c_{\psi}^{21} + c_{\psi}^{22}
\right)~,\label{eq:Id1}& \quad&
c_{VA}^{\,\psi} \equiv  \frac{1}{4}\left(
-c_{\psi}^{11} - c_{\psi}^{12} + c_{\psi}^{21} + c_{\psi}^{22}
\right)~,\\
c_{AV}^{\,\psi} \equiv  \frac{1}{4}\left(
-c_{\psi}^{11} + c_{\psi}^{12} - c_{\psi}^{21} + c_{\psi}^{22}
\right)~,&\quad&
c_{AA}^{\,\psi} \equiv  \frac{1}{4}\left(
c_{\psi}^{11} + c_{\psi}^{12} + c_{\psi}^{21} + c_{\psi}^{22}
\right)~.\label{eq:Id4}
\end{eqnarray}
Eq.~(\ref{not}) shows the clear advantage of our notation,
since one matrix structure embodies four different chiral operators.
From the CP transformation properties
\begin{equation}
CP\left\{
\bar{\Psi}_i\Gamma^{\mu}\Psi_i
\right\} = (-1)(-1)^{\mu}\bar{\Psi}_i\Gamma^{\mu}\Psi_i~,
\end{equation}
with both $\Gamma^{\mu} = \gamma^{\mu}, \gamma^{\mu}\gamma^5$
follows that the four operators in Eq.~(\ref{not}) are CP-invariant.
Finally, note that if $\pmb{\chi}$ is a Majorana fermion, $\pmb{\chi}^{C} = \pmb{\chi}$,   the vector bilinear 
vanishes since $\bar{\pmb{\chi}}\gamma^{\mu}\pmb{\chi} = - \bar{\pmb{\chi}}\gamma^{\mu}\pmb{\chi}$, and 
the only surviving operators are $c_{AV}$ and $c_{AA}$.
In the following we refer to the four operators  with the notation $[_{6}\mathcal{F}_{\psi}^V]^{\mathcal{D},\mathcal{M}}_{AV,VV,VA,AA}$, 
in which the additional lower index refers to the chiral structure of the corresponding operator 
(first letter for the DM current, second for the SM current). 
The upper index refers to the Majorana ($\mathcal{M}$) or Dirac ($\mathcal{D}$) nature of the DM particle.
\\

\noindent
{{\bf Relic density.}}
We complement the information given in the text with an explicit calculation of the annihilation cross section for massive fermions; we find
  \begin{eqnarray}
   \left.\sigma v_{\rm rel}\right|_{[_{6}\mathcal{F}^{V}_{\psi}]_{AV}^{\mathcal{D}}} &=& \frac{ c^{\,\psi~2}_{AV} g_{*}^4}{M^4}\left[
   \frac{m_{\chi}^2\sqrt{1-x_{\psi}^2}(2+x_{\psi}^2)v_{\rm rel}^2}{4\pi} + \mathcal{O}(v_{\rm rel}^4)\right]~,\label{eq:F1}  \\
   \left.\sigma v_{\rm rel}\right|_{[_{6}\mathcal{F}^{V}_{\psi}]_{VV}^{\mathcal{D}}} &=&  \frac{c^{\,\psi~2}_{VV} g_{*}^4}{M^4}
   \left[
\underbrace{\frac{3m_{\chi}^2\sqrt{1-x_ {\psi}^2}(2+x_ {\psi}^2)}{2\pi}}_{{\rm unsuppressed~s-wave}}
 + \frac{m_{\chi}^2(8 - 4x_ {\psi}^2 +5 x_ {\psi}^4) v_{\rm rel}^2}{16\pi \sqrt{1-x_ {\psi}^2}} + \mathcal{O}(v_{\rm rel}^4)
\right]~,\label{eq:F2}\\
   \left.\sigma v_{\rm rel}\right|_{[_{6}\mathcal{F}^{V}_{\psi}]_{VA}^{\mathcal{D}}} &=&  \frac{c^{\,\psi~2}_{VA} g_{*}^4}{M^4}
   \left[
\underbrace{\frac{3m_{\chi}^2(1-x_ {\psi}^2)^{3/2}}{\pi}}_{{\rm unsuppressed~s-wave}}
 + \frac{m_{\chi}^2\sqrt{1-x_ {\psi}^2}(4 + 5x_ {\psi}^2 ) v_{\rm rel}^2}{8\pi} + \mathcal{O}(v_{\rm rel}^4)
\right]~,\label{eq:F3}
    \\
\left.\sigma v_{\rm rel}\right|_{[_{6}\mathcal{F}^{V}_{\psi}]_{AA}^{\mathcal{D}}} &=& 
\frac{c^{\,\psi~2}_{AA} g_{*}^4}{M^4}
\left[
\underbrace{\frac{3 m_\psi^2}{2\pi}\sqrt{1-x_\psi^2}}_{{\rm suppressed~s-wave}} + 
\frac{m_{\chi}^2 (8 - 22 x_\psi^2 + 17 x_\psi^4) v_{\rm rel}^2}{16\pi \sqrt{1-x_\psi^2}} + \mathcal{O}(v_{\rm rel}^4)
 \right]~,\label{eq:F4}
  \end{eqnarray}
  with $x_{\psi} \equiv m_{\psi}/m_{\chi}$. This confirms our arguments based on conserved chiral symmetry, given in section~\ref{sec:RD}.
Eqs.~(\ref{eq:F1}-\ref{eq:F4}) are valid for a Dirac DM particle; for Majorana DM, eqs.~(\ref{eq:F2}-\ref{eq:F3}) are equal to zero while
  the annihilation cross sections corresponding to the operators $[_{6}\mathcal{F}_{\psi}^{V}]_{AV}$ and $[_{6}\mathcal{F}_{\psi}^{V}]_{AA}$ must be multiplied by a factor of $4$ since the number of diagrams in the scattering amplitude -- fermions being equivalent to anti-fermions -- doubles.\\

\begin{figure}[!htb!]
  \begin{center}
\fbox{F1 -- Composite fermions -- AA, AV Structures}\vspace{-0.4cm}
\end{center}
\minipage{0.5\textwidth}
  \includegraphics[width=.975\linewidth]{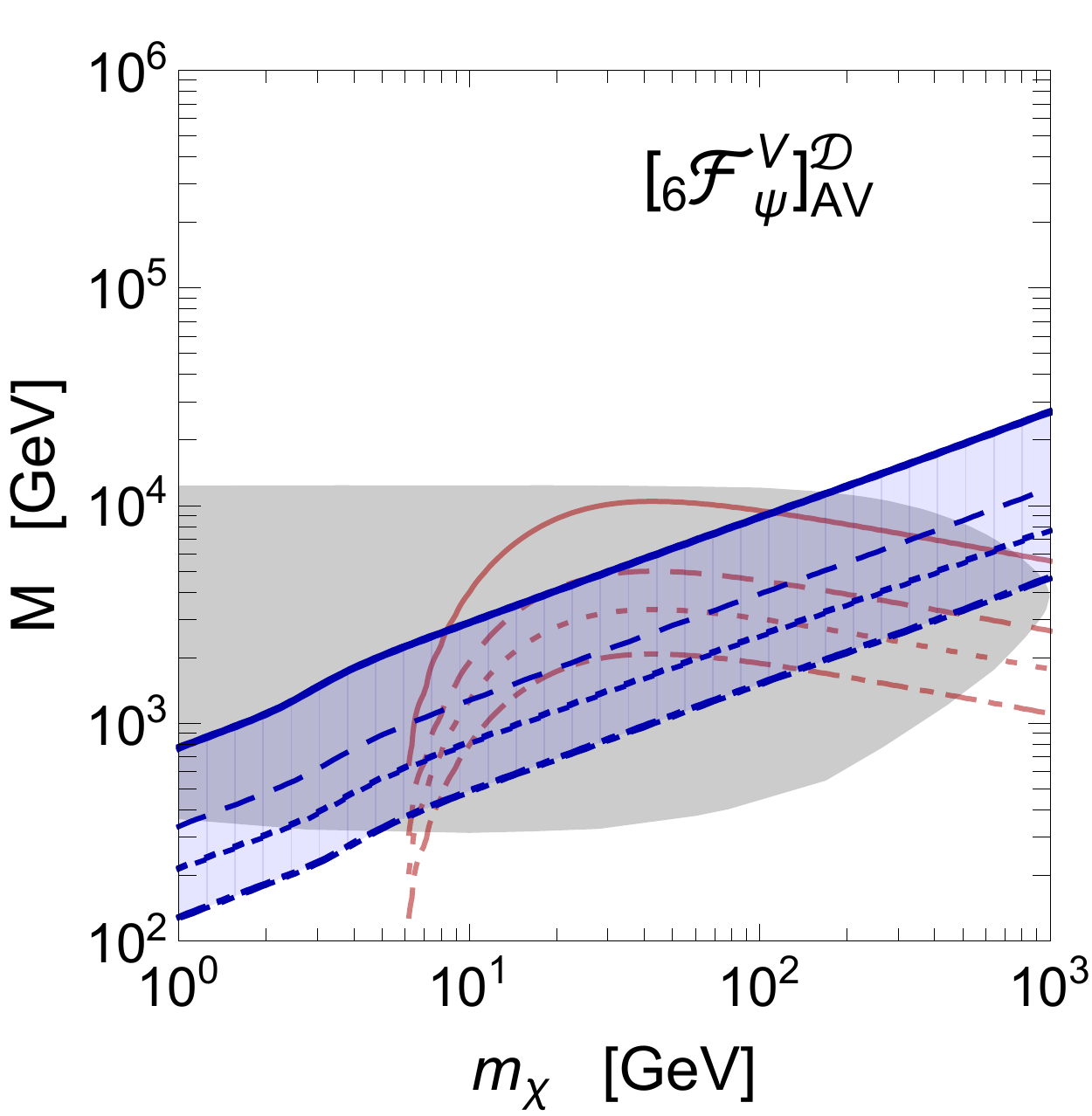}
\endminipage\hfill
  \minipage{0.5\textwidth}
  \includegraphics[width=.975\linewidth]{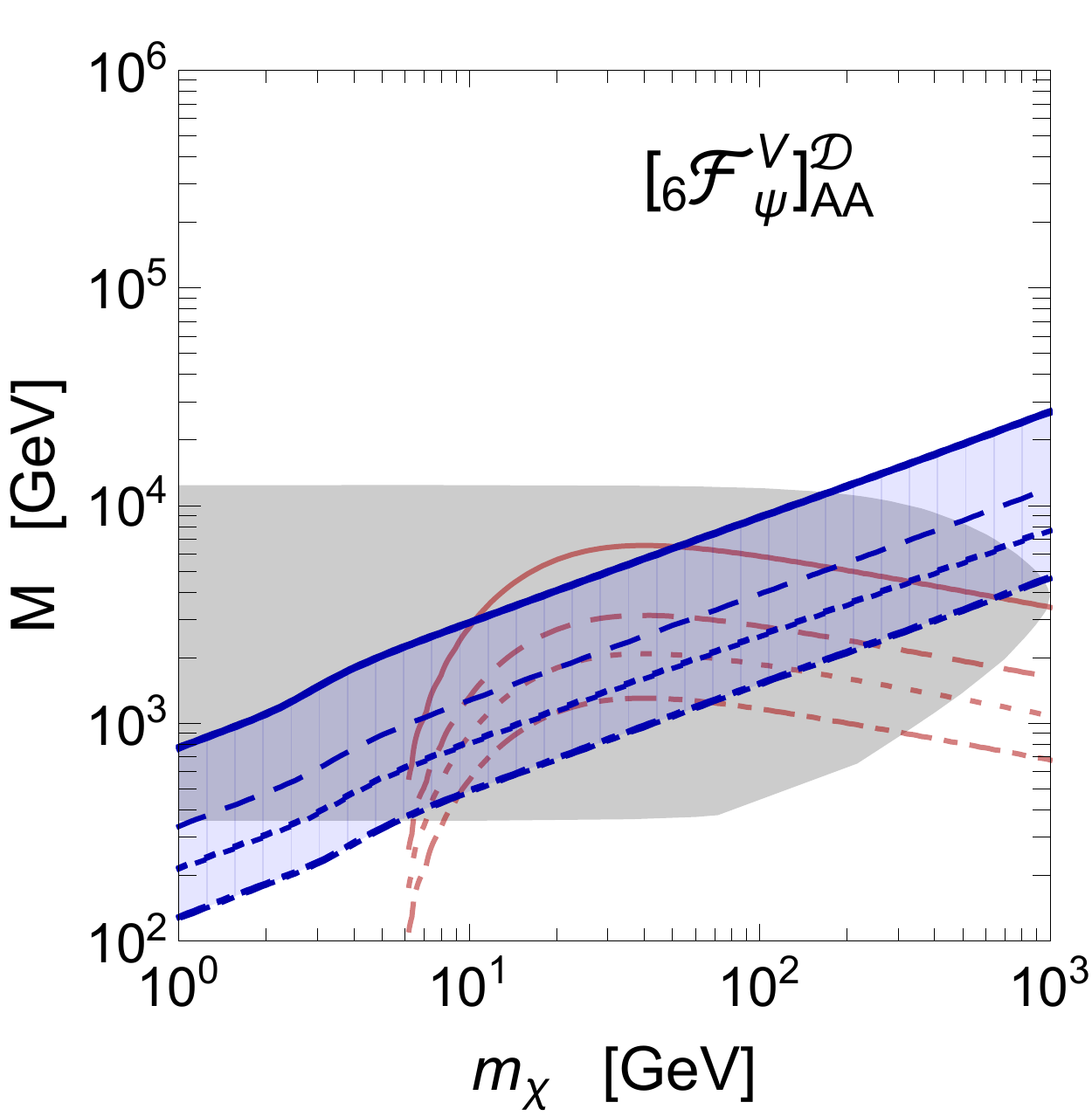}
\endminipage\hfill
\caption{\em 
Interplay between LHC, RD, and DD for the operator $[_6\mathcal{F}^{V}_{\psi}]_{AV}^{\mathcal{D}}$ (left panel) and $[_6\mathcal{F}^{V}_{\psi}]_{VV}^{\mathcal{D}}$ (right panel). The region shaded in blue reproduces the observed RD while the red lines correspond to DD bounds at  90\,\% C.L. 
(left panel: LUX experiment~\cite{LUX}, spin-independent DM-nucleon cross section; right panel: LUX experiment~\cite{Akerib:2016lao}, spin-dependent DM-neutron cross section).
Different lines correspond to different values of strong coupling: $g_* = 4\pi\,(\hbox{solid}),\,6\,(\hbox{dashed}),\,4\,(\hbox{dotted}),\,2.5\,(\hbox{dot-dashed})$.
The region shaded in gray is excluded by LHC mono-jet searches. 
 }
\label{fig:DDExtra}
\end{figure}

\noindent
{{\bf Direct  detection.}}
At the nucleon level, the amplitudes for the DM-nucleon scattering in the non-relativistic limit are 
\begin{eqnarray}
\left.\mathcal{M}_N\right|_{[_6\mathcal{F}^{V}_{\psi}]_{AV}} &=& \frac{8c_{AV}g_*^2 m_{\chi}}{M^2}
\left[
m_N \vec{s}_{\chi} \cdot \vec{v}^{\perp} + i \vec{s}_{\chi}\cdot (\vec{s}_{N} \times \vec{q})
\right]\times
\left\{
\begin{array}{cc}
   2c_{AV}^{\,u} + c_{AV}^{\,d}  & (N = p)   \\
   c_{AV}^{\,u} + 2c_{AV}^{\,d} &  (N = n)
\end{array}
\right.\label{eq:DD1}
\\
\left.\mathcal{M}_N\right|_{[_6\mathcal{F}^{V}_{\psi}]_{VV}} &=& \frac{4g_*^2 m_{\chi} m_N}{M^2}\times
\left\{
\begin{array}{cc}
   2c_{VV}^{\,u} + c_{VV}^{\,d}  & (N = p)   \\
   c_{VV}^{\,u} + 2c_{VV}^{\,d} &  (N = n)
\end{array}
\right.\label{eq:DDV}
\\
\left.\mathcal{M}_N\right|_{[_6\mathcal{F}^{V}_{\psi}]_{VA}} &=&  \frac{8g_*^2m_N}{M^2}
\left[
-m_{\chi}\vec{s}_{N}\cdot \vec{v}^{\perp} + i\vec{s}_{\chi}\cdot (\vec{s}_N \times \vec{q})
\right]
\left(
\sum_{\psi = u,d,s} c_{VA}^{\,\psi} \Delta_{\psi}^{(N)}
\right)~,\label{eq:DDVA}\\
\left.\mathcal{M}_N\right|_{[_6\mathcal{F}^{V}_{\psi}]_{AA}} &=& -\frac{16g_*^2 m_{\chi} m_N}{M^2}
\left(\vec{s}_{\chi}\cdot \vec{s}_N \right)
\left(
\sum_{\psi = u,d,s} c_{AA}^{\,\psi} \Delta_{\psi}^{(N)}
\right)~,\label{eq:DD2}
\end{eqnarray}
where the coefficients $\Delta_{\psi}^{(N)}$, implicitly 
defined through 
 the nuclear matrix element $2\Delta_{\psi}^{(N)} s^{\mu} = \langle N| \bar{\psi}\gamma^{\mu}\gamma^5\psi |N\rangle$  
  parametrize the quark spin content of the nucleon $N$ ($s^{\mu}$ is the spin nucleon four-vector).
 We refer to \cite{DelNobile:2013sia} for the corresponding numerical values. 
As discussed in the text, the operator  $[_6\mathcal{F}^{V}_{\psi}]_{VV}^{\mathcal{D}}$ generates a non-vanishing 
DM-nucleon spin-independent elastic cross section that is not suppressed by  small DM velocity or momentum transfer, while $\left.\mathcal{M}_N\right|_{[_6\mathcal{F}^{V}_{\psi}]_{VA}} $
is certainly poorly constrained since it leads to a spin-dependent cross section suppressed also by $\vec{v}^{\perp} $ or $\vec{q}$.
The amplitude $\left.\mathcal{M}_N\right|_{[_6\mathcal{F}^{V}_{\psi}]_{AV}}$, on the contrary, is characterized by a spin-independent contribution suppressed by $\vec{v}^{\perp}$ only.
Given the remarkable constraining power of DD experiments in the presence of a spin-independent cross section, 
it is worth studying this contribution in more detail (this has been ignored in the main text since it is always subleading to VV, when both are present).
This is shown in the left panel of Fig.~\ref{fig:DDExtra},
in which the region shaded in blue reproduces the observed RD (for different values of $g_*$, see caption) while DD bounds (from~\cite{LUX}, at 90\,\% C.L.) correspond to  
lines in red (for each $g_*$, the region below the corresponding red line is excluded). 
Finally, the amplitude $\left.\mathcal{M}_N\right|_{[_6\mathcal{F}^{V}_{\psi}]_{AA}}$ 
leads to a spin-dependent cross section that is not suppressed neither by $\vec{v}^{\perp} $ nor $\vec{q}$.
The LUX experiment set in~\cite{Akerib:2016lao} the strongest bound on spin-dependent DM-neutron\footnote{
DM is coupled to the net spin of a nucleus, generated by its unpaired nucleon. 
For a Xenon detector, constraints on  DM-neutron spin dependent cross section  
are stronger 
since there are two naturally occurring Xenon isotopes with
an  odd  number  of  neutrons, $^{129}$Xe  and $^{131}$Xe.} cross section, 
and we use this result, at 90\,\% C.L., to constrain the operator $[_6\mathcal{F}^{V}_{\psi}]_{AA}$.
We show the corresponding exclusion regions in the right panel of Fig.~\ref{fig:DDExtra}.
Compared to Fig.~\ref{fig:SNR}, DD bounds are much weaker but they still place meaningful constraints on the parameter space in the presence of a strong coupling.
However, LHC mono-jet searches set the strongest constraints in the region favored by the observed relic abundance. 
Fig.~\ref{fig:DDExtra} strengthens the importance of complementarity between LHC, RD and DD constraints for the effective operator~$_6\mathcal{F}^{V}_{\psi}$.
\\

\noindent
$\boxed{
_{8}\mathcal{F}_{\psi}^{V}}$
We recast this operator in matrix form as 
\begin{gather}\label{eq:MatrixOperator1}
C_{\psi}\left[\chi^{\dag}\bar{\sigma}^{\mu}(\partial^{\nu}\chi)\right]\left[
\psi^{\dag}\bar{\sigma}_{\mu}(\partial_{\nu}\psi)
\right]  \equiv \nonumber 
\left(
\begin{array}{cc}
 \xi_1^{\dag}\bar{\sigma}^{\mu}(\partial^{\nu}\xi_1)  &    \eta_1^{\dag}\bar{\sigma}^{\mu}(\partial^{\nu}\eta_1)
\end{array}
\right)
\underbrace{\left(
\begin{array}{cc}
 C_{\psi}^{11}  &  C_{\psi}^{12}   \\
 C_{\psi}^{21} & C_{\psi}^{22}       
\end{array}
\right)}_{C_{\psi}  \in \mathbb{C}}
\left(
\begin{array}{c}
  \xi_2^{\dag}\bar{\sigma}_{\mu}(\partial_{\nu}\xi_2)   \\
 \eta_2^{\dag}\bar{\sigma}_{\mu}(\partial_{\nu}\eta_2) 
\end{array}
\right)  +h.c.\\
= C_{VV} \left[
\bar{\pmb{\chi}}\gamma^{\mu}(\partial^{\nu}\pmb{\chi})
\right]\left[
\bar{\Psi}\gamma_{\mu}(\partial_{\nu}\Psi)
\right]\nonumber 
+ C_{VA} \left[
\bar{\pmb{\chi}}\gamma^{\mu}(\partial^{\nu}\pmb{\chi})
\right]\left[
\bar{\Psi}\gamma_{\mu}\gamma^5(\partial_{\nu}\Psi)
\right]\nonumber \\ 
+ C_{AV} \left[
\bar{\pmb{\chi}}\gamma^{\mu}\gamma^5(\partial^{\nu}\pmb{\chi})
\right]\left[
\bar{\Psi}\gamma_{\mu}(\partial_{\nu}\Psi)
\right]
+ C_{AA} \left[
\bar{\pmb{\chi}}\gamma^{\mu}\gamma^5(\partial^{\nu}\pmb{\chi})
\right]\left[
\bar{\Psi}\gamma_{\mu}\gamma^5(\partial_{\nu}\Psi)
\right] + h.c.~,\nn
\end{gather}
where the coefficients are defined as in Eqs.~(\ref{eq:Id1}-\ref{eq:Id4}) with $c \to C$ (with the difference that now we are dealing with complex numbers).

Let us now assume that $\pmb{\chi}$ is a Majorana fermion as for Goldstino DM, with $\pmb{\chi}^C = \pmb{\chi}$.
For a Majorana fermion we have\footnote{\label{footnote:MajoranaCJustTheNote}From 
the definition of charge conjugation we have $\pmb{\chi}^C = \mathcal{C}\bar{\pmb{\chi}}^T$, $\bar{\pmb{\chi}}^C = \pmb{\chi}^T\mathcal{C}$, 
with $\mathcal{C} \equiv i\gamma^0\gamma^2$. As a consequence we write
\begin{eqnarray}\label{footnote:MajoranaC}
\bar{\pmb{\chi}}\gamma^{\mu}\gamma^5(\partial^{\nu}\pmb{\chi}) &=& \bar{\pmb{\chi}}^C\gamma^{\mu}\gamma^5(\partial^{\nu}\pmb{\chi}^C) 
= \pmb{\chi}^T\mathcal{C}\gamma^{\mu}\gamma^5\mathcal{C}(\partial^{\nu}\bar{\pmb{\chi}}^T) \nonumber \\
&=& - (\partial^{\nu}\bar{\pmb{\chi}})\mathcal{C}^T (\gamma^{\mu}\gamma^5)^T \mathcal{C}^T\pmb{\chi}
=  (\partial^{\nu}\bar{\pmb{\chi}})\mathcal{C}^{-1} (\gamma^{\mu}\gamma^5)^T \mathcal{C}\pmb{\chi}
=   (\partial^{\nu}\bar{\pmb{\chi}})\gamma^{\mu}\gamma^5\pmb{\chi}~,
\end{eqnarray}
where in the last line the first minus sign, coming from the transposition of spinors, 
gets absorbed by $\mathcal{C}^T = \mathcal{C}^{-1} = - \mathcal{C}$; in the final step
 we made use of $\mathcal{C}^{-1} (\gamma^{\mu}\gamma^5)^T \mathcal{C} = \gamma^{\mu}\gamma^5$.
 A similar relation can be derived for $\bar{\chi}\gamma^{\mu}(\partial^{\nu}\chi)$. The difference in sign comes from 
 $\mathcal{C}^{-1} (\gamma^{\mu})^T \mathcal{C} = - \gamma^{\mu}$.
}
$\bar{\pmb{\chi}}\gamma^{\mu}\gamma^5(\partial^{\nu}\pmb{\chi}) = (\partial^{\nu}\bar{\pmb{\chi}})\gamma^{\mu}\gamma^5\pmb{\chi}$ and $\bar{\pmb{\chi}}\gamma^{\mu}(\partial^{\nu}\pmb{\chi}) = - (\partial^{\nu}\bar{\pmb{\chi}})\gamma^{\mu}\pmb{\chi}$.
Separating real and imaginary part in $C_{ij} \equiv C_{ij}^{\mathbb{Re}} + i C_{ij}^{\mathbb{Im}}$, and using that
\begin{eqnarray}
CP\left\{\bar{\pmb{\chi}}\gamma^{\mu}(\partial^{\nu}\pmb{\chi})
\right\} &=& (-1)^{\mu}(-1)^{\nu}\bar{\pmb{\chi}}\gamma^{\mu}(\partial^{\nu}\pmb{\chi})~,\\
CP\left\{\bar{\pmb{\chi}}\gamma^{\mu}\gamma^5(\partial^{\nu}\pmb{\chi})
\right\} &=& (-1)(-1)^{\mu}(-1)^{\nu}\bar{\pmb{\chi}}\gamma^{\mu}\gamma^5(\partial^{\nu}\pmb{\chi})~,\\
CP\left\{
\bar{\Psi}\Gamma^{\mu}(\partial^{\nu}\Psi) \pm (\partial^{\nu}\bar{\Psi})\Gamma^{\mu}\Psi
\right\} &=& (\mp 1)(-1)^{\mu}(-1)^{\nu}[\bar{\Psi}\Gamma^{\mu}(\partial^{\nu}\Psi) \pm (\partial^{\nu}\bar{\Psi})\Gamma^{\mu}\Psi]~, 
\end{eqnarray}
with both $\Gamma^{\mu}= \gamma^{\mu},\gamma^{\mu}\gamma^5$,
we end up with the following interactions
\begin{eqnarray}
&&
\begin{array}{ccc}
 C_{VV}^{\mathbb{Re}} [\bar{\pmb{\chi}}\gamma^{\mu}(\partial^{\nu}\pmb{\chi})] \times
 \left[
 \bar{\Psi}\gamma_{\mu}(\partial_{\nu}\Psi) - (\partial_{\nu}\bar{\Psi})\gamma_{\mu}\Psi
 \right]    & & \hbox{CP-preserving}  \\
  iC_{VV}^{\mathbb{Im}} [\bar{\pmb{\chi}}\gamma^{\mu}(\partial^{\nu}\pmb{\chi})] \times
 \left[
 \bar{\Psi}\gamma_{\mu}(\partial_{\nu}\Psi) + (\partial_{\nu}\bar{\Psi})\gamma_{\mu}\Psi
 \right]     &  & \hbox{CP-violating}  \vspace{.2cm} \\
 C_{VA}^{\mathbb{Re}} [\bar{\pmb{\chi}}\gamma^{\mu}(\partial^{\nu}\pmb{\chi})] \times
 \left[
 \bar{\Psi}\gamma_{\mu}\gamma^5(\partial_{\nu}\Psi) - (\partial_{\nu}\bar{\Psi})\gamma_{\mu}\gamma^5\Psi
 \right] &   & \hbox{CP-preserving}  \\
  iC_{VA}^{\mathbb{Im}} [\bar{\pmb{\chi}}\gamma^{\mu}(\partial^{\nu}\pmb{\chi})] \times
 \left[
 \bar{\Psi}\gamma_{\mu}\gamma^5(\partial_{\nu}\Psi) + (\partial_{\nu}\bar{\Psi})\gamma_{\mu}\gamma^5\Psi
 \right]   &  &  \hbox{CP-violating}  \vspace{.2cm} \\ C_{AV}^{\mathbb{Re}} [\bar{\pmb{\chi}}\gamma^{\mu}\gamma^5(\partial^{\nu}\pmb{\chi})] \times
 \left[
 \bar{\Psi}\gamma_{\mu}(\partial_{\nu}\Psi) + (\partial_{\nu}\bar{\Psi})\gamma_{\mu}\Psi
 \right]   & & \hbox{CP-preserving}  \\
  iC_{AV}^{\mathbb{Im}} [\bar{\pmb{\chi}}\gamma^{\mu}\gamma^5(\partial^{\nu}\pmb{\chi})] \times
 \left[
 \bar{\Psi}\gamma_{\mu}(\partial_{\nu}\Psi) - (\partial_{\nu}\bar{\Psi})\gamma_{\mu}\Psi
 \right]   &  &  \hbox{CP-violating}   \vspace{.2cm} \\ C_{AA}^{\mathbb{Re}} [\bar{\pmb{\chi}}\gamma^{\mu}\gamma^5(\partial^{\nu}\pmb{\chi})] \times
 \left[
 \bar{\Psi}\gamma_{\mu}\gamma^5(\partial_{\nu}\Psi) + (\partial_{\nu}\bar{\Psi})\gamma_{\mu}\gamma^5\Psi
 \right] &   & \hbox{CP-preserving}  \\
  iC_{AA}^{\mathbb{Im}} [\bar{\pmb{\chi}}\gamma^{\mu}\gamma^5(\partial^{\nu}\pmb{\chi})] \times
 \left[
 \bar{\Psi}\gamma_{\mu}\gamma^5(\partial_{\nu}\Psi) - (\partial_{\nu}\bar{\Psi})\gamma_{\mu}\gamma^5\Psi
 \right]    & &  \hbox{CP-violating}    
\end{array}
\end{eqnarray}
We assume CP as a fundamental symmetry, then
\begin{equation}
\mathbb{Im}\left(
\begin{array}{cc}
 C_{\psi}^{11}  &  C_{\psi}^{12}   \\
 C_{\psi}^{21} & C_{\psi}^{22}       
\end{array}
\right) = 0~.
\end{equation}
In the text we assume coefficients to be real.\\

\noindent{\bf Relic density.} Goldstino annihilation cannot  proceed via s-wave. 
In the initial state a system of two Majorana particles with $L = 0$ is forced to stay in a state with
 total spin $S = 0$ (since $C = (-1)^{L + S} \overset{!}{=} 1$). As a consequence, $J = 0$ considering s-wave annihilation.
 However, the tensor current on the SM side has $J = 2$, thus forcing the s-wave to vanish to conserve total angular momentum.
The same argument shows that p-wave annihilation is allowed since two Majorana particles 
with total orbital angular momentum $L = 1$ are forced to stay in a state with total spin $S=1$ (and hence $J = 2$ is an allowed eigenvalue).
In the massless limit for the final state quarks, we find \eq{eq:Goldstino3}.\\

\noindent
$\boxed{
_{8}\mathcal{F}^{V\,\prime}_{\psi}}$
Here we have\begin{eqnarray}\label{eq:MatrixOperator2}
&&C_{\psi}^{\prime}\chi^{\dag}\bar{\sigma}^{\mu}\chi\left[
(\partial^{\nu}\psi^{\dag})\bar{\sigma}_{\mu}(\partial^{\nu}\psi)
\right]  \equiv 
\left(
\begin{array}{cc}
 \xi_1^{\dag}\bar{\sigma}^{\mu}\xi_1  &    \eta_1^{\dag}\bar{\sigma}^{\mu}\eta_1
\end{array}
\right)
\underbrace{\left(
\begin{array}{cc}
 C_{\psi}^{\prime 11}  &  C_{\psi}^{\prime 12}   \\
 C_{\psi}^{\prime 21} & C_{\psi}^{\prime 22}       
\end{array}
\right)}_{C_{\psi}^{\prime}\in \mathbb{R}}
\left(
\begin{array}{c}
  (\partial^{\nu}\xi_2^{\dag})\bar{\sigma}_{\mu}(\partial_{\nu}\xi_2)   \\
 (\partial^{\nu}\eta_2^{\dag})\bar{\sigma}_{\mu}(\partial_{\nu}\eta_2) 
\end{array}
\right)~,\nn \\
&&= C^{\prime}_{VV} 
\bar{\pmb{\chi}}\gamma^{\mu}\pmb{\chi}\left[
(\partial^{\nu}\bar{\Psi})\gamma_{\mu}(\partial_{\nu}\Psi)
\right]
+ C^{\prime}_{VA} 
\bar{\pmb{\chi}}\gamma^{\mu}\pmb{\chi}
\left[
(\partial^{\nu}\bar{\Psi})\gamma_{\mu}\gamma^5(\partial_{\nu}\Psi)
\right]\nonumber \\ 
&&~~+ C^{\prime}_{AV} 
\bar{\pmb{\chi}}\gamma^{\mu}\gamma^5\pmb{\chi}\left[
(\partial^{\nu}\bar{\Psi})\gamma_{\mu}(\partial_{\nu}\Psi)
\right]
+ C^{\prime}_{AA} 
\bar{\pmb{\chi}}\gamma^{\mu}\gamma^5\pmb{\chi}
\left[
(\partial^{\nu}\bar{\Psi})\gamma_{\mu}\gamma^5(\partial_{\nu}\Psi)
\right]~,
\end{eqnarray}
where the coefficients defined as in Eqs.~(\ref{eq:Id1}-\ref{eq:Id4}) with $c \to C^{\prime}$.
This  simplifies in the case of Goldstino DM, where only the two CP-preserving combinations $C_{AV}^{\prime}$ and $C_{AA}^{\prime}$ survive\footnote{The CP transformation properties follow from 
\begin{eqnarray}
CP\left\{\bar{\pmb{\chi}}\gamma^{\mu}\gamma^5\pmb{\chi}
\right\} &=& (-1)(-1)^{\mu}\bar{\pmb{\chi}}\gamma^{\mu}\gamma^5\pmb{\chi}~,\\
CP\left\{
(\partial_{\nu}\Psi)\Gamma_{\mu}(\partial^{\nu}\Psi)
\right\} &=&  (-1)(-1)^{\mu}(\partial_{\nu}\Psi)\Gamma_{\mu}(\partial^{\nu}\Psi)~,
\end{eqnarray}
with both $\Gamma^{\mu} = \gamma^{\mu},\gamma^{\mu}\gamma^5$.}.
In the following, we refer to the two associated effective operators with the notation $[_{8}\mathcal{F}^{V\,\prime}_{\psi}]_{AV,AA}$.\\

\noindent
$\boxed{_8\opf_{V}^{\not \,s}}$
Interactions without derivatives in the DM current involve the following operator in matrix form
\begin{eqnarray}\label{eq:EffGluon1}
C_{V}^{\not \,s}\chi\chi V_{\mu\nu}^aV^{a\,\mu\nu} &\equiv&
\left(C_{V,1}^{\not \,s} \eta_1\xi_1 + C_{V,1}^{\not \,s} \xi_1^{\dag}\eta_1^{\dag}\right) V_{\mu\nu}^aV^{a\,\mu\nu}\\
&=&
\left[
\frac{1}{2}\left(
C_{V,1}^{\not \,s} + C_{V,2}^{\not \,s}
\right)\bar{\pmb{\chi}}\pmb{\chi} + 
\frac{1}{2}\left(
-C_{V,1}^{\not \,s} + C_{V,2}^{\not \,s}
\right)\bar{\pmb{\chi}}\gamma^5\pmb{\chi}
\right]V_{\mu\nu}^aV^{a\,\mu\nu}~.\nn
\end{eqnarray}
Clearly, only the first term is CP-preserving. CP-invariance, therefore, is preserved by imposing 
in Eq.~(\ref{eq:EffGluon1}) the condition $C_{V,1}^{\not \,s} = C_{V,2}^{\not \,s}$.
We refer to this operator with the notation $[_8\opf_{V}^{\not \,s}]_V$. Here, the only subscript refers to chirality in the DM current.
 In the following, we focus on the interactions with gluons, $V = G$.\\

\noindent
$\boxed{_8\mathcal{F}_V}$ This is a novel structure to this analysis and we shall discuss it more extensively.\begin{eqnarray}\label{eq:GluonsDerivative}
&&C_{V}\chi^\dagger\bar\sigma^{\mu}(\partial^{\nu}\chi) V_{\mu\rho}^aV^{a\,\rho}_\nu
\equiv
\left[C_{V,1} \xi_1^{\dag}\bar{\sigma}^{\mu}(\partial^{\nu}\xi_1) + C_{V,2} \eta_1^{\dag}\bar{\sigma}^{\mu}(\partial^{\nu}\eta_1)\right] V_{\mu\rho}^a V^{a\,\rho}_\nu + h.c.~\\
&&=\left[
\frac{1}{2}\left(C_{V,1} - C_{V,2}\right)\bar{\pmb{\chi}}\gamma^{\mu}(\partial^{\nu}\pmb{\chi}) + 
\frac{1}{2}\left(-C_{V,1} - C_{V,2}\right)\bar{\pmb{\chi}}\gamma^{\mu}\gamma^5(\partial^{\nu}\pmb{\chi})
\right] V_{\mu\rho}^a V^{a\,\rho}_\nu + h.c.~.
\end{eqnarray}
Considering explicitly the hermitian conjugation, we have the following structures
\begin{equation}
\begin{array}{ccc}
\left[C_V \bar{\pmb{\chi}}\gamma^{\mu}(\partial^{\nu}\pmb{\chi}) + C_V^*(\partial^{\nu}\bar{\pmb{\chi}})\gamma^{\mu}\pmb{\chi}\right]V_{\mu\rho}^a V^{a\,\rho}_\nu~,
 &   &  C_V \equiv \frac{1}{2}(C_{V,1} - C_{V,2})   \\ 
\left[C_A \bar{\pmb{\chi}}\gamma^{\mu}\gamma^5(\partial^{\nu}\pmb{\chi}) + C_A^*(\partial^{\nu}\bar{\pmb{\chi}})\gamma^{\mu}\gamma^5\pmb{\chi}\right]
V_{\mu\rho}^a V^{a\,\rho}_\nu~,
 &   &  C_A \equiv -\frac{1}{2}(C_{V,1} + C_{V,2})
\end{array}
\end{equation}
Now, under CP, $CP\{ V_{\mu\rho}^a V^{a\,\rho}_\nu\} = (-1)^{\mu}(-1)^{\nu}V_{\mu\rho}^a V^{a\,\rho}_\nu$ and $CP\{
\bar{\pmb{\chi}}\Gamma^{\mu}(\partial^{\nu}\pmb{\chi}) \mp (\partial^{\nu}\bar{\pmb{\chi}})\Gamma^{\mu}\pmb{\chi} 
\} = (\pm)(-1)^{\mu}(-1)^{\nu}[\bar{\pmb{\chi}}\Gamma^{\mu}(\partial^{\nu}\pmb{\chi}) \mp (\partial^{\nu}\bar{\pmb{\chi}})\Gamma^{\mu}\pmb{\chi}]$, with both $\Gamma^{\mu} = \gamma^{\mu}, \gamma^{\mu}\gamma^5$.
Imposing the restrictions $C_A = -C_A^* \to \mathbb{Re}(C_A) = 0$
and $C_V = -C_V^* \to \mathbb{Re}(C_V) = 0$ (which amount to take pure imaginary coefficients $C_{V,i=1,2}$ in Eq.~(\ref{eq:GluonsDerivative})),
we have, in the case of Dirac DM, two possible CP-invariant combinations
\begin{equation}
\begin{array}{ccc}
iC_V\left[\bar{\pmb{\chi}}\gamma^{\mu}(\partial^{\nu}\pmb{\chi}) - (\partial^{\nu}\bar{\pmb{\chi}})\gamma^{\mu}\pmb{\chi}\right]V_{\mu\rho}^a V^{a\,\rho}_\nu~,
 &   &   \hbox{CP-preserving, Dirac and Majorana DM} \\ 
iC_A\left[\bar{\pmb{\chi}}\gamma^{\mu}\gamma^5(\partial^{\nu}\pmb{\chi}) - (\partial^{\nu}\bar{\pmb{\chi}})\gamma^{\mu}\gamma^5\pmb{\chi}\right]
V_{\mu\rho}^a V^{a\,\rho}_\nu~,
 &   &   \hbox{CP-preserving, Dirac DM}\label{eq:GluonD2}
\end{array}
\end{equation}
The effective operator in Eq.~(\ref{eq:GluonsDerivative}) is present also for the case of Goldstino DM.
Because of the Majorana nature of the Goldstino (see footnote~\ref{footnote:MajoranaCJustTheNote}) 
we are left with only one CP-invariant combination in Eq.~(\ref{eq:GluonD2}).
In the following, we refer to Eq.~(\ref{eq:GluonD2})  with the notation $[_8\mathcal{F}_V]_{V,A}^{\mathcal{D}}$ for the two operators with Dirac DM, and 
 $[_8\mathcal{F}_V]_{V}^{\mathcal{M}}$ for the only structure present in the Majorana case.\\
 
{\bf Relic density.} Let us discuss in more detail why for the vector operators  $[_8\mathcal{F}_V]_{V}^{\mathcal{D},\mathcal{M}}$ annihilation in p-wave is allowed,
while for the operator $[_8\mathcal{F}_V]_{A}^{\mathcal{D}}$ only annihilation in d-wave is possible, as mentioned in the text.
Our argument goes as follows.
First, note that the traceless tensor gluon operator has $J = 2$. Conservation of total angular momentum imposes $J = 2$ also in the initial state.
Since the total spin of the two annihilating DM particle is either $S= 0$ or $S=1$, we have three possibilities. 
If $S=0$, from $J=2$ it follows that the only allowed value of total orbital angular momentum is $L=2$ (d-wave). 
Note that in this case charge conjugation and parity are, respectively, $C = 1$, $P = 1$. 
If $S = 1$, 
we have two cases since the condition $J = 2$ restricts the value of total orbital angular momentum to $L = 1$ (p-wave, with $C = 1$, $P = 1$), $L = 2$ (d-wave, with $C = 1$, $P = -1$). Now, the 
DM current in the operator  Eq.~(\ref{eq:GluonD2}), describes the annihilation of two DM particles with ingoing momenta $k_{1,2}$. In momentum space
\begin{equation}\label{referto}
\left[\bar{\pmb{\chi}}\Gamma^{\mu}(\partial^{\nu}\pmb{\chi}) - (\partial^{\nu}\bar{\pmb{\chi}})\Gamma^{\mu}\pmb{\chi}\right]~~\Longrightarrow~~ (k_1 - k_2)^{\nu}\bar{\pmb{\chi}}\Gamma^{\mu}\pmb{\chi}~,~~
~~~\hbox{with}~~\Gamma^{\mu} = \gamma^{\mu}, \gamma^{\mu}\gamma^5~.
\end{equation}
It is possible to extract the velocity-dependence of the two factor $(k_1 - k_2)^{\nu}$ and $\bar{\pmb{\chi}}\Gamma^{\mu}\pmb{\chi}$ separately. 
The kinematic in the initial state implies $k_{1,2} = (\sqrt{s}/2,0,0, \pm \sqrt{s}v_{\rm rel}/4)$, with $s = 4m_{\chi}^2/(1- v_{\rm rel}^2/4)$.
As a consequence, only the spatial part of $(k_1 - k_2)^{\nu}$ is non-zero, and we have $(k_1 - k_2)^{i} \sim v_{\rm rel}$.
As far as the DM current is concerned, 
in the non-relativistic limit the velocity-independent terms are $\bar{\pmb{\chi}}\gamma^j\pmb{\chi}$ and  $\bar{\pmb{\chi}}\gamma^0\gamma^5\pmb{\chi}$. All in all,
a p-wave contribution to the annihilation cross section  can be generated only by the combinations
$(k_1 - k_2)^i \bar{\pmb{\chi}}\gamma^{j}\pmb{\chi}$ and $(k_1 - k_2)^i \bar{\pmb{\chi}}\gamma^{0}\gamma^5\pmb{\chi}$.
Under CP transformation we have $CP\{\bar{\pmb{\chi}}\gamma^j\pmb{\chi}\} = 1$, $CP\{\bar{\pmb{\chi}}\gamma^0\gamma^5\pmb{\chi}\} = -1$; 
as a consequence, the axial-vector structure cannot give a p-wave cross section (since, as discussed above, it would require $CP = 1$). We thus reproduce Eqs. (\ref{eq:RelicGluon1}) and (\ref{eq:RelicGluon2}).

\subsection{Scalar DM: Effective operators in matrix form}\label{app:GoldstiniEffOp}
\noindent
$\boxed{_6{\cal S}^V_{\psi}}$ At $D=6$, the effective coupling with SM fermions is
\begin{eqnarray}\label{eq:D6Scalar}
c_{\psi}^V\phi^\dagger\partial_{\mu}\phi \psi^\dagger\bar\sigma^\mu \psi \!\!&\equiv& \!\!
\phi^\dagger(\partial_{\mu}\phi)
\left(
c_{\psi,1}^V \xi_2^{\dag}\bar{\sigma}^{\mu}\xi_2 + c_{\psi,2}^V \eta_2^{\dag}\bar{\sigma}^{\mu}\eta_2
\right) + h.c.\\
&=&\!\!\!\left[
c_V \phi^{\dag}(\partial_{\mu}\phi) + (c_V)^* (\partial_{\mu}\phi^{\dag})\phi
\right]\bar{\Psi}\gamma^{\mu}\Psi 
+
\left[
c_A \phi^{\dag}(\partial_{\mu}\phi) + (c_A)^* (\partial_{\mu}\phi^{\dag})\phi
\right]\bar{\Psi}\gamma^{\mu}\gamma^5\Psi~,\nn
\end{eqnarray}
where we defined $c_V \equiv (c_{\psi,1}^V - c_{\psi,2}^V)/2$, $c_A \equiv (-c_{\psi,1}^V - c_{\psi,2}^V)/2$.
We now impose CP-invariance: as already discussed, on the SM side we have  $CP\left\{
\bar{\Psi}\Gamma^{\mu}\Psi
\right\} = (-1)(-1)^{\mu}\bar{\Psi}\Gamma^{\mu}\Psi$, and on the DM side, $CP\left\{
\phi^{\dag}(\partial^{\mu}\phi) - (\partial^{\mu}\phi^{\dag})\phi
\right\} = (-1)(-1)^{\mu}[\phi^{\dag}(\partial^{\mu}\phi) - (\partial^{\mu}\phi^{\dag})\phi]$. 
Imposing  $c_V = -(c_V)^* \to \mathbb{Re}(c_V) = 0$
and $c_A = -(c_A)^* \to \mathbb{Re}(c_A) = 0$,
we have two possible CP-invariant combinations
\begin{eqnarray}
 ic_V \left[
\phi^{\dag}(\partial^{\mu}\phi) - (\partial^{\mu}\phi^{\dag})\phi
\right]  \bar{\Psi}\gamma^{\mu}\Psi~,  \label{eq:ComplexScalarEff}\,\quad
 ic_A \left[
\phi^{\dag}(\partial^{\mu}\phi) - (\partial^{\mu}\phi^{\dag})\phi
\right]  \bar{\Psi}\gamma^{\mu}\gamma^5\Psi~,  \nn
\end{eqnarray}
We refer to these two operators with the notation $[_6{\cal S}^V_{\psi}]_{V,A}^{\mathcal{C}}$ (the additional upper
index $\mathcal{C}$ refers to the complex scalar nature of DM); these operators vanish if DM is a real scalar.\\

\noindent 
$\boxed{_8\mathcal{S}_{\psi}^T }$ There are two possible combinations
\begin{equation}\label{eq:ScalarD8}
\partial^{[\mu}\phi^\dagger\partial^{\nu]}\phi \psi^\dagger\bar\sigma_\mu D_\nu\psi~,~~~\partial^{\{\mu}\phi^\dagger\partial^{\nu\}}
\phi \psi^\dagger\bar\sigma_\mu D_\nu\psi~,
\end{equation}
with $\partial^{[\mu}\phi^\dagger\partial^{\nu]}\phi \equiv \partial^{\mu}\phi^{\dag}\partial^{\nu}\phi - \partial^{\nu}\phi^{\dag}\partial^{\mu}\phi$, and 
 $\partial^{\{\mu}\phi^\dagger\partial^{\nu\}}\phi \equiv \partial^{\mu}\phi^{\dag}\partial^{\nu}\phi + \partial^{\nu}\phi^{\dag}\partial^{\mu}\phi$.
 Using the EoM and some algebra of Dirac matrices it is possible to recast the antisymmetric operator in the following form~\cite{Grzadkowski:2010es}
\begin{equation}
\partial^{[\mu}\phi^\dagger\partial^{\nu]}
\phi \psi^\dagger\bar\sigma_\mu D_\nu\psi~~\Longrightarrow~~
(\partial_{\rho}\phi^{\dag})\lra{\partial^{\mu}}(\partial^{\rho}\phi)\psi^{\dag}\bar{\sigma}_{\mu}\psi~.
\end{equation}
This operator shares the same symmetries as the  operator $\phi^\dagger\partial_{\mu}\phi \psi^\dagger\bar\sigma^\mu \psi$ (in particular, note that both vanish  considering the case of real DM -- that is the case S2 in section~\ref{sec:analysis}).
Even without restricting the analysis to the special case of real DM, it is  
easy to realize that the two operators 
$\phi^\dagger\partial_{\mu}\phi \psi^\dagger\bar\sigma^\mu \psi$ and 
$(\partial_{\rho}\phi^{\dag})\lra{\partial^{\mu}}(\partial^{\rho}\phi)\psi^{\dag}\bar{\sigma}_{\mu}\psi$
contribute to a given observable in the same way. 
For definiteness, let us consider the scattering process $\bar{\psi}\psi \to \phi^{\dag}\phi$, with outgoing momenta $p_{1,2}$ in the final state.
At $D=6$ -- considering the first operator above -- the amplitude is proportional to $(p_1 - p_2)_{\mu}\psi^{\dag}\bar{\sigma}^{\mu}\psi$; 
on the contrary, at $D=8$, the operator 
$(\partial_{\rho}\phi^{\dag})\lra{\partial^{\mu}}(\partial^{\rho}\phi)\psi^{\dag}\bar{\sigma}_{\mu}\psi$ generates an amplitude 
proportional to $p_1\cdot p_2 (p_1 - p_2)_{\mu}\psi^{\dag}\bar{\sigma}^{\mu}\psi$. As anticipated, it follows that the two operators contribute to the analyzed observable 
in the same way -- the operator at $D=8$ being a $\mathcal{O}(E^2/M^2)$ correction if compared to the one at $D=6$. 

We now turn our attention to the symmetric operator in Eq.~(\ref{eq:ScalarD8}).
In component, we have
\begin{equation}
C^T_{\psi}\partial^\mu\phi^\dagger\partial^\nu\phi \psi^\dagger\bar\sigma_\mu D_\nu\psi \equiv 
\partial^{\{\mu}\phi^\dagger\partial^{\nu\}}\phi \left[
C^T_{\psi,1} \xi^{\dag}\bar{\sigma}_{\mu}(D_{\nu}\xi) + 
C^T_{\psi,2} \eta^{\dag}\bar{\sigma}_{\mu}(D_{\nu}\eta)
\right] + h.c.~.
\end{equation}
In more familiar four-component notation, we have 
\begin{equation}
\begin{array}{ccc}
\partial^{\{\mu}\phi^\dagger\partial^{\nu\}}\phi
\left[C_V \bar{\Psi}\gamma_{\mu}(\partial_{\nu}\Psi) + C_V^*(\partial_{\nu}\bar{\Psi})\gamma_{\mu}\Psi\right]~,
 &   &  C_V \equiv \frac{1}{2}(C^T_{\psi,1} - C^T_{\psi,2})   \\ 
\partial^{\{\mu}\phi^\dagger\partial^{\nu\}}\phi
\left[C_A \bar{\Psi}\gamma_{\mu}\gamma^5(\partial_{\nu}\Psi) + C_A^*(\partial_{\nu}\bar{\Psi})\gamma_{\mu}\gamma^5\Psi\right]~,
 &   &  C_A \equiv -\frac{1}{2}(C^T_{\psi,1} + C^T_{\psi,2})
\end{array}
\end{equation}
and two possible CP-invariant combinations
\begin{equation}\label{eq:ComplexScalarD8}
\begin{array}{ccc}
iC_V\partial^{\{\mu}\phi^\dagger\partial^{\nu\}}\phi \left[\bar{\Psi}\gamma_{\mu}(\partial_{\nu}\Psi) - (\partial_{\nu}\bar{\Psi})\gamma_{\mu}\Psi  \right]~,
 &   &   \hbox{CP-preserving, complex scalar DM} \\ 
iC_A \partial^{\{\mu}\phi^\dagger\partial^{\nu\}}\phi \left[\bar{\Psi}\gamma_{\mu}\gamma^5(\partial_{\nu}\Psi) - (\partial_{\nu}\bar{\Psi})\gamma_{\mu}\gamma^5\Psi  \right]~,
 &   &   \hbox{CP-preserving, complex scalar  DM}
\end{array}
\end{equation}
with real coefficients $C$.
In the case of real scalar DM, $\partial^{\{\mu}\phi^\dagger\partial^{\nu\}}\phi = 2 (\partial^{\mu}\phi)(\partial^{\nu}\phi)$.
We refer to the two operators in Eq.~(\ref{eq:ComplexScalarD8}) with the notation $[_8{\cal S}^T_{\psi}]_{V,A}$.
As a rule of thumb, the cross section for real scalar DM is four time larger if compared with the complex case.\\

\noindent{\bf Relic density.} The annihilation cross section has a d-wave suppression since
the tensor structure in the SM current implies $J = 2$ 
while 
the two annihilating scalar particles have $S = 0$, thus forcing $L = 2$ in the initial state in order to conserve the total angular momentum.
By direct computation, we find \eq{superspacial}.

\section{The Event Generation and Analysis workflow}\label{sec:event-gen}
As opposed to previous studies, where pure monojet events were studied \cite{Racco:2015dxa}, we recast a recent analysis by ATLAS \cite{Aad:2015zva} which allows for multiple jets. The cuts require at least one jet with a $p_T>120 \GeV$ and allow for any number of additional jets. In particular it allows for jets with a lower $p_T$ i.e. soft and/or collinear jets. It is well known that a generation at Matrix Element (ME) level cannot accurately generate events with soft or collinear jets. A two step generation is therefore inevitable. This section gives a detailed account of the data generation workflow. We will also discuss the implementation of the analysis and we will discuss some subtleties that are specific to a consistent EFT analysis. 

We used FeynRules 2.0 \cite{Alloul:2013bka} to create the model files. For Dirac fermions and scalars this is very straight forward. There are however difficulties with four fermion vertices including Majorana fermions. This problem has two possible solutions, both giving the same result. We can introduced a new very heavy mediator with zero decay width. Choosing the mass of this new particle very high we assure that introducing this new particle in the model does not alter the interaction and that our effective field theory picture still applies. Note that in this case we have to absorb the mass of the mediator in the vertex in order to be left with exactly the same structure as for the EFT. Or, we can run the simulation with Dirac fermions instead and keep track of the factor $4$ that arise as a difference in the cross section between Dirac and Majorana fermions. For the present work we chose the latter option.

The model files are passed to MADGRAPH5 \cite{Alwall:2011uj} interfaced with PYTHIA-6 \cite{Sjostrand:2006za}. MADGRAPH5 generates events at ME level which are then passed to PYTHIA-6 for parton showering and hadronisation.  As multiple jets are allowed by the selection cuts chosen in \cite{Aad:2015zva} we have to generate events with an arbitrary number of jets. In \cite{Aad:2015zva} it was pointed out that it is sufficient to only generate 0-, 1- and 2-jet events at ME level and let PYTHIA generate the remaining arbitrary number of jets. The matching and merging procedure which is necessary under this circumstances will veto a substantial part of the created events. In order to still get enough statistics and because the signal regions used in \cite{Aad:2015zva} include very high missing $E_T$ we have perform a binned data generation. For the present study we chose three bins in $H_T$: $H_T^1<250 \GeV<H_T^2<600 \GeV<H_T^3$. We simulate 50k events in each bin, resulting in 150k events per DM mass. The scan over the mass of the DM is performed in an interval from $1\GeV$ to $1\TeV$. 

For the analysis we need to have access to the events at parton and reconstructed level. We need the event at parton level because we have to determine the relevant energy of the event, which, as discussed in the text, we chose to be $\sqrt{\hat{s}}$. The information at the reconstructed level is used for the recast of the ATLAS search \cite{Aad:2015zva}, where the selection cuts and the signal regions are defined in terms of the observables at the reconstructed level. MadAnalyis5 \cite{Conte:2012fm} provides the framework needed for this analysis. In MadAnalysis expert mode we can have access to both levels of information at the same time. An example of the main and analyzer C++ files can be found on \href{https://github.com/tarendo/LastGaspDMEFT}{GitHub}.

\begin{figure}[htb]
      \begin{center}
     \includegraphics[width=0.5\textwidth]{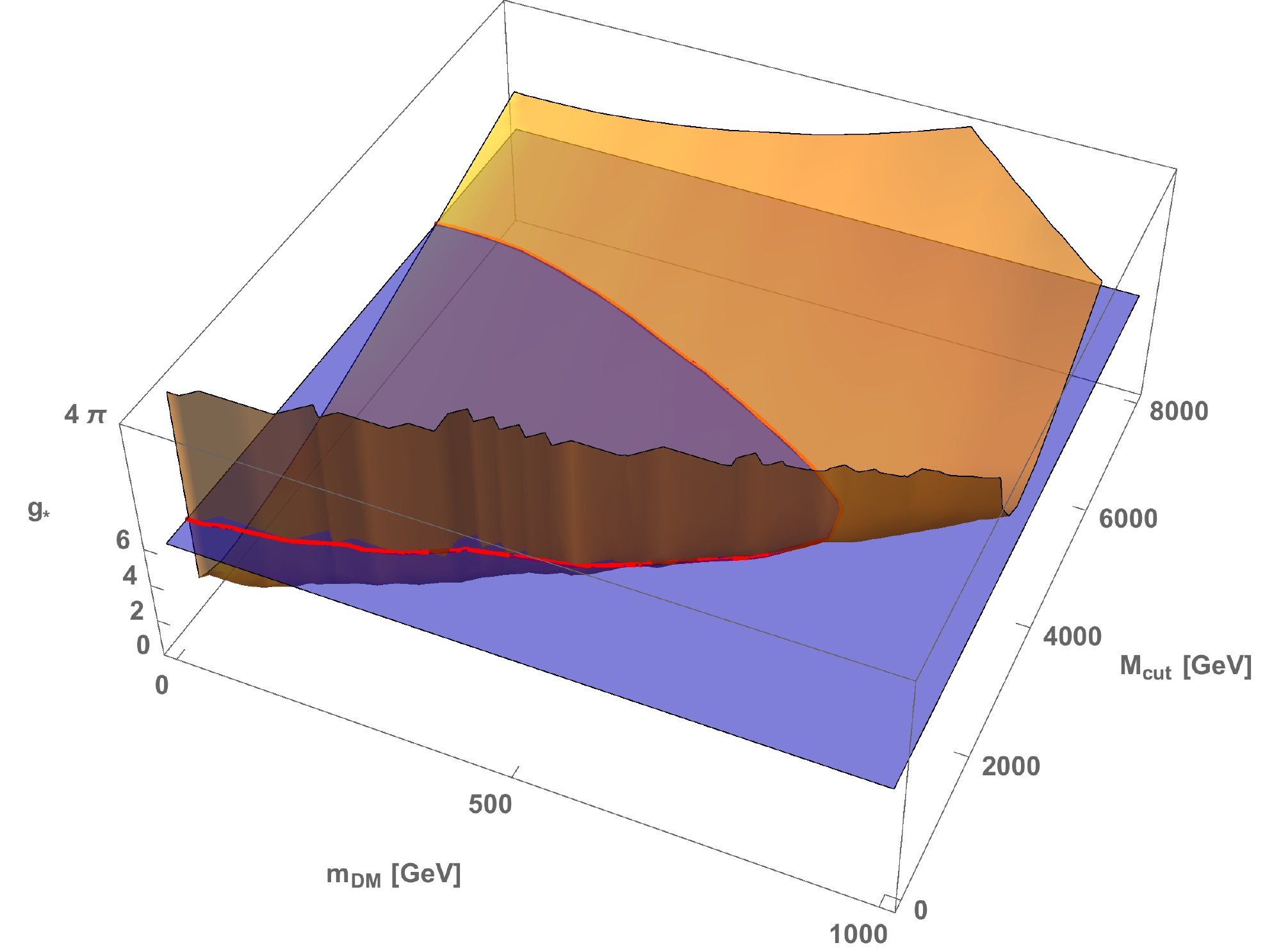}
      \end{center}
    \caption[K-factor]{\em The yellow surface shows the constraints on $g_*$. Values larger (above) the yellow surface are excluded. We can infer the constraints on the physical scale of the dark mediator by intersecting the yellow surface with the blue surface representing a fixed value of $g_*$.}
    \label{fig:ExclusionProcedure}
\end{figure}

 For each value of the DM mass we scan over $M_{cut}$ ranging from $10\GeV$ to $8\TeV$. This results in a consistent limit on $g_*$ for each value of $m_{DM}$ and $M_{cut}$ which can then easily be inverted. By assuming that the physical scale of new physics $M$ is also the scale $M_{cut}$ up to which we can trust the EFT expansion (i.e. we saturate the EFT validity requirement), we obtain a consistent limit on $M$ for each value of the DM mass and the coupling $g_*$. Fig.~\ref{fig:ExclusionProcedure} gives a graphical account of the procedure. The figure also shows very nicely the need for strongly coupled theories in the context of collider DM searches.

\end{document}